\documentclass[aps,pra,twocolumn,groupedaddress,nofootinbib,notitlepage,showpacs,floatfix,superscriptaddress]{revtex4-1}
\usepackage{graphicx,graphics,epsfig,subfigure,times,bm,bbm,amssymb,amsmath,amsfonts,amsthm,mathrsfs,MnSymbol}
\usepackage{epstopdf}
\usepackage[matrix,frame,arrow]{xypic}
\usepackage[pdfstartview=FitH]{hyperref}
\usepackage{overpic}
\hypersetup{
    colorlinks=true,       	% false: boxed links; true: colored links
    linkcolor=blue,          	% color of internal links
   citecolor=blue,        % color of links to bibliography
    filecolor=magenta,      	% color of file links
    urlcolor=cyan,           	% color of external links
    runcolor=cyan
}
\usepackage{braket}%Dirac Notation in QM
\usepackage{enumerate}
\usepackage[normalem]{ulem}
\usepackage{subfigure}
\usepackage{overpic}
\usepackage{dsfont}

\usepackage{multirow}
\newcommand{\be}{\begin{equation}}
\newcommand{\ee}{\end{equation}}
\newcommand{\ba}{\begin{eqnarray}}
\newcommand{\ea}{\end{eqnarray}}
\newcommand\tr{{\mbox{Tr\,}}}
\newcommand{\ignore}[1]{}
\newcommand{\avg}[1]{\left< #1 \right>}

\begin{document}

\title{Optimal control of Rydberg lattice gases}

\author{Jian Cui}
\affiliation{ Institute for complex quantum systems \& Center for Integrated Quantum Science and Technology (IQST),Universit\"at Ulm, Albert-Einstein-Allee 11, D-89075 Ulm, Germany}

\author{Rick van Bijnen }
\affiliation{  Max-Planck-Institut f\"ur Physik komplexer Systeme, N\"othnitzer Strasse 38, 01187 Dresden, Germany}
\affiliation{Institut f\"ur Quantenoptik und Quanteninformation, Technikerstr. 21a, 6020 Innsbruck, Austria}

\author{Thomas Pohl}
\affiliation{ Max-Planck-Institut f\"ur Physik komplexer Systeme, N\"othnitzer Strasse 38, 01187 Dresden, Germany}
\affiliation{Department of Physics and Astronomy, Aarhus University, Ny Munkegade 120, DK 8000 Aarhus C, Denmark}

\author{Simone Montangero}
\affiliation{ Institute for complex quantum systems \& Center for Integrated Quantum Science and Technology (IQST),Universit\"at Ulm, Albert-Einstein-Allee 11, D-89075 Ulm, Germany}
\affiliation{Theoretische Physik, Universit\"at des Saarlandes, D-66123 Saarbr\"ucken, Germany}

\author{Tommaso Calarco}
\affiliation{ Institute for complex quantum systems \& Center for Integrated Quantum Science and Technology (IQST),Universit\"at Ulm, Albert-Einstein-Allee 11, D-89075 Ulm, Germany}

\date{\today }

\begin{abstract}

We present optimal control protocols to prepare different many-body quantum states of Rydberg atoms in optical lattices. Specifically, we show how to prepare highly ordered many-body ground states, GHZ states as well as some superposition of symmetric excitation number Fock states, that inherit the translational symmetry from the Hamiltonian, within sufficiently short excitation times minimising detrimental decoherence effects. For the GHZ states, we propose a two-step detection protocol to experimentally verify the optimized preparation of the target state based only on standard measurement techniques. Realistic experimental constraints and imperfections are taken into account by our optimisation procedure making it applicable to ongoing experiments.
\end{abstract}

\maketitle

\section{Introduction}
Quantum simulation and quantum information processing crucially rely on the ability to create precisely controllable multipartite quantum systems, with designed Hamiltonians and low decoherence rates compared to experimental time scales. 
Ultracold atoms in optical lattices, laser-coupled to high-lying Rydberg states, provide an appealing platform for engineering such quantum systems. 
Optical potentials trapping the atoms provide highly flexible control over spatial geometries~\cite{Bloch08, Nogrette14}, with lattice sites that can be loaded with single atoms with near-unit fidelity~\cite{Barredo16, Endres16}. Quantum gas microscopes represent an established technology for observing the quantum state of individual atoms within the lattices~\cite{Qgas}. 

Strong and tunable long-range interactions between atoms across lattice sites can be established by laser-coupling them to Rydberg states, with interaction strengths that can be far in excess of all other energy scales in the system~\cite{SaffmanReview, LoewReview}.
A striking consequence is the so-called Rydberg blockade~\cite{Jaksch00, Lukin01}, 
which was succesfully employed to entangle pairs of atoms~\cite{Gaetan09, Urban09, Jau16}, as well as ensembles of atoms~\cite{Heidemann07, Dudin12, Ebert14, Ebert15, Weber15, Zeiher15, Labuhn16}.
Rydberg-excited atoms in lattice geometries can be described with Ising spin models~\cite{Weimer08, Sela11,Pohl10, Lesanovsky11}, which have recently seen impressive experimental confirmation~\cite{Barredo14, Labuhn16, Zeiher16}. Extended spin models can be realised by adding exchange interactions through coupling of multiple Rydberg levels~\cite{Ditzhuijzen08, Ravets14, Fahey15,Barredo15,Guenter13, Glaetzle15, Bijnen15}, or by introducing controlled dissipation~\cite{Lee11, Hoening13, Marcuzzi14, Sanders14, Hoening14, Helmrich16, Overbeck16, Roghani16}. Finally, even a general purpose Rydberg quantum simulator~\cite{Weimer10qsim} and quantum annealer~\cite{Annealer} have been proposed.

Evidently, Rydberg atoms hold high promise for applicability in quantum information processing and quantum simulation.
Yet, thus far most experimental investigations have been limited to studying \textit{dynamics} of Rydberg-excited systems, while previously predicted interesting ground state physics and associated quantum phase transitions~\cite{Weimer08, Weimer10, Pohl10, Lesanovsky11, Sela11, Glaetzle14} remain largely unexplored. 
The primary limiting factor preventing observation of many-body ground states is the finite lifetime of the Rydberg states~\cite{LoewReview}. Although Rydberg atoms boast relatively long lifetimes of up to  tens of microseconds~\cite{Beterov09}, it is still a very stringent requirement that the typically complex ground state preparation scheme is executed well before a single decay event occurs. 
Preliminary experimental success has been achieved in preparing `crystalline' states of regularly spaced Rydberg excitations on a 1D chain of atoms~\cite{Schauss15}. These experiments effectively probed the first few steps of a full Devil's staircase, i.e. the stepwise increase of the Rydberg atom number in the
many-body ground state with increasing laser detuning or system size, that characterises the ground state phase diagram of a lattice gas with power-law interactions~\cite{BakBruinsma}. The experiment in Ref.~\cite{Schauss15} employed a carefully designed adiabatic pulse scheme~\cite{Pohl10, Schachenmayer10, Bijnen11, Petrosyan16}, slowly evolving the initial ground state with no Rydberg excitations into the desired crystalline state. 

An adiabatic state preparation scheme, however, has some inherent limitations. Firstly, it has to be executed slowly compared with the minimum energy gap \textit{by definition}, which is directly at odds with the previously stated neccessity of performing the state preparation as fast as possible. Secondly, many-body states  that are not adiabatically connected to a trivial initial state are out of reach of adiabatic preparation.
To overcome these limitations, we turn to the tools of Optimal Control (OC)~\cite{Khaneja2005296, krotov,controlmethod,OC_tutorial, reviewOC}. Stimulated by earlier succeses of OC in quantum information processing~\cite{OC_quantuminformation, Gate, Montangero2007, PhysRevA.70.012306,BoseSpinChain,Cui_Mintert,ESU,Koch11,Koch14}, and the design of many-body quantum dynamics~\cite{PhysRevLett.103.240501, CRAB, PhysRevA.84.012312}, as well as the successful applications in experiments~\cite{ExpClosedLoop,Exp_atom-chip,Exp_Mintert,Exp_atom}, especially those with Rydberg atoms \cite{Koch11,RydbergOC2011,Koch14,RydbergOptimalControl16,RydbergOC2016},
we adopt the ``chopped random basis" (CRAB) and dressed CRAB (dCRAB) optimal control method~\cite{CRAB, Lloyd2014a,dCRAB} for quantum state preparation in Rydberg lattice gases.
We will showcase three typical examples: (i) crystalline states of regularly spaced excitations~\cite{Schauss15} as a prominent and experimentally relevant example of the Rydberg blockade effect, (ii) GHZ states with maximal multipartite entanglement, relevant for quantum information processing tasks~\cite{GHZ_teleportation,GHZ_teleportationExp,
GHZ_teleportationanddensecoding,GHZ_QKD,GHZ_swapping,}, and (iii) an arbitrary superposition state, for which no other preparation method is known so far, demonstrating the generality of our method.

The paper is organized as follows. In Sec.~\ref{sec:Basic} we provide a description of the Rydberg system under study, as well as an outline of the relevant experimental considerations. Sec.~\ref{sec:Crystalline} demonstrates the results for the Rydberg crystalline state preparation and the obtained excitation staircase. In Sec.~\ref{sec:GHZ} we show the optimized dynamics for creating and detecting a GHZ state which encodes the qubits in groups of atoms collectively sharing an excitation, complemented by an arbitrary quantum superposition state preparation scheme described in Sec.~\ref{sec:Arbitrary}. Finally, Sec.~\ref{sec:Discussion} summarizes the paper and provides an outlook on exploring the so-called quantum speed limit of state preparation in Rydberg atoms.

\section{Basic description}
\label{sec:Basic}
\begin{figure}[h!]
\includegraphics[width=0.8\columnwidth]{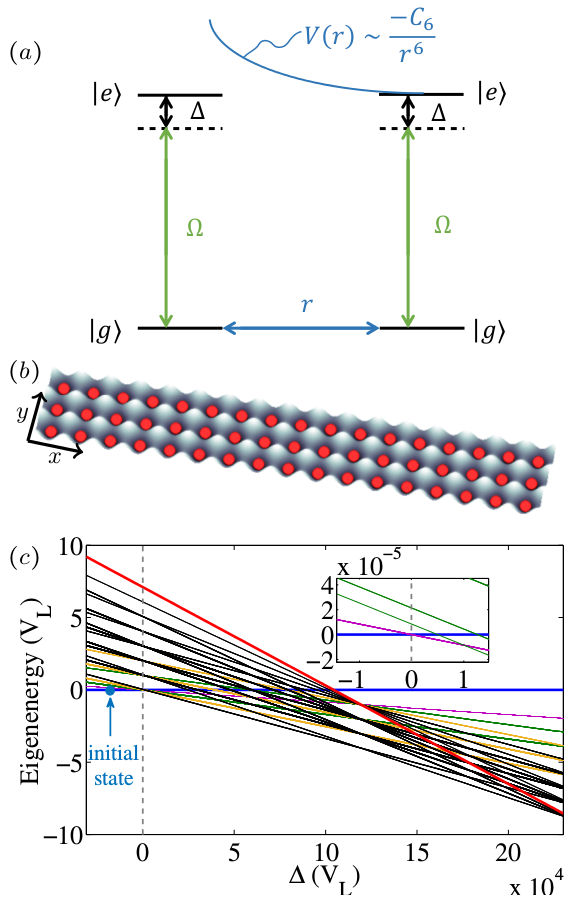}
\caption{Rydberg atomic gas. (a) Level scheme of two $^{87}Rb$ atoms in optical lattice sites. The atomic ground state $\ket{g}$ is coupled, with Rabi frequency strength $\Omega$, to an excited Rydberg state $\ket{e}$. The laser is detuned by $\Delta$. The two atoms are separated by a distance $r$; the blue curve represents their mutual energy shift due to van der Waals interaction.
(b) The unit-filling optical lattice is tailored into a $3\times N$ bar shape for the crystalline state preparation. The strong repulsive van der Waals interactions result in the Rydberg blockade effect with a blockade radius approximately $8a$ such that each group of 3 atoms on the $y$ axis effectively forms a super-atom. Such a system can be described as a one-dimensional chain along the $x$ direction of $N$ super-atoms with $\sqrt{3}$ enhanced Rabi coupling. 
(c) Energy spectrum of an $N=8$ atom chain in the classical limit ($\Omega = 0$), plotted as a function of the detuning in units of $V_L = C_6 / L^6$, which is the interaction energy between atoms located at opposite ends of the chain, with $L=(N-1)a$ the length of the chain. The dashed vertical line marks the phase transition point $\Delta=0$. Each eigenstate of the system has a well-defined total number of excitations $N_{\rm e}$, indicated by a color code: blue ($N_{\rm e} = 0$), magenta ($N_{\rm e} = 1$), green ($N_{\rm e} = 2$), yellow ($N_{\rm e} = 3$), black ($N_{\rm e} = 4$ to $7$), red ($N_{\rm e} = N$). For the ground states, these excitations are regularly spaced, minizing the interaction energy and forming a crystalline state. The inset shows a zoom of the low-lying spectrum near the quantum phase transition point and the first ground-state level-crossing point.}
\label{fig:sketch}
\end{figure}

The system we consider is composed of a two-dimensional lattice with one atom per site, which can be realized experimentally either in an optical lattice~\cite{NatExp,Schauss15} or in an array of optical dipole traps~\cite{Nogrette14}, or even in dense disordered gases by targeted laser excitation \cite{Bijnen15b}. Given the short time scales considered in this paper and other works in the literature \cite{Younge09,Pohl10,Lesanovsky11,RydbergGHZ_Pohl,Barredo14,Hoening13,Hoening14,Bijnen15,Schauss15,Zeiher16,Lee11,Marcuzzi14,Sanders14,NatExp}, only the internal electronic degrees of freedom are considered. Initially the system is prepared in the Mott insulating phase in which every atom is in its electronic ground state $\ket{g}$. Laser light couples the atomic ground state $|g\rangle$ to a high-lying Rydberg state $|e\rangle$ with a Rabi frequency $\Omega$ and frequency detuning $\Delta$, as illustrated in Fig.~\ref{fig:sketch}(a). Experimentally, such Rydberg state transitions can either be driven by a two-photon transition via a low-lying intermediate state~\cite{LoewReview} or by a direct single-photon transition~\cite{Weber15, Zeiher16, Jau16}.
In the present calculations we focus on the specific situation of previous lattice experiments~\cite{NatExp,Schauss15} where Rubidium atoms have been excited to $43S_{1/2}$ Rydberg states via a far detuned intermediate $5P_{3/2}$ state with two laser beams. This essential state picture is well justified, as near-resonant state mixing \cite{Reinhard08,Younge09} can be neglected \cite{Pohl10}.

If two atoms at different lattice sites with positions ${\bf r}_i$ and ${\bf r}_j$ are excited to the Rydberg level, they experience strong van der Waals interactions, $V_{ij}=C_6/|{\bf r}_{i}-{\bf r}_j|^6$. For the selected $43S$ state the corresponding $C_6=1.625 \times 10^{-60}\mathrm{J}\mathrm{m}^6$~\cite{C6,Schauss15}. 
The interaction between two ground-state atoms or between one ground- and one Rydberg atom is negligible \cite{InteractionComparat,Browaeys2016_reviewRydExp,SaffmanReview}. In the interaction picture, this system can be described by the Hamiltonian~\cite{Schauss15, Labuhn16}
\ba
H(t) &= &\frac{\hbar}{2}\Omega(t)\sum_{i}\big(\hat{\sigma}_{eg}^{(i)} + \hat{\sigma}_{ge}^{(i)} \big) + \sum_{i\neq j}\frac{V_{ij}}{2}\hat{\sigma}_{ee}^{(i)} \hat{\sigma}_{ee}^{(j)}  \nonumber \\
& & -\hbar\Delta(t)\sum_{i}\hat{\sigma}_{ee}^{(i)} ,
\label{Eq:Hamiltonian}
\ea
where the operators $\hat{\sigma}_{\alpha\beta}^{(i)} = |\alpha_i\rangle\langle \beta_i | $ denote the atomic transition and projection operators for the $i$th atom at position ${\bf r}_i$. We investigate the Rydberg atom excitation dynamics by integrating the Schr\"odinger equation governed by $H(t)$, employing a numerical approach described in ~\cite{Pohl10,Bijnen11}.

In Fig.~\ref{fig:sketch}(c) we show the spectrum of eigen-energy levels of the system described by the Hamiltonian~(\ref{Eq:Hamiltonian}) in the classical limit $\Omega=0$. In this case, all eigenstates are tensor products of excitation number Fock states on each site, i.e., many-body Fock states corresponding to a given spatial configuration of site-localized Rydberg excitations.

Increasing the laser detuning lowers the energy of the excited atomic state and, therefore, favours the excitation of Rydberg atoms as seen in Fig.~\ref{fig:sketch}(c). Therefore, the low-energy sector of the spectrum is composed of ordered Rydberg atom configurations which minimize the total interaction energy~\cite{Pohl10}.  

Accurate pulse shaping of the Rydberg excitation laser provides precise experimental control of both $\Omega(t)$ and $\Delta(t)$. This permits to steer the many-body quantum dynamics of the atomic lattice and to prepare specific many-body states starting from the simple initial state $|gg\ldots g\rangle$, with all atoms in their ground state. While the basic idea of this approach~\cite{Pohl10,Schachenmayer10, Bijnen11} has been demonstrated in recent experiments~\cite{NatExp,Schauss15}, preparation fidelities have remained limited by lattice imperfections and unavoidable transitions between the ground state and the low-lying excited many-body eigenstates of eq. (\ref{Eq:Hamiltonian}).
Here, we use optimal control techniques to mitigate such limitations.

We apply the dCRAB method to the preparation of crystalline states, GHZ states as well as an arbitrary superposition state in Rydberg atom lattices. In general the dCRAB method identifies the optimal temporal shapes of the control parameters, which have been expanded on a randomized truncated Fourier basis, through iteratively updating the coefficients of the basis functions using a numerical minimization (e.g. simplex) method, which enables to obtain better fidelities from iteration to iteration. 
In order to draw a close connection to ongoing experiments, we incorporate typical parameter constraints, limiting the Rabi frequency to $\Omega/2\pi\le 400$kHz~\cite{NatExp,Schauss15}, and imposing a truncation on the highest Fourier frequency for synthesising $\Omega(t)$ and $\Delta(t)$ at $8.3$MHz and $0.5$MHz,
which translate into a
minimum rise and fall time for $\Omega$ and $\Delta$ of $60$ns and $1000$ns, respectively. In this paper, we constrain the amplitude of $\Delta$ to be within $\pm 2$MHz as in the experiments~\cite{Schauss15}; however, this cutoff is not a fundamental limit. We will see later that even with this limitation we can prepare high-fidelity crystalline states and GHZ states, and if we allow for larger detunings $\Delta$ in the optimization, the results can only improve.  

Finally, in order to account for lattice defects, we consider an ensemble of $N_r=50$ realizations with a lattice filling fraction of $0.9$ in the optimization. We use the average fidelity, $F^C \equiv \overline{|\bra{\psi^C}{\psi(\tau)}\rangle|^2}$ and $F^G\equiv \overline{|\bra{\psi^G}{\psi(\tau)}\rangle|^2}$ for crystalline state $\ket{\psi^C}$ and the GHZ state $\ket{\psi^G}$, respectively, as the figure of merit for the optimization. Here the bars represent the ensemble average over $N_r$ realizations, and $\ket{\psi(\tau)}$ is the final state at time $\tau$. This choice for the figure of merit ensures that while the obtained control parameters do not just optimize certain individual configurations but yield an optimized average dynamics with a high degree of robustness with regards to lattice defects.  In this paper we neglect other sources of imperfections, such as dephasing due to instrumentation or stray fields, which were found to be of minor relevance under typical experimental conditions \cite{NatExp,Schauss15}.

\section{Crystalline State preparation}
\label{sec:Crystalline}
In order to prepare a crystalline state with a given number $N_{\rm e}$ of Rydberg excitations, one can drive the system through a sequence of level crossings by chirping the frequency detuning from negative to positive values as shown in Fig.~\ref{fig:sketch}(c).
Such a near-adiabatic modification of the low-energy many-body states~\cite{Pohl10,Schachenmeyer10, Bijnen11} has been demonstrated experimentally in~\cite{Schauss15}. However, a strictly adiabatic preparation of the absolute ground state is hampered by the finite lifetime of the excited Rydberg atoms, which limits the available evolution times. Consequently, slight crystal defects emerge from unavoidable transitions between the ground state and the low-lying excited many-body Fock states. In~\cite{Schauss15} the employed excitation pulses allowed to prepare an ordered quantum state of slightly delocalized Rydberg excitations, rather than the actual ground-state crystal consisting of a single Fock state component. 

\begin{figure}[h]
\includegraphics[width=0.49\columnwidth]{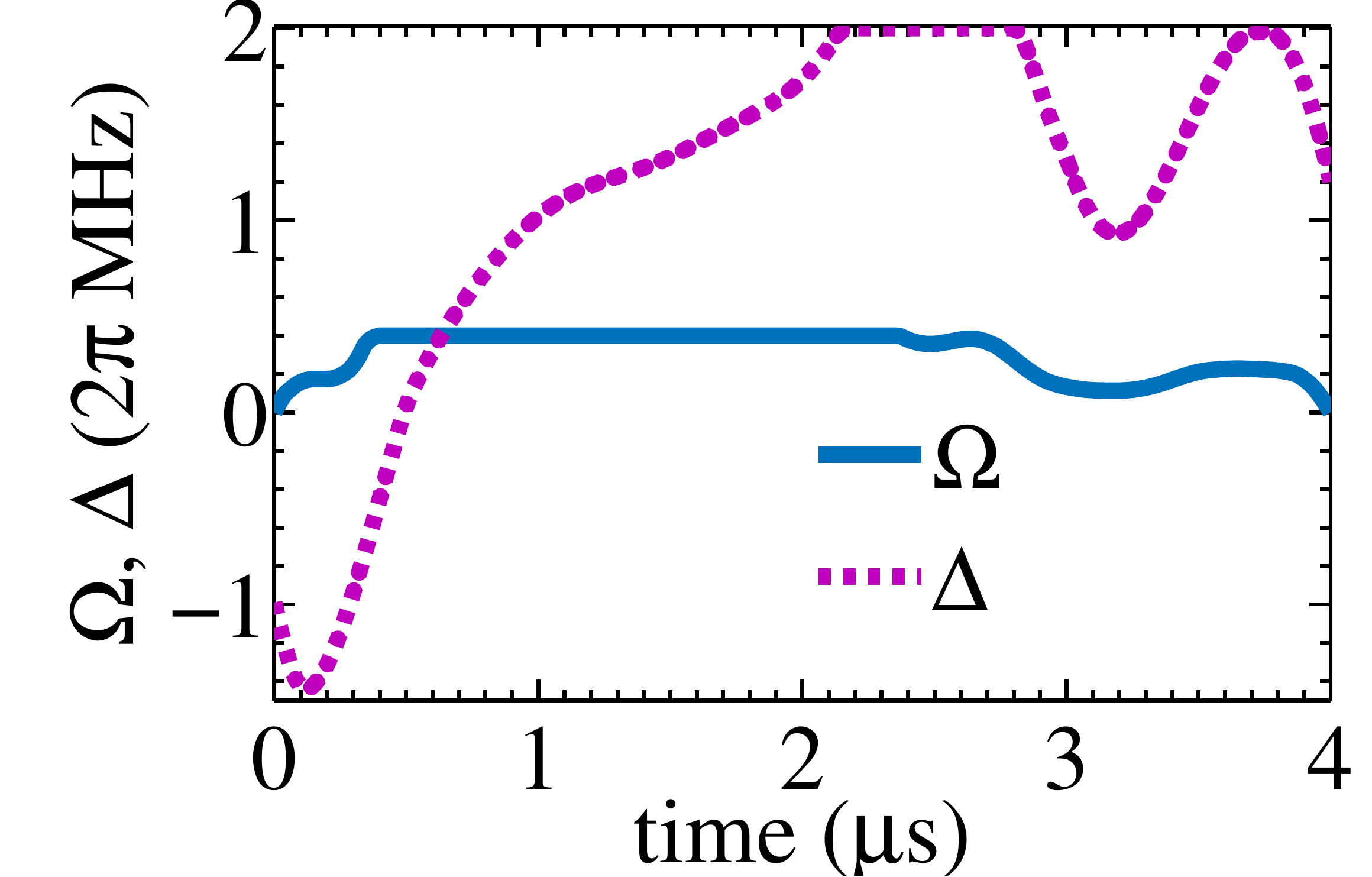}
\includegraphics[width=0.49\columnwidth]{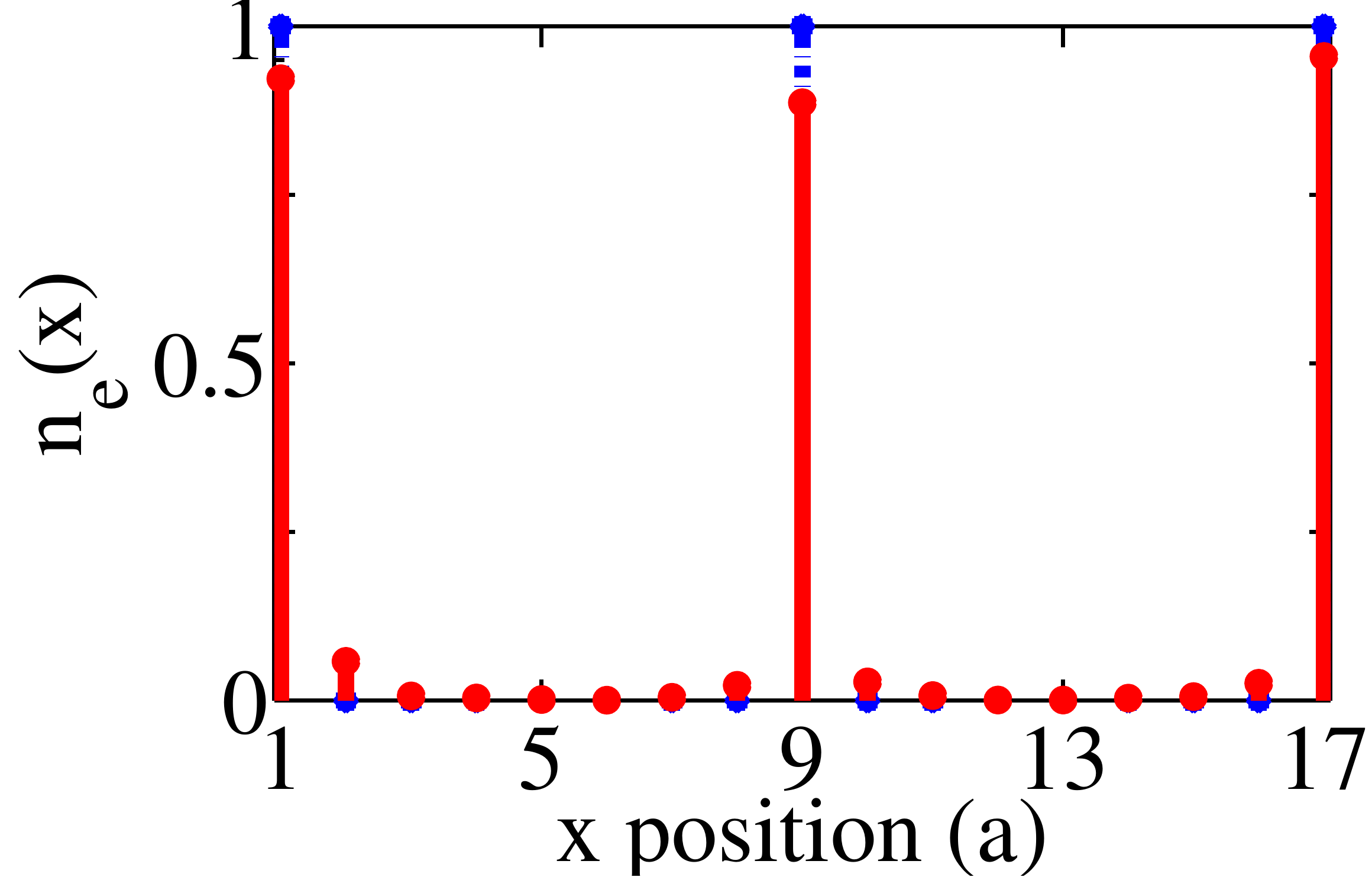}
\includegraphics[width=0.49\columnwidth]{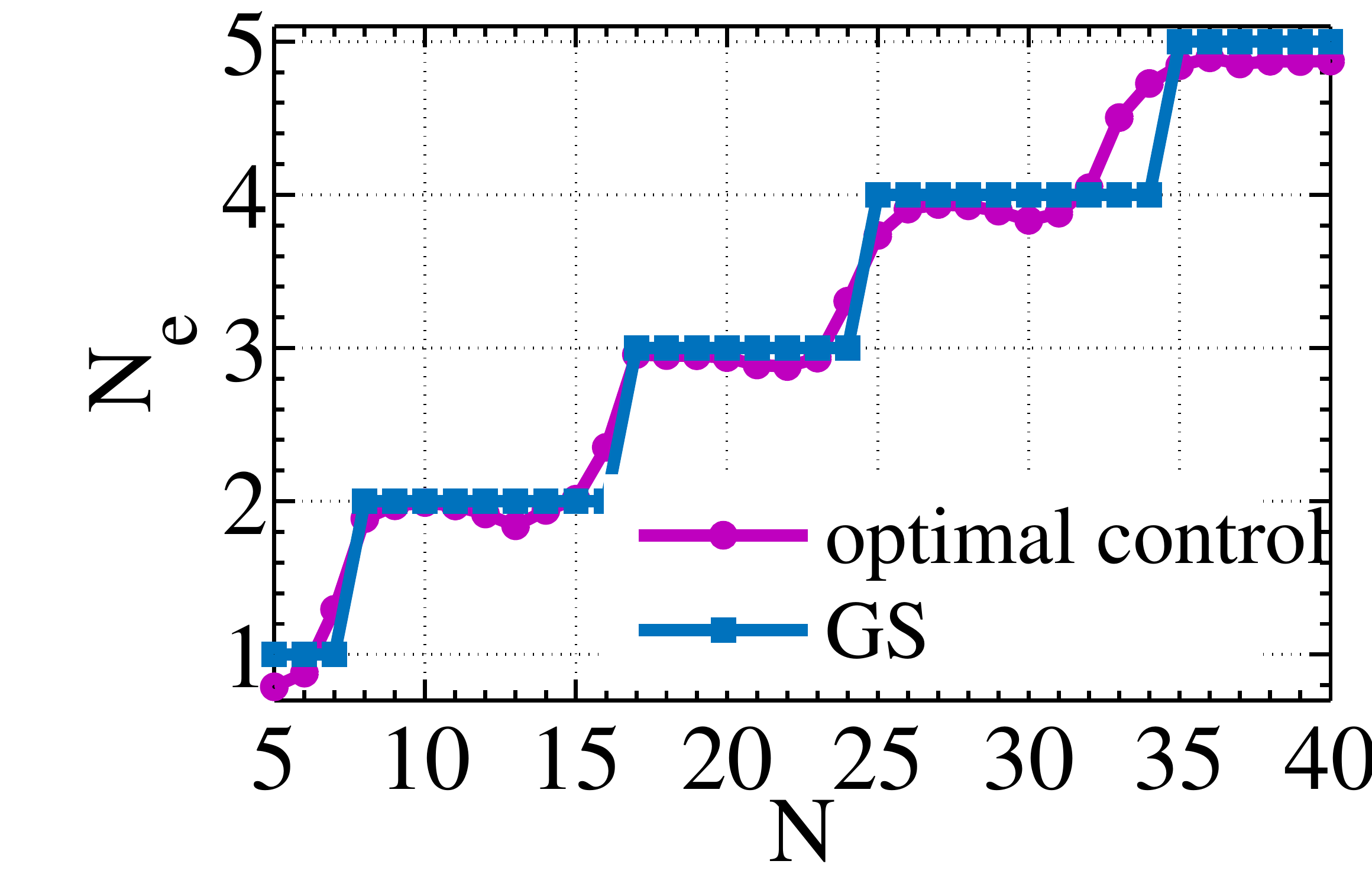}
\includegraphics[width=0.49\columnwidth]{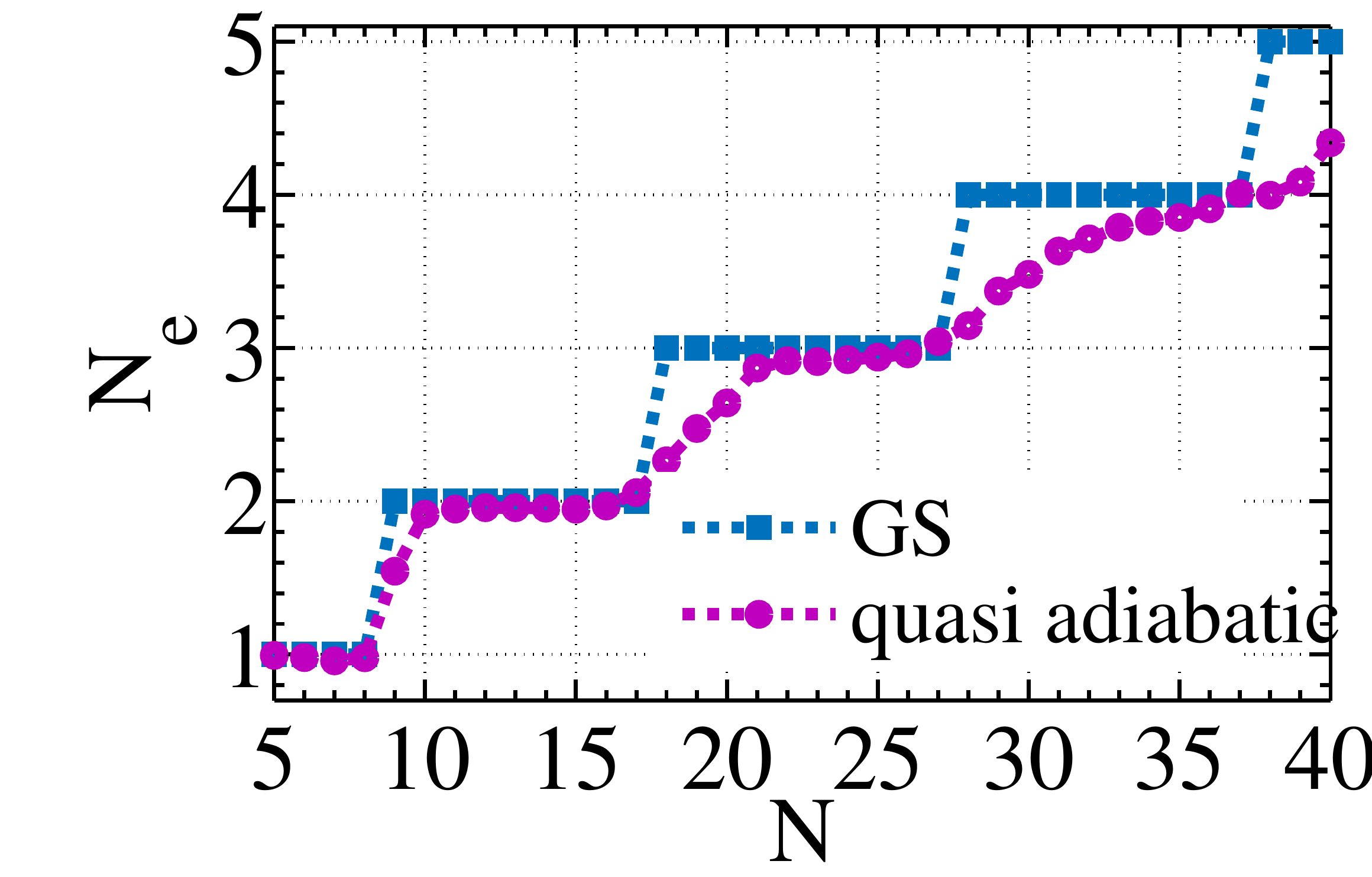}
  \begin{picture}(0,-1)
    \put(-219,145){$(a)$}
    \put(-97,145){$(b)$}
    \put(-219,65){$(c)$}
    \put(-97,65){$(d)$}
  \end{picture}
\caption{Rydberg crystalline state preparation. (a) Optimized control driving parameters. (b) Excitation density $n_e(x)\equiv \bra{e_x} \rho \ket{e_x} $  of the 3-excitation crystalline state (blue) and the average final state from dCRAB optimal control (red). The average fidelity $F^C$ is larger than $0.85$. (c,d) Total excitation number staircase as a function of the  lattice length for the classical states (blue curves with square markers), i.e., the ground states (GS) for $\Omega \rightarrow 0$ and at fixed final detuning $\Delta$, and the prepared states (purple curves with circle markers),  obtained by applying the pulses from optimized control (c) as well as from the quasi-adiabatic method of Ref. \cite{Schauss15} (d) to systems with different size $N$. }
\label{fig:N=17_10p_excitation}
\end{figure}
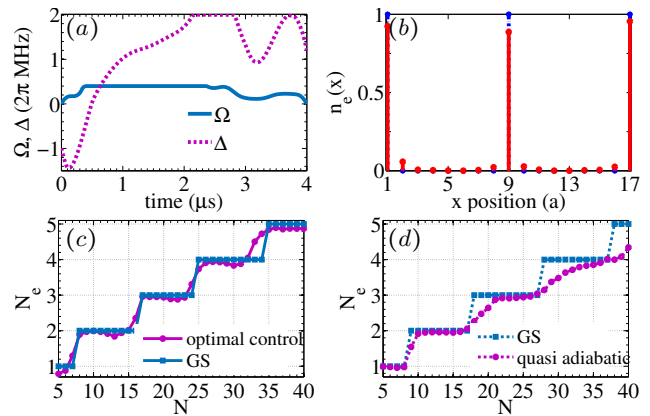

Below we demonstrate theoretically high-fidelity ground state preparation within experimentally relevant preparation times using optimal control. Following the experimental scenario of~\cite{Schauss15}, we consider a quasi-one-dimensional geometry in the form of a $3\times N$ lattice as illustrated in Fig.~\ref{fig:sketch}(b), where the lattice spacing $a=532$nm. Since the transverse extent is considerably smaller than the Rydberg blockade radius, this geometry behaves as a one-dimensional chain $N$ super-atoms and of length $L = (N - 1) a$ with a collectively enhanced Rabi frequency $\sqrt{3}\Omega$~\cite{Schauss15}.
As described in the previous section, our method accounts for possible lattice defects and therefore includes resulting fluctuations of the effective Rabi frequency the fluctuating number of atoms per super-atom.

In Fig.~\ref{fig:N=17_10p_excitation}(a) we show the pulse shape optimised via the dCRAB optimal control method~\cite{dCRAB} for the generation of a $3$-excitation crystal in a chain of $N=17$ qubits for an excitation pulse duration of $4\mu$s. 
The resulting Rydberg excitation density is nearly identical to that of a perfect three-atom crystal, as shown in Fig.~\ref{fig:N=17_10p_excitation}(b) where only very weak fluctuations around the optimal Rydberg atom positions occur. The quality of a prepared Rydberg crystalline state has been quantified through the total population of Fock states with given excitation number $n$, i.e., $P_n \equiv \avg{\sum_{i_1,\dots,i_n}\ket{e_{i_1}}\bra{e_{i_1}}\otimes \cdots \otimes \ket{e_{i_n}}\bra{e_{i_n}}}$~\cite{Schauss15}. 
A more stringent evaluation than $P_n$ is the state fidelity $F^{\rm G}$. 
 Our optimal control scheme reaches a high ground state average fidelity over $N_r$ imperfect realizations of $F^{\rm G}>0.85$ and a high final population $P_3=0.97$ of $3$-excitation Fock states. Notice that for this Rydberg lattice gas system the quasi-adiabatic scheme employed in Ref.~\cite{Schauss15} tends to obtain states with low fidelity but relatively high $P_n$, because of unavoidable transitions to the low-lying excited many-body Fock states. While these states have the correct number of excitations $n$, the excitations can be slightly displaced with respect to positions of the actual ground state.
Even though our protocol is run in an imperfectly prepared lattice with defects, the achieved fidelity yields a significant improvement over previous work for $n=3$, where $P_3 = 0.91$ could be achieved for an ideal lattice~\cite{Schauss15}. Note that these numbers can be further increased for higher Rabi frequencies, which are now available for single-photon Rydberg excitation as recently demonstrated in~\cite{Zeiher16}.

The enabled high preparation fidelity shows up most prominently in the so-called Rydberg blockade staircase~\cite{Pohl10}. As shown in Fig.~\ref{fig:N=17_10p_excitation}(c), this staircase appears as a stepwise increase of the Rydberg atom number of the many-body ground state, when increasing the system length while keeping all other parameters fixed.
In order to obtain the staircase, we apply the optimized control fields for the case of $N=17$ to systems of different length $N$. As detailed in \cite{Pohl10,Schauss15}, varying the chain length is practically equivalent to a rescaling of the applied detuning, $\Delta$, via a change of $V_L$( see Fig.~\ref{fig:sketch}c). Hence, one can effectively target many-body ground states with different excitation numbers upon changing the chain length for fixed  parameters of the excitation pulse.
As shown in Fig.~\ref{fig:N=17_10p_excitation}(c), our optimised preparation pulse yields sharp transitions between the different excitation numbers $N_{\rm e}$ and enables the high-fidelity preparation of ordered Fock states with $N_{\rm e}=5$. Both features represent significant improvements with respect to the excitation pulses employed in both theory and the experiment of Ref.~\cite{Schauss15}. For comparison, Fig.~\ref{fig:N=17_10p_excitation}(d) shows the numerical excitation staircase obtained by using the adiabatic pulse employed in Ref.~\cite{Schauss15}). 

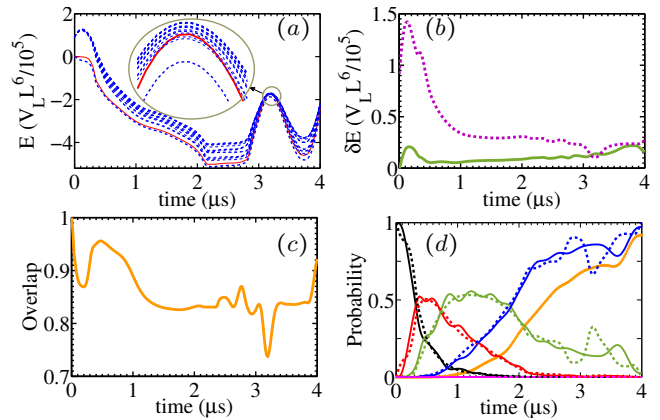
\begin{figure}[h]
\includegraphics[width=0.49\columnwidth]{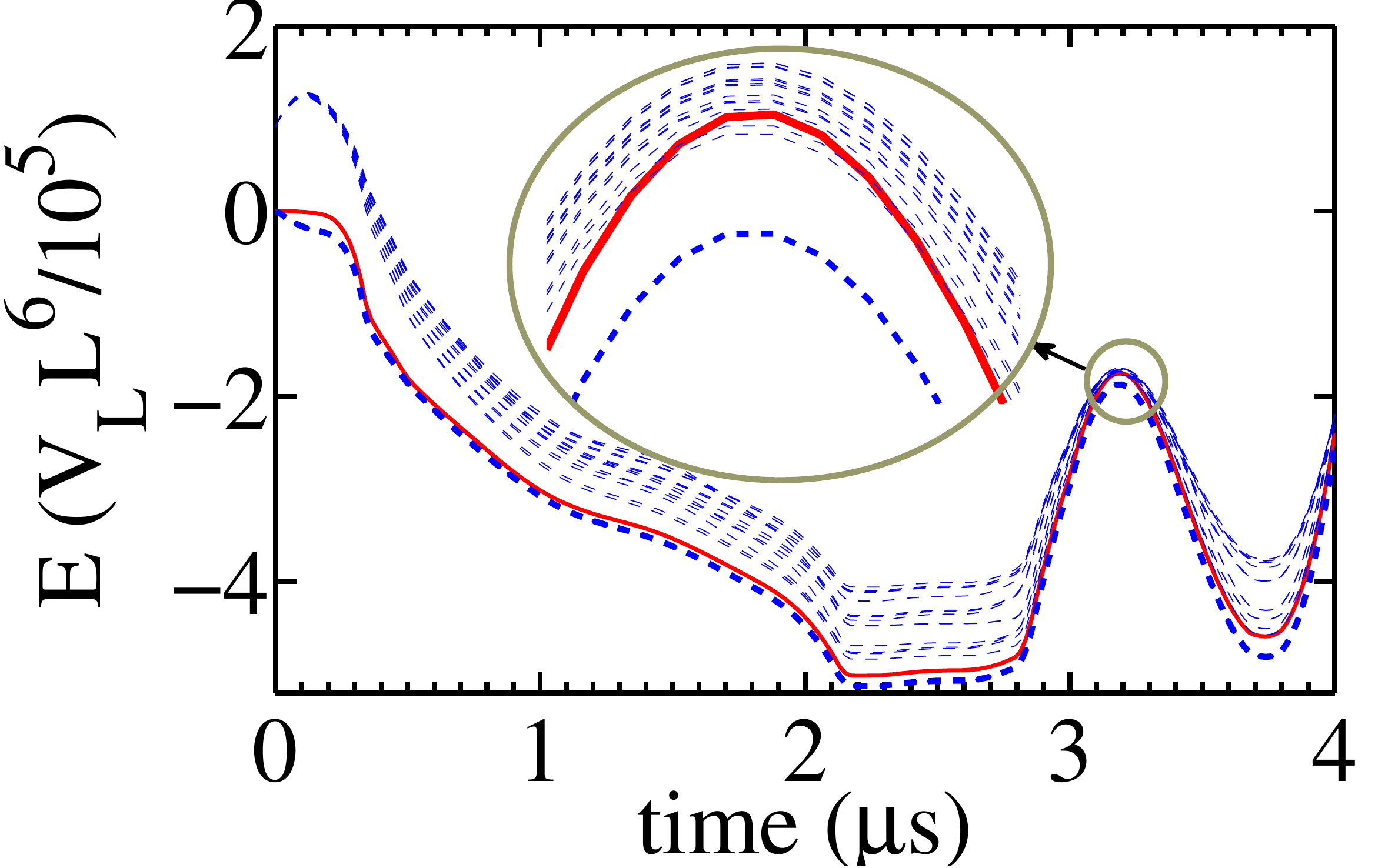}
\includegraphics[width=0.49\columnwidth]{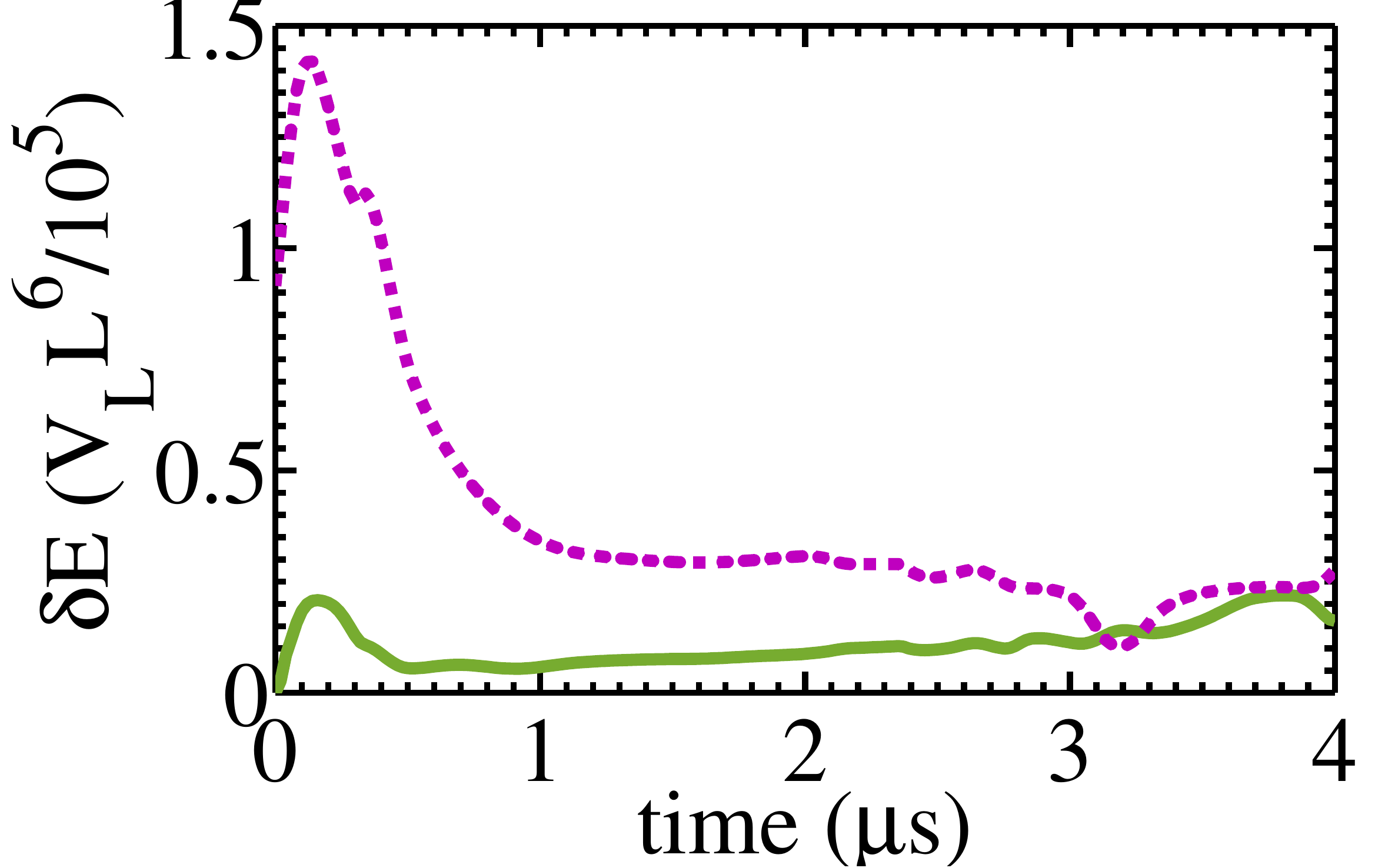}
\includegraphics[width=0.49\columnwidth]{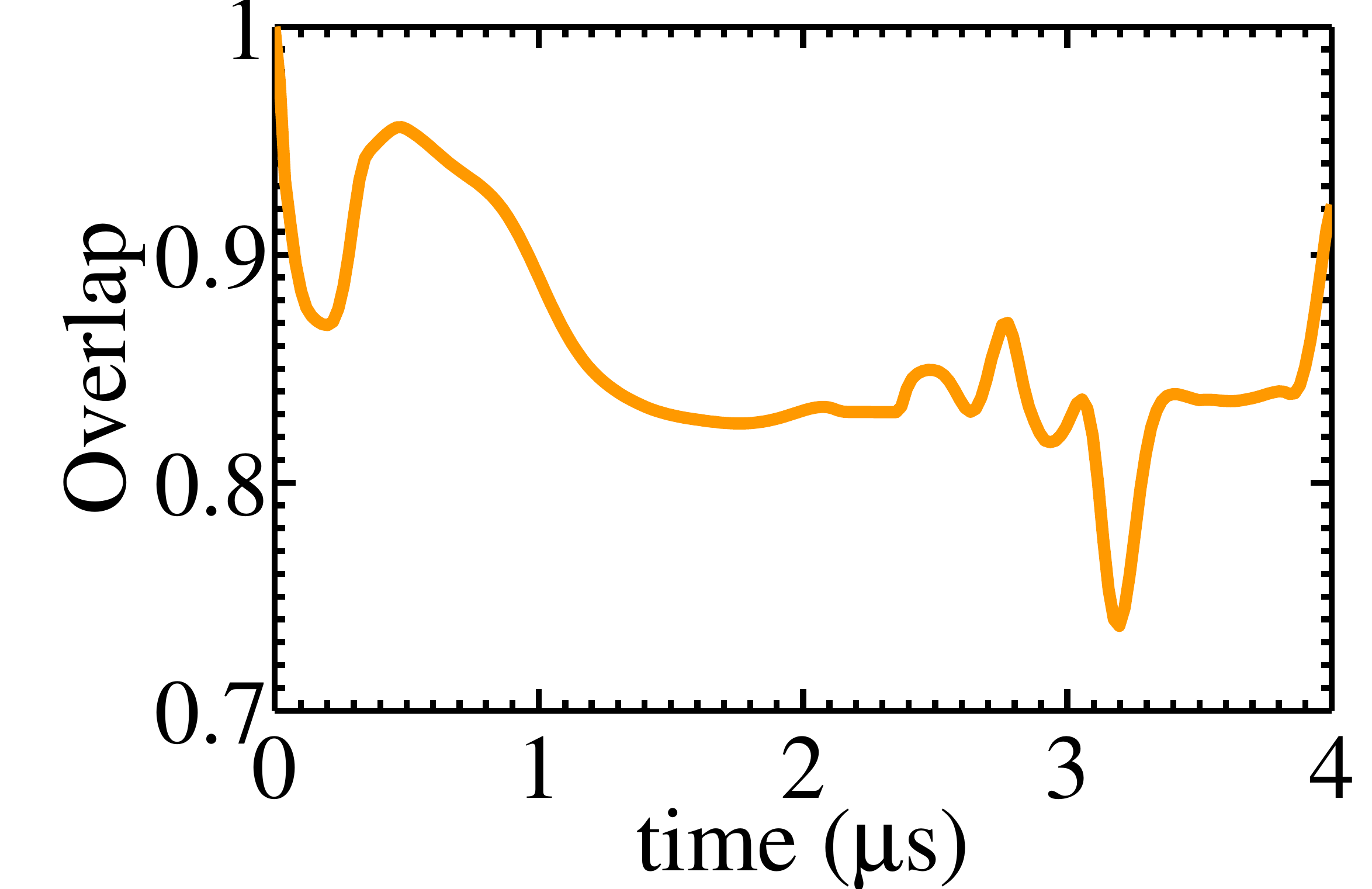}
\includegraphics[width=0.49\columnwidth]{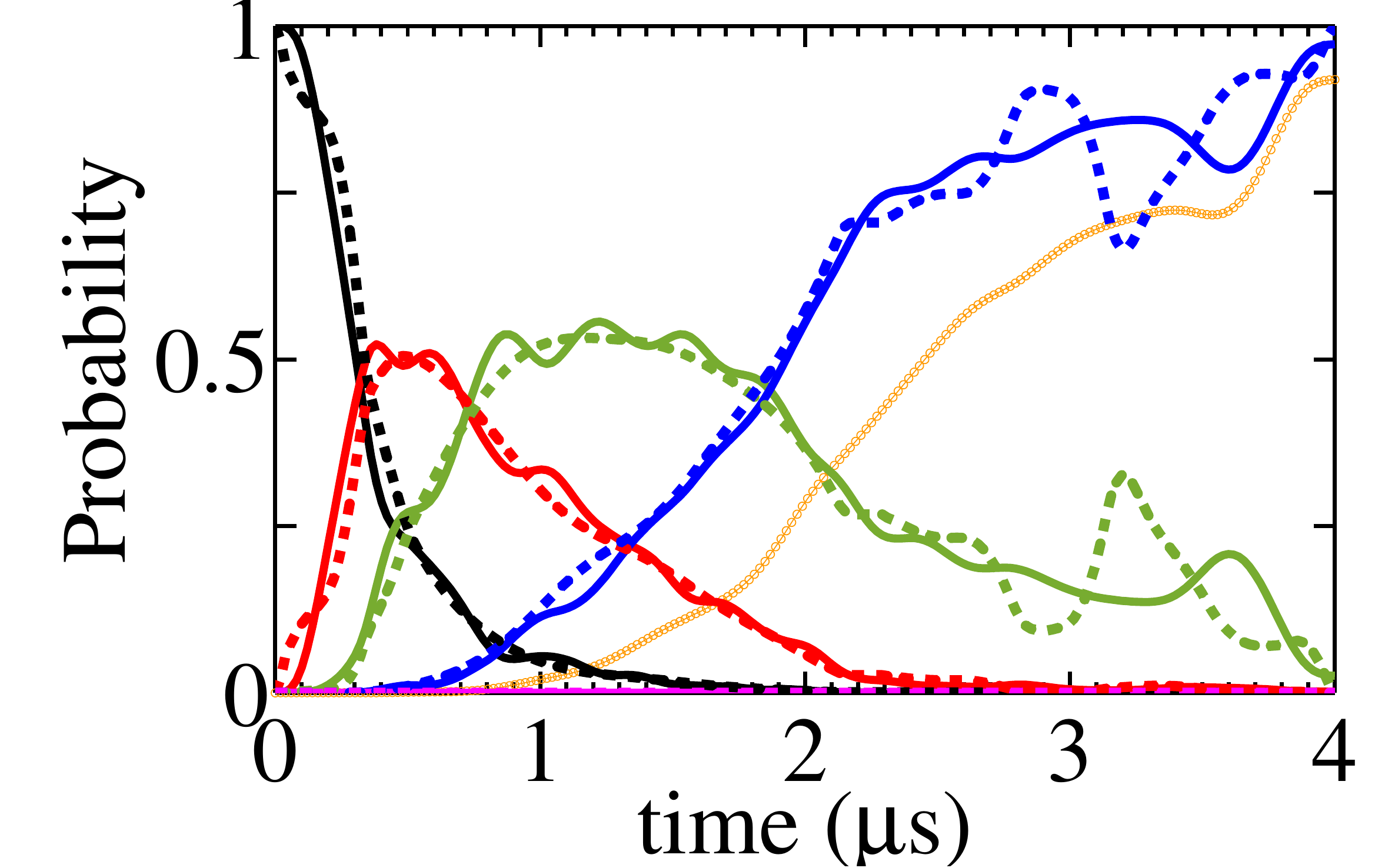}
  \begin{picture}(0,-1)
    \put(-144,145){$(a)$}
    \put(-88,145){$(b)$}
    \put(-144,63){$(c)$}
    \put(-88,63){$(d)$}
  \end{picture}
\caption{Dynamics of the Rydberg crystallization. 
(a) Low-lying energy spectrum of the laser-dressed system (blue dashed curve) and the energy of the instantaneous state from the optimized dynamics (red solid curve). We zoom in at the curves between $3.1$ and $3.3$ $\mu s$ in the inset, during which the energy of the optimized dynamics goes into the excited spectrum.
(b) The energy gap between the first excited state and ground state (magenta dashed curves) as well as the energy difference between the instantaneous state and the ground state (green solid curve).  In the time window from $3.1$ to $3.3$ $\mu s$ the red curve is above the lowest blue curves, which means the energy of the instantaneous state is higher than the first excited state.
(c) The overlap between the instantaneous state $\ket {\psi(t)}$ and the time-local ground state $\ket{\psi_G(t)}$. This overlap measures how close the optimized dynamics is to the adiabatic evolution.
(d) Fidelity ( orange curve with circle marks) and the probability of excitations with given numbers, $P_n $, for the instantaneous state (solid curves) and the time-local ground state (dotted curves) by color code: black, red, green, blue, magenta correspond to $n$ from $0$ to $4$.
}
\label{fig:crystallineDynamics}
\end{figure}

Fig.~\ref{fig:crystallineDynamics} illustrates the Rydberg excitation dynamics induced by our optimised laser pulse. As demonstrated by the time evolution of the energy [Fig.~\ref{fig:crystallineDynamics}(a)], energy gap [Fig.~\ref{fig:crystallineDynamics}(b)], the overlap between the instantaneous state and time-local ground state [Fig.~\ref{fig:crystallineDynamics}(c)] as well as the excitation number distribution and the instantaneous state fidelity [Fig.~\ref{fig:crystallineDynamics}(d)], the optimized system dynamics indeed remains near adiabatic and closely follows the instantaneous many-body ground state during the first $3\mu$s. This suggests that adiabatic preparation methods~\cite{Pohl10, Schachenmayer10, Bijnen11}  indeed provide a useful strategy for preparing low-energy many-body states~\cite{Schauss15}. However, the final stage of the optimised system dynamics significantly deviates from adiabaticity, which ultimately yields the enhanced ground state fidelity described above. Notice that the optimized control pulses presented here are robust against the lattice imperfections arising from non-unity filling of atoms. 
Decoherence process, e.g. Rydberg state radiative decay, only plays a minor role on a time scale of $4 \mu s$, as can be seen from the total decay probability $P_d(t) \equiv \int_0^{t} \Gamma N_{\rm e}(t')dt'$, where $N_{\rm e}(t') = \avg {\sum_i \ket{e_i}\bra{e_i}}$ is the total excitation number of the state at time $t'$, and $\Gamma = 11.8$kHz is the single atom radiative decay rate for the $43S$ state of $^{87}Rb$~\cite{decayrate}. For the optimized evolution the total decay probability at the final time is only $P_d(\tau)=0.1$.

Recent numerical work~\cite{Petrosyan16} pointed out that the preparation scheme employed in~\cite{Schauss15} would yield a rather low ground state fidelity $F^C \lesssim 0.2$ for the short pulse duration of $4\mu$s used in the experiment~\cite{Schauss15}. It was, hence, concluded that adiabatic crystal state preparation requires substantially longer excitations times at which dissipative processes would inevitably start to play a significant role~\cite{Petrosyan16}. The above results (see Fig.~\ref{fig:N=17_10p_excitation} and Fig.~\ref{fig:crystallineDynamics}), however, demonstrate that optimal control allows to alleviate this problem by facilitating high-fidelity ground state preparation for time scales for which the excitation dynamics remains highly coherent.

\section{GHZ state preparation and detection}
\label{sec:GHZ}
\begin{figure}[ht]
\includegraphics[width=0.99\columnwidth]{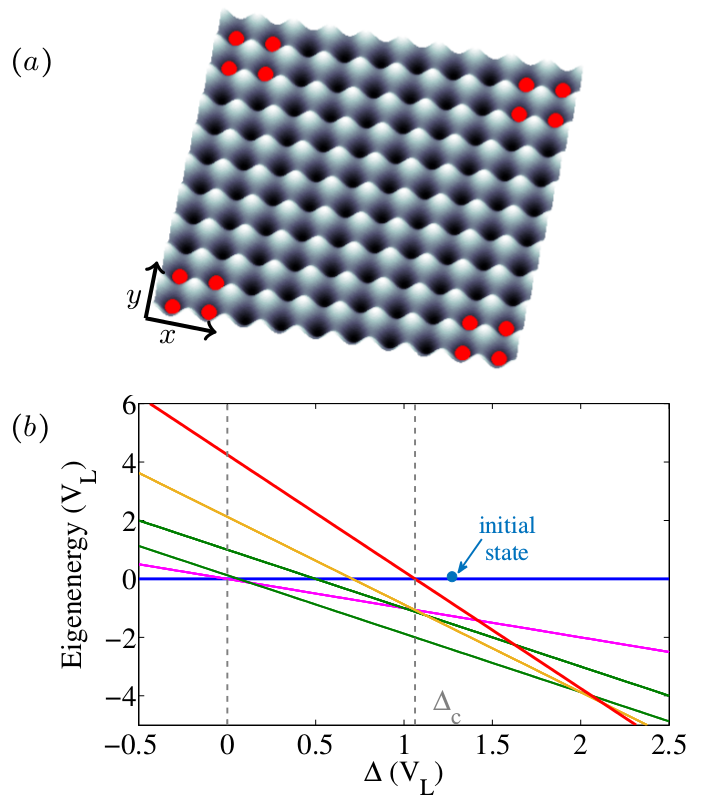}
\caption{Lattice geometry and spectrum for the GHZ state. 
(a) The lattice is tailored in such a way that four nearest neighbouring sites are filled with atoms to form a super-atom at each of the corners encoding a quantum bit, with the centers of the super-atoms separated by $8a$. 
(b) Energy spectrum of a $2 \times 2$ atomic array in the classical limit, as a function of detuning $\Delta$, with $V_L = C_6 / L^6$ the interaction energy along one edge of the lattice of length $L = 8a$. Color coding as in Fig.~\ref{fig:sketch}(c). The states $\ket{0000}$ and $\ket{1111}$ become resonant at a detuning $\Delta_c$ (defined in the main text). The second  gray line is guideline for the crossing detuning point $\Delta_c$.}
\label{fig:GHZsketch}
\end{figure}

Having demonstrated the power of optimal control techniques for preparing ordered low-energy states of Rydberg excitations, we now consider the high-energy region of the many-body energy spectrum. One area of particular interest lies around $\Delta_{\rm c }=N^{-1}\sum_{i < j}  V_{ij}/ \hbar$, as marked in Fig.~\ref{fig:GHZsketch}(b), where the $N$-atom ground state, $|G\rangle\equiv|g_1,g_2,...,g_{\rm N}\rangle$, becomes degenerate with the fully excited state $|E\rangle\equiv|e_1,e_2,...,e_{\rm N}\rangle$, which allows to generate maximally entangled GHZ states, $\ket{\psi^G}=(|G\rangle+{\rm e}^{i\theta}|E\rangle)/\sqrt{2}$~\cite{RydbergGHZ_Pohl}.

Due to the strong Rydberg-Rydberg atom interaction the preparation of such high energy states requires a different lattice geometry than that of the previous section. Specifically, we consider an optical lattice with the aforementioned parameters but filled in such a way~\cite{Weitenberg11} as to obtain $4$ qubits, each of which located at one corner of a $10 \times 10$ square lattice, see Fig. \ref{fig:GHZsketch}(a). In every corner, only $2\times 2$ lattice sites are filled with one atom each, in which only one Rydberg excitation can exist and be shared coherently by the $2\times 2$ sites because of the blockade effect thus encoding the $\ket{1}$ state for the qubit. The $\ket{0}$ state of one qubit corresponds to all its $4$ constituent atoms in the ground state. A collection of $N_{\rm bl}$ atoms ($N_{\rm bl} = 4$ in our example) in a blockade sphere is also called a ``superatom", featuring in addition a collective enhancement of the effective Rabi frequency with a factor of $\sqrt{N_{\rm bl}}$ ~\cite{SA2014,SA2015Fleischhauer,Zeiher15,SA2016Browaeys}.   
The large qubit spacing ensures a moderate interaction energy of $\hbar^{-1}C_{6}/(8a)^6=0.4125\times 2\pi$MHz for the $43S_{1/2}$ Rydberg state used in~\cite{NatExp,Schauss15}, while the use of multiple adjacent atoms reduces the detrimental effects of lattice defects as described above.

Because of their highly entangled nature, the preparation of GHZ states is much more sensitive to decoherence processes than that of the classical crystalline states discussed in the previous section. In particular, a single Rydberg state decay would completely decohere a prepared GHZ state and project the system onto a separable state. Avoiding such undesired effects once more requires very short operation times, i.e. it calls for optimised preparation pulses.

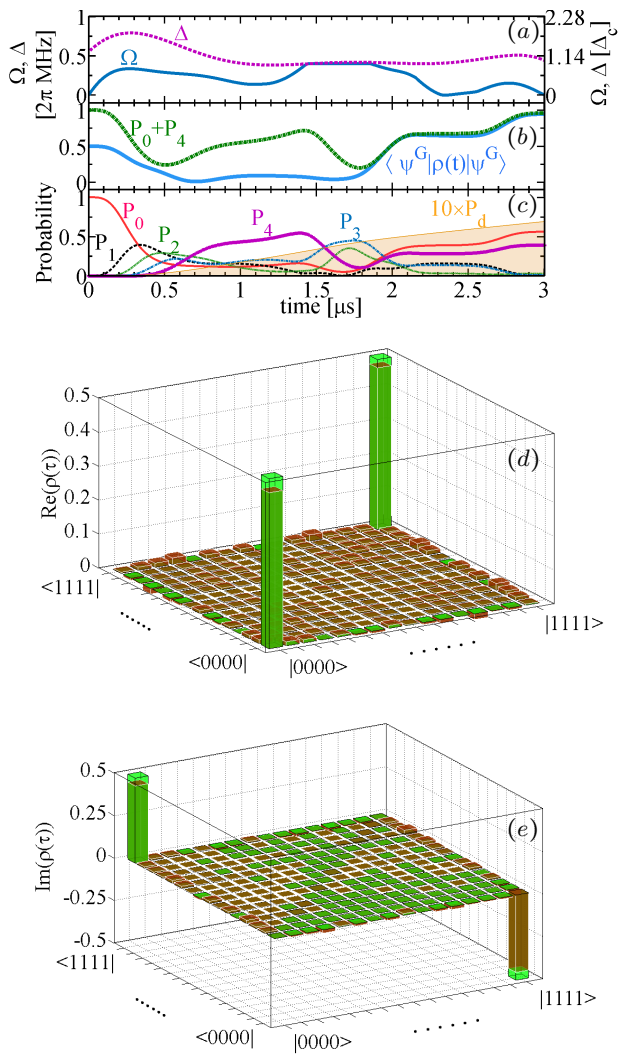
\begin{figure}[t]
\includegraphics[width=\columnwidth]{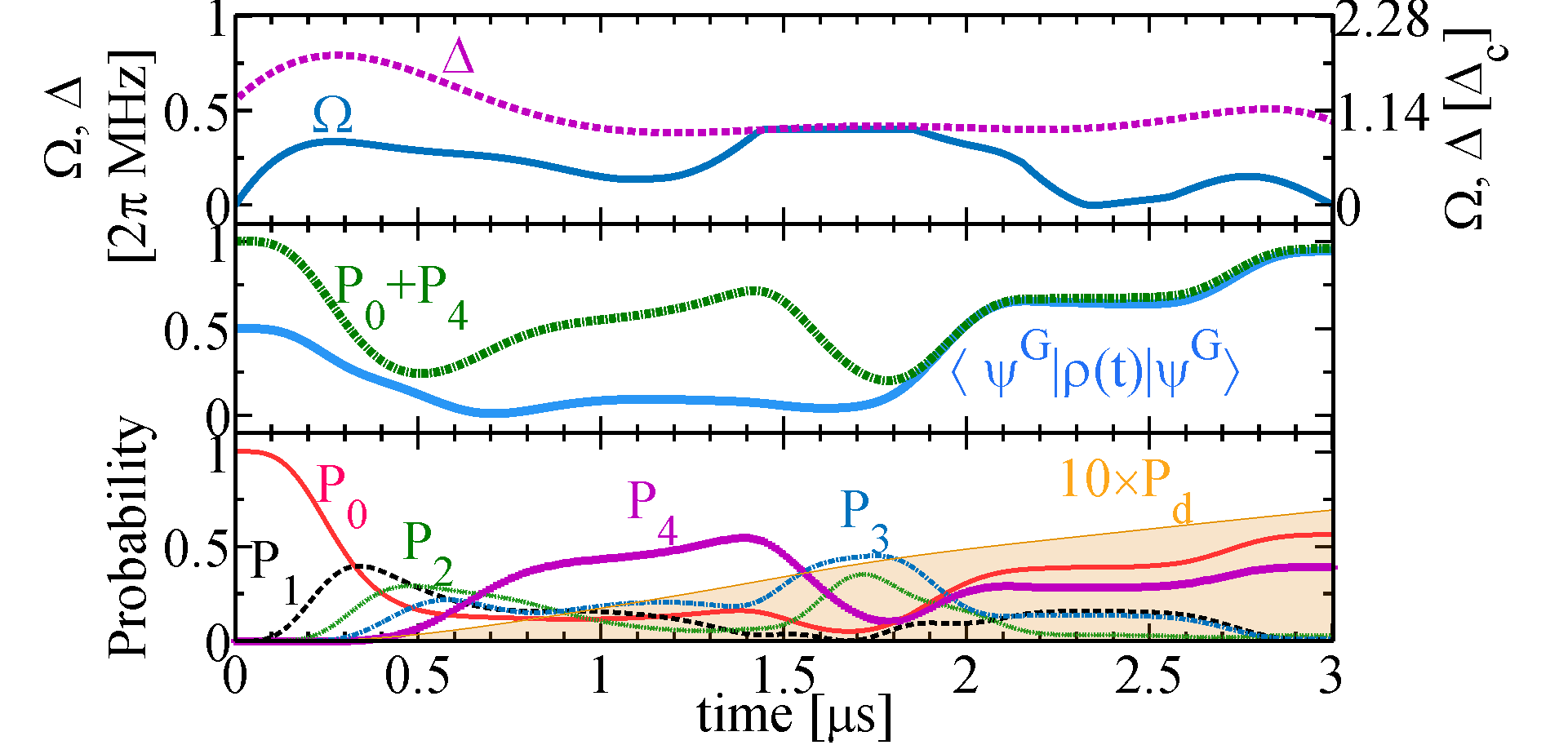}
\includegraphics[width=0.9\columnwidth]{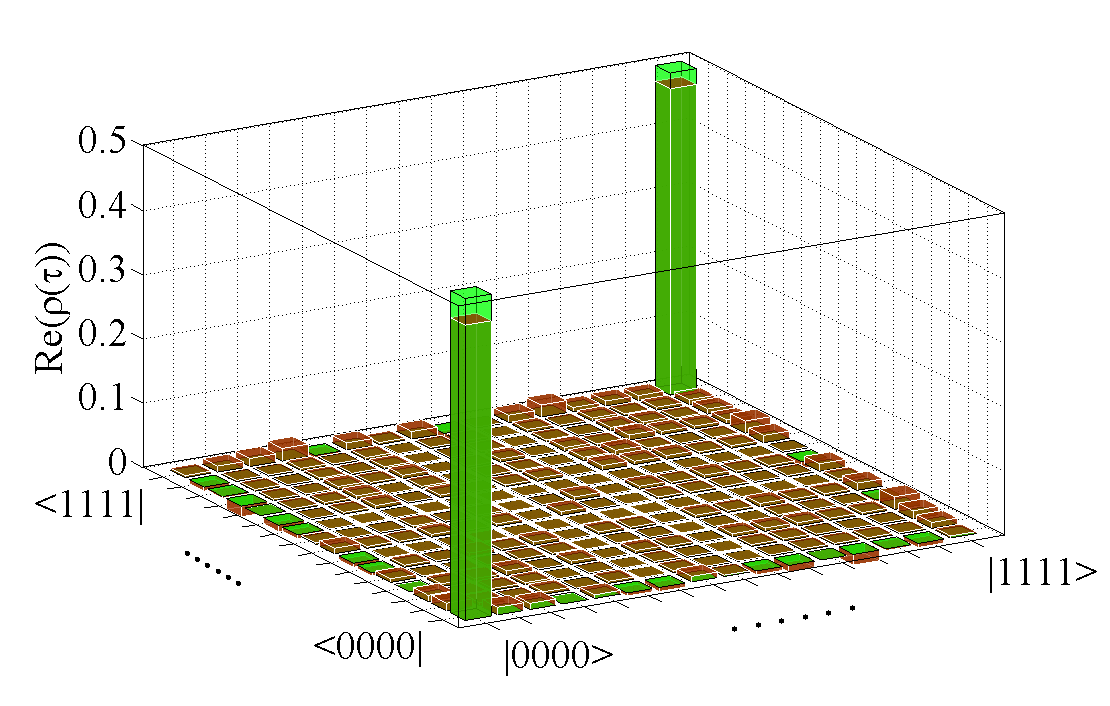}
\includegraphics[width=0.9\columnwidth]{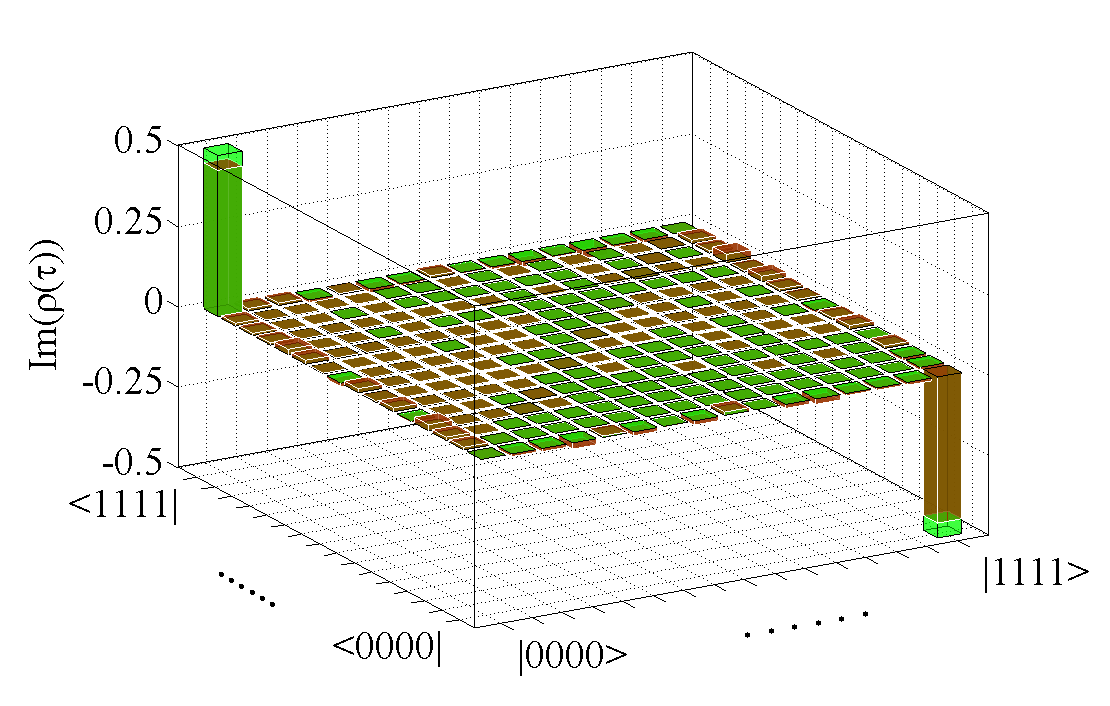}
  \begin{picture}(0,-1)
    \put(-40,390){$(a)$}
    \put(-40,350){$(b)$}
    \put(-40,324){$(c)$}
    \put(-40,228){$(d)$}
    \put(-40,88){$(e)$}
  \end{picture}
\caption{GHZ state preparation. (a) Optimized control parameters. The detuning $\Delta$ at the final time has been fixed to be $\Delta_C$ in Fig.~\ref{fig:sketch}, see the right vertical axis. (b) Fidelity for the GHZ state with $\theta = 0.5\pi$ (cyan solid curve), and population in the subspace spanned by $\ket{0000}$ and $\ket{1111}$ (green dotted curve), when applying the optimized control in panel (a) to one realization without lattice defects. The fidelity between the final state of this realization and the GHZ state is approximately $0.94$. (c) Probability $P_n \equiv \avg{\sum_{i_1,\dots,i_n}\ket{1_{i_1}}\bra{1_{i_1}}\otimes \cdots \otimes \ket{1_{i_n}}\bra{1_{i_n}}}$ of excitations with given numbers $n$ and the $10$ times amplified decay probability $P_d $ for the instantaneous state $\rho(t)$.
(d) Real part and (e) imaginary part of the average final states (brown) and the GHZ state (green). }
\label{fig:GHZpreparation}
\end{figure}

Fig.~\ref{fig:GHZpreparation}(a) shows such an optimised pulse for a targeted GHZ state with $\theta=\pi/2$ and a chosen pulse duration of $3\mu$s, and requiring a vanishing initial and final Rabi frequency as well as a detuning of $\Delta(\tau)=\Delta_c$ at the end of the pulse.
The time evolution of the corresponding fidelity is depicted in Fig.~\ref{fig:GHZpreparation}(b) (cyan solid curve) and yields a final average value of $F^G=0.92$. Note that such high fidelities are indeed obtained despite a significant fraction of lattice defects around $10\%$. Remarkably, the fidelity that can be obtained for a defect-free atomic lattice is virtually perfect with infidelity $9\times 10^{-5}$. Such conditions and geometries can, for example, be realized with optical dipole-trap arrays as demonstrated in a number of recent experiments~\cite{Nogrette14, Labuhn16, Barredo16, Endres16}.
As can be seen from the $P_0+P_4$ curve in the panel (b), the optimized quantum dynamics differs significantly from the preparation protocol proposed in Ref.\cite{RydbergGHZ_Pohl}, where the accessible many-body states are constrained to $|G\rangle$ and $|E\rangle$, and GHZ states are generated by inducing Landau-Zener transitions between them. As shown in Fig.~\ref{fig:GHZpreparation}(b), the optimised preparation pulses presented here, on the contrary, exploit a significantly larger fraction of the underlying Hilbert space for high-fidelity generation of GHZ states within a short preparation time.
Indeed the chosen $3\mu$s preparation time of Fig.~\ref{fig:GHZpreparation} is sufficiently short to ensure a total decay probability $P_d(T)$ of less than $0.07$ [see Fig.~\ref{fig:GHZpreparation}(c) the orange shaded area plotting the $10$ times amplified $P_d(t)$]. The final value $P_d(\tau)$ provides an upper bound on infidelity caused by Rydberg state decay, assuming that any decay prevents the target state preparation. The overall preparation fidelity can thus be estimated as $F^G \times (1-P_d(T))=0.86$. The real part and the imaginary part of the final density matrix for the prepared state (brown) and the targeted GHZ state (green) are shown in Fig.\ref{fig:GHZpreparation}(d) and (e), respectively.

The experimental detection method for this system is limited to the excitation probability on each site, which is sufficient to probe the crystalline state~\cite{NatExp,Schauss15} but not enough to demonstrate the presence of the GHZ state directly. Here we propose to apply a sequence of measurements to probe GHZ states, exploiting the fact that information on the coherence present in the state can be extracted from the free time evolution of the system~\cite{FlorenceDetection}.

We start from the natural assumption that many copies of identical final states can be obtained simply by repeating the experiment, as is routinely done to improve measurement statistics. 
The first step is then to perform a standard excitation measurement~\cite{Schauss15} on many copies of the final states $\ket{\psi(\tau)}$. 
If the system is in the GHZ state, 50\% of the measurement outcomes will result in no excitations while the other 50\% will result in 4 excitations. No other configuration should appear for any individual measurement.
That shows that the final state (not necessary pure) lives in the subspace spanned by the states $\ket {0000} $  and $\ket {1111}$ as 
$\rho_{\psi}^\mathrm{sub}=\left(\begin{smallmatrix}
1/2& \gamma e^{i\alpha} \\
\gamma e^{-i\alpha}& 1/2
\end{smallmatrix}\right)$ with $0\leq\gamma\leq1/2$.
Clearly, the GHZ states $\ket{\psi^G}$ are described by $\rho_{\psi}^\mathrm{sub}$ for $\gamma=1/2$ and $\alpha=\theta$.

In the second step, we still need to distinguish between $\ket{\psi^G}$ and the other states in $\rho_{\psi}^\mathrm{sub}$.
{One intuitive way to distinguish between them is of course to measure the purity of the final state. Recently, the Greiner group has shown an experimental method to probe the purity of the state for cold atoms in an optical lattice through measuring the average parity of the atomic interference between identical two-copy states~\cite{Exp_purity}. However, this parity scheme is not particularly suitable for  many-body Rydberg systems, since the long-range interactions between Rydberg atoms from the same copy are difficult to switch off in the interference.} Hence, we propose a free-evolution scheme in which one can distinguish them by
simply evolving the systems with a detection Hamiltonian $H_d = \frac{\hbar}{2}\Omega_{M}\sum_{i}\big(\hat{\sigma}_{eg}^{(i)} + \hat{\sigma}_{ge}^{(i)} \big) + \sum_{i\neq j}\frac{V_{ij}}{2}\hat{\sigma}_{ee}^{(i)} \hat{\sigma}_{ee}^{(j)}$, where $\Omega_{M}$ is the maximal Rabi coupling generated by the control lasers. 
The coherence $\gamma$ as well as the phase factor $\alpha$ for each individual initial state in $\rho_{\psi}^\mathrm{sub}$ will result in unique dynamics. The difference between the targeted GHZ state and any others states in $\rho_{\psi}^\mathrm{sub}$ is thus detectable from the differing dynamics of the excitation probabilities for one qubit, $ E_i(\rho(t)) \equiv \tr[\rho(t) {\ket {1_i} \bra{1_i}}]$. 

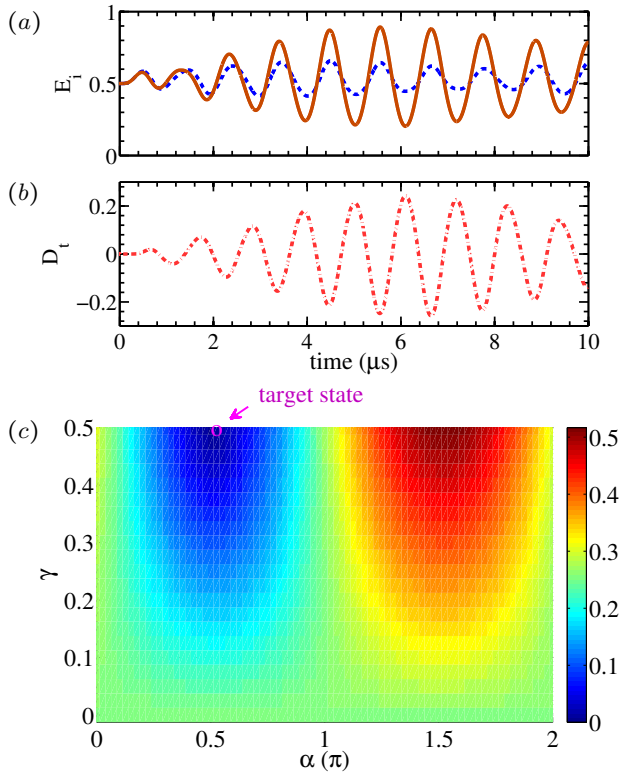
\begin{figure}[h]
\includegraphics[width=0.9\columnwidth]{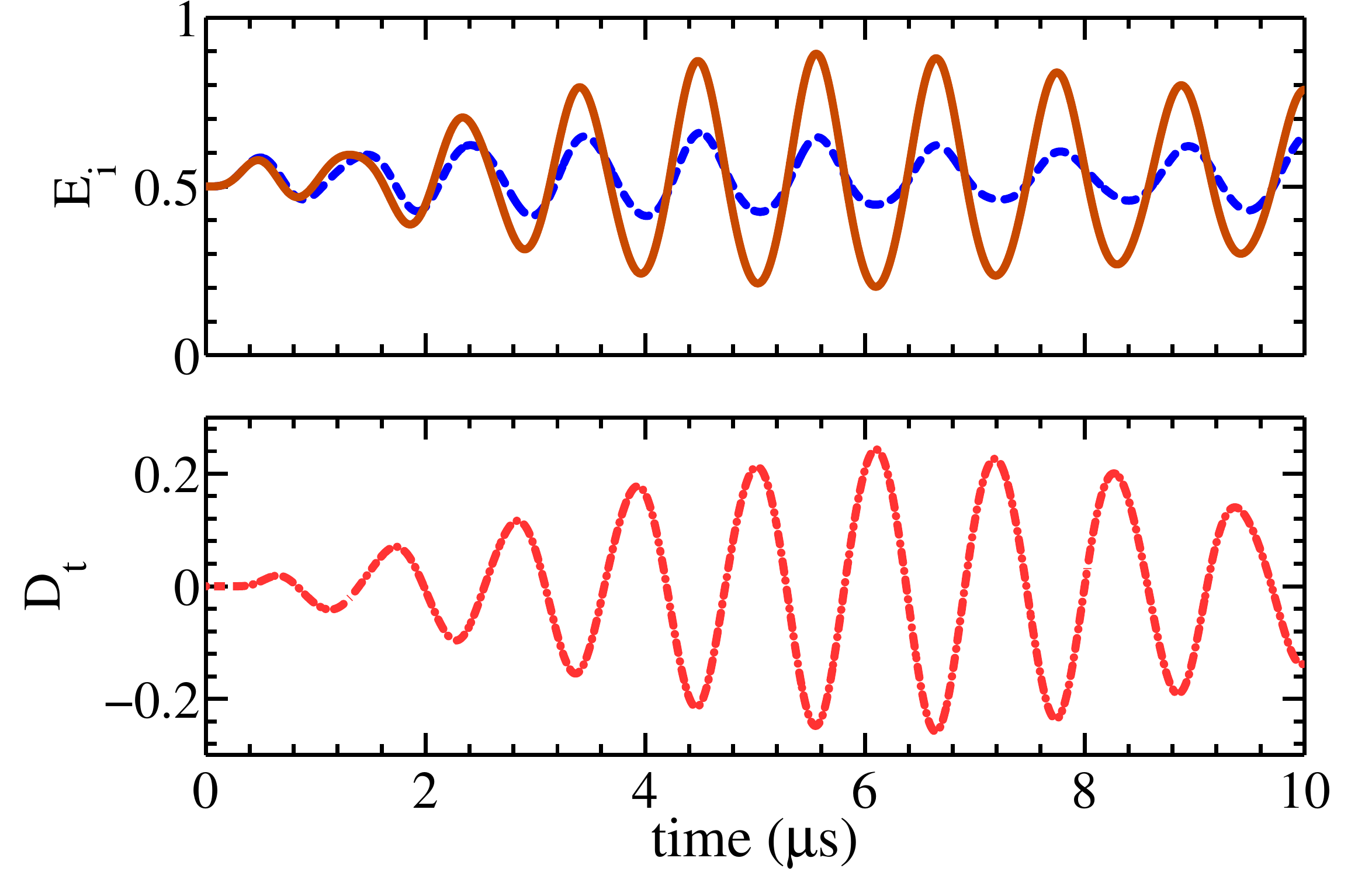}
\includegraphics[width=\columnwidth]{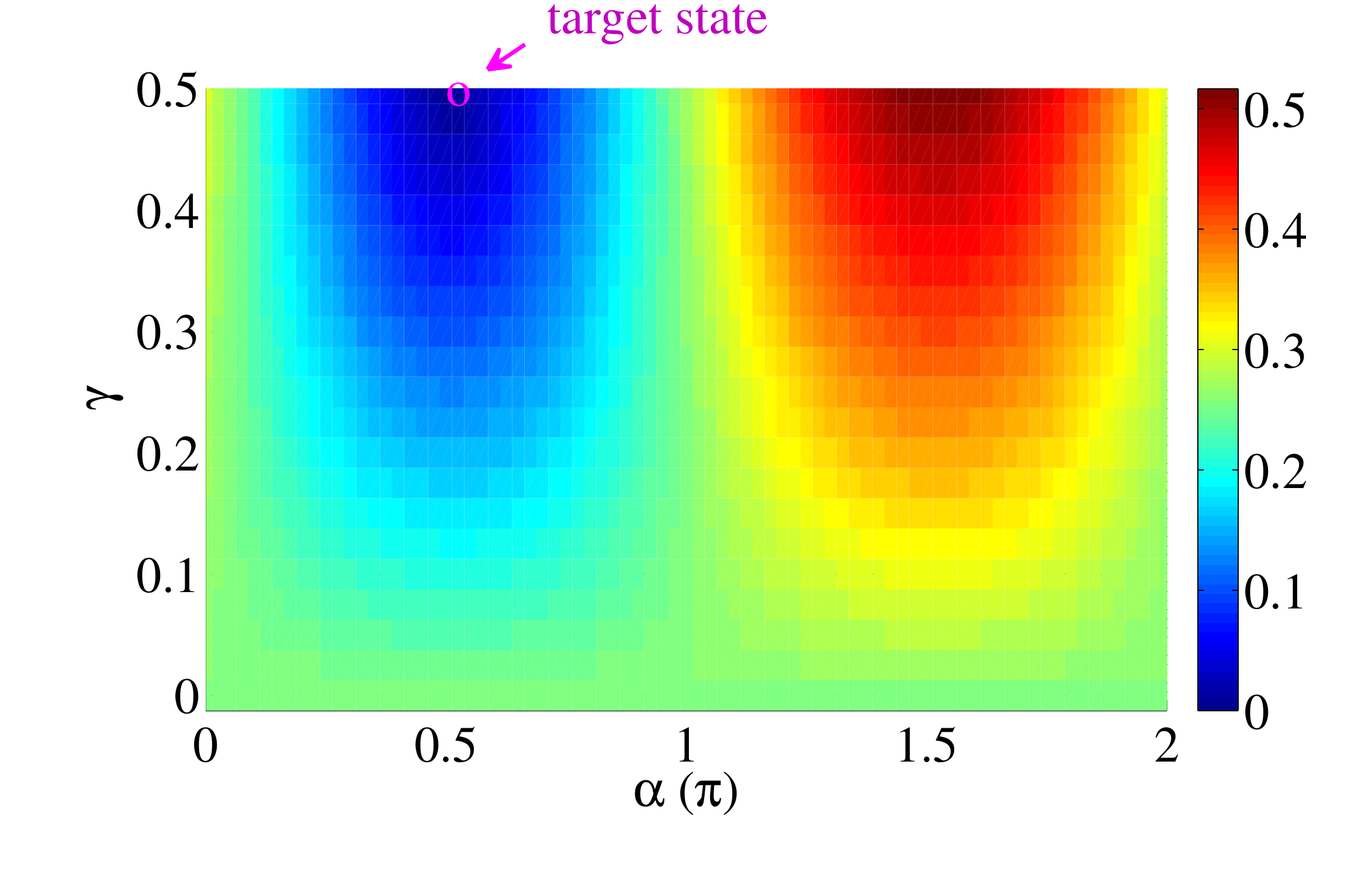}
  \begin{picture}(0,-1)
    \put(-120,305){$(a)$}
    \put(-120,240){$(b)$}
    \put(-120,150){$(c)$}
  \end{picture}
\caption{GHZ state detection. Evolution of the single superatom excitation (a) under the detection Hamiltonian with different initial state: GHZ state $  (\ket {0000} + i\ket {1111})/\sqrt 2$ (solid brown curve), equally mixed state $\rho_{\mathrm{mix}} = (\ket {0000} \bra{0000} + \ket {1111} \bra{1111} )/2$ (dashed blue curve), and (b) the deviation $D_t = E_i(\rho_{\mathrm{mix}}) - E_i(\rho_{\mathrm{GHZ}}(t)) $(red dot-dash curve). Due to the symmetry of the lattice geometry all the superatoms behave the same. (c) The time-maximal deviation $|D|$ as a function of $\gamma$ and $\alpha$. The target state is highlighted with purple circle.}
\label{fig:GHZDetection}
\end{figure}

As an example, Fig.~\ref{fig:GHZDetection}(a) shows that the excitation dynamics of the targeted GHZ initial state (the $\gamma=0.5,\alpha=0.5\pi$ state in $\rho_{\psi}^\mathrm{sub}$) differs from that of a fully mixture state, labelled as $\rho_{\mathrm{mix}}$, in $\rho_{\psi}^\mathrm{sub}$ with $\gamma=0$.
The excitation difference $D_t \equiv  E_i(\rho_{\psi}^\mathrm{sub}(t)) - E_i(\rho_{\mathrm{GHZ}}(t)) $ is a function of $\gamma$ and $\alpha$ for a general initial state in $\rho_{\psi}^\mathrm{sub}$, where the parameter $t$ in the brackets represents the evolution of the corresponding state from time $0$ to time $t$. We use a notation without $t$ to denote the time-maximal deviation within the experimental time as $|D| = {\mathrm{max}}_{t} |D_t|$. Fig.~\ref{fig:GHZDetection}(b) depicts $|D_t|$ for $\rho_{\mathrm{mix}}$. In this example $|D|$ occurs at about 6 $\mu s$. This time only varies slightly by changing parameters. Fig.~\ref{fig:GHZDetection}(c) depicts $|D|$ for different $\gamma$ and $\alpha$. In the small $\gamma$ limit, $\rho_{\psi}^\mathrm{sub}$ is close to the fully mixed state, so that $|D|$ is insensitive to the phase factors $\alpha$. For $\gamma=0.5$, $\rho_{\psi}^\mathrm{sub}$ consists of the GHZ states with different phase factor, and therefore $|D|$ significantly depends on $\alpha$.  In general, every state differs from each other in terms of $D_t$, and $|D|$ is a good measure of the difference.

Thus, the detection scheme we propose is firstly measuring the excitation profile of the prepared state and then evolving the prepared state under the detection Hamiltonian to compare the dynamics of a single qubit excitation $D_t$ with respect to that of the targeted GHZ state.
The total experimental time, which is composed of the preparation time ($t_p=3\mu s$) and the free evolution time in the second step ($t_d=6\mu s$) plus the excitation detection time ($t_e=10\mu s$~\cite{NatExp}), is shorter than the lifetime of the Rydberg state.  

\section{Arbitrary state preparation}
\label{sec:Arbitrary}

Let us finally demonstrate the general applicability of the method by studying the preparation of arbitrary many-body states in a Rydberg lattice. As a specific example we choose the same lattice geometry as in section \ref{sec:GHZ} and consider symmetric target states, $\ket{\psi_s} = \sum_{n} a_n \ket{s_n}$, spanned by the number states $|s_0\rangle=\ket{0000}$, $|s_1\rangle=\left(\ket{0001}+\ket{0010}+\ket{0100}+\ket{1000}\right)/2$, etc.. Again we implement realistic experimental constraints for the excitation pulse and account for random lattice defects by performing an ensemble average over $50$ random spatial configurations. 

\begin{figure}
\includegraphics[width=0.9\columnwidth]{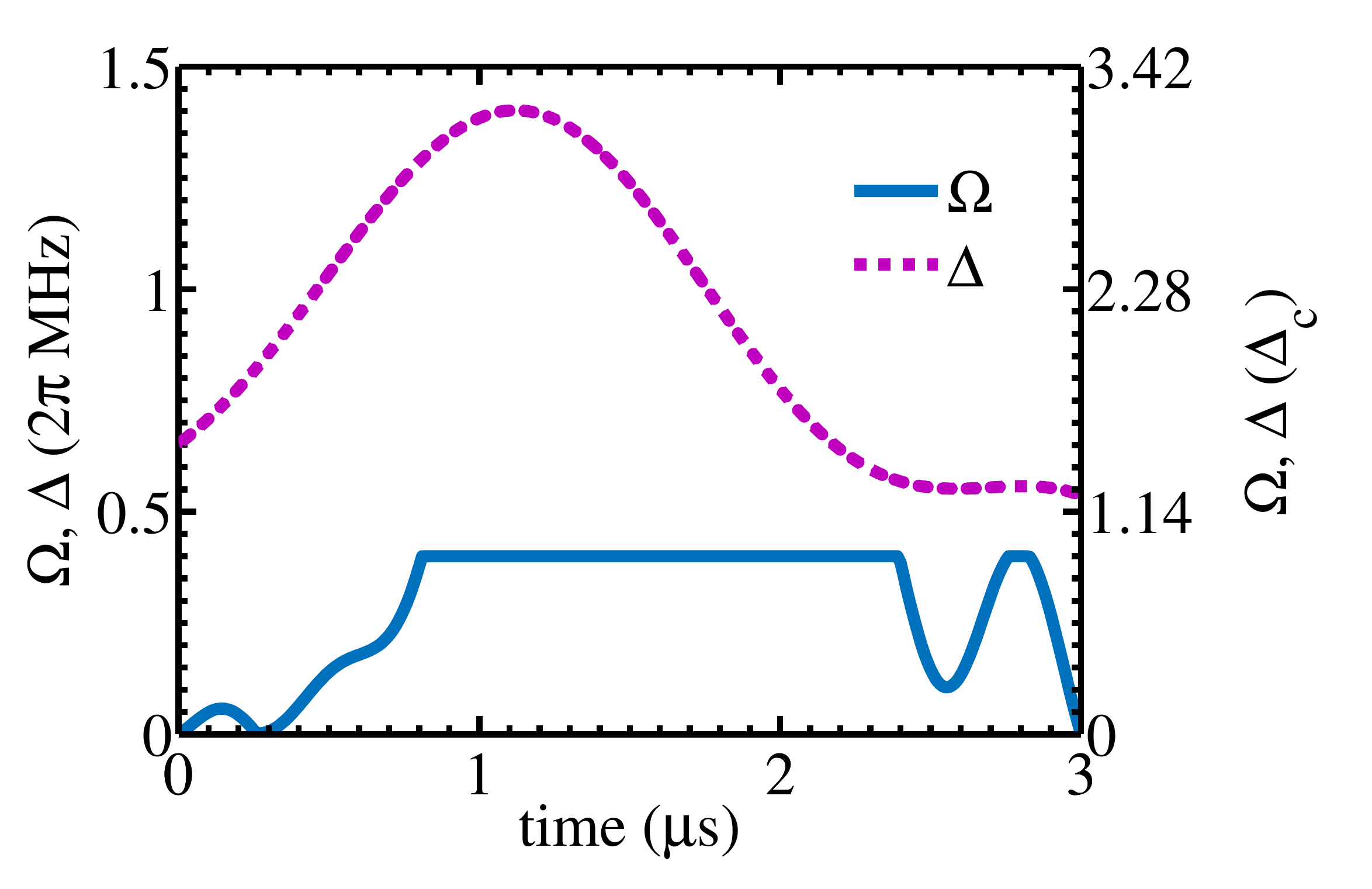}
\includegraphics[width=0.9\columnwidth]{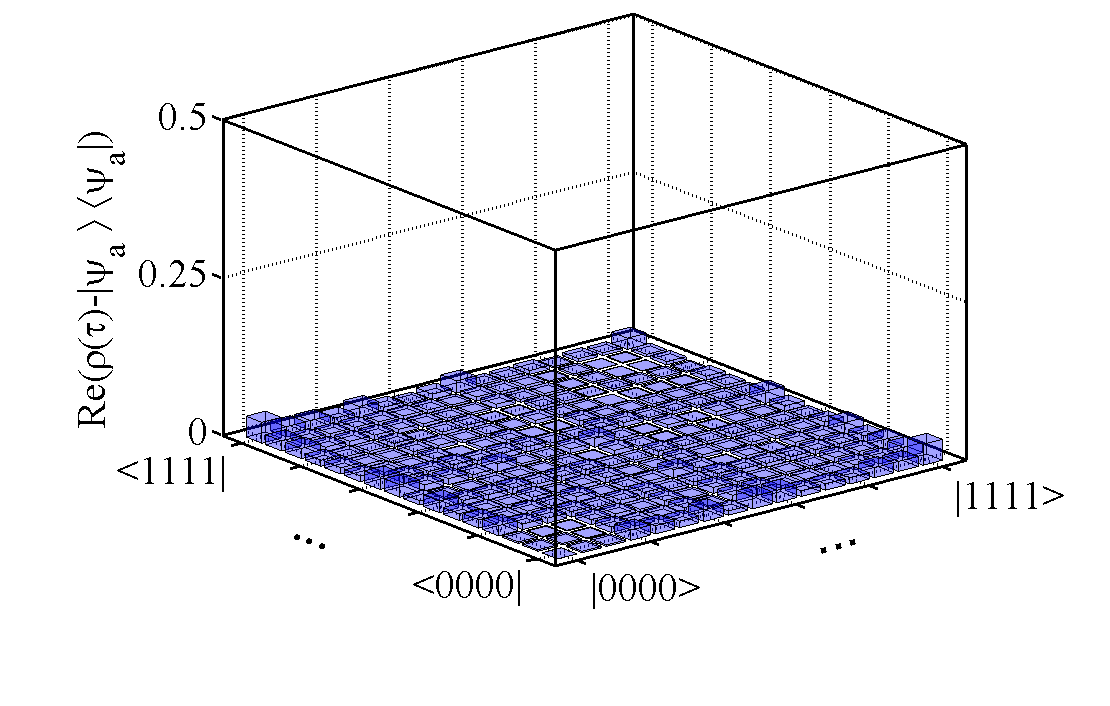}
  \begin{picture}(0,-1)
    \put(-225,275){$(a)$}
    \put(-225,130){$(b)$}
  \end{picture}
\caption{Arbitrary state preparation. (a) Optimized control parameters. Panel (b) shows the elements of the difference $\rho(\tau)-\ket{\psi_a}\bra{\psi_a}$ between the generated and the target state. Here $\rho(\tau)$ is the average final state over $50$ realizations of the system with lattice defects. Since the elements of the target state are real by construction we only show the real part in panel (b).}
\label{fig:arbitrary}
\end{figure}
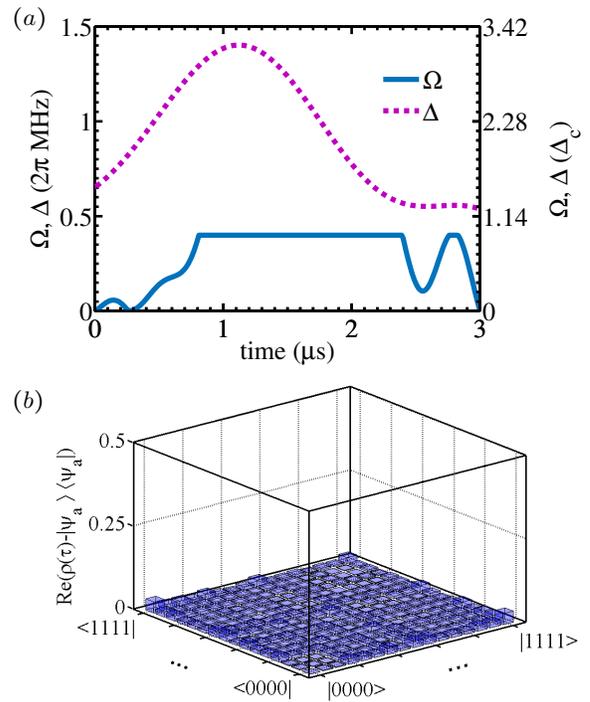

In Fig.\ref{fig:arbitrary}(a) we show the optimised excitation pulse for preparing the state $|\psi_a\rangle$ with randomly generated coefficients, $a_0=0.5737$, $a_1=0.5586$, $a_2=0.3399$, $a_3=0.3500$ and $a_4=0.3475$. Even for a short preparation time of $3\mu$s the optimized pulse allows to generate the target state with a high fidelity of $0.975$. This is illustrated in Fig.\ref{fig:arbitrary}(b) where we show the difference $\rho(\tau) - \ket{\psi_a}\bra{\psi_a}$ between the target state and the generated state. It's elements are very small throughout demonstrating the high quality of the optimised state preparation approach. 

\section{Discussion and summary}
\label{sec:Discussion}
In this work, we have investigated the applicability of optimal control approaches for the dynamical preparation of many-body states in a lattice of interacting Rydberg atoms. We have demonstrated that this opens up the fidelity preparation of ordered ground states, highly entangled GHZ states and even arbitrary, randomly chosen, many-body states under realistic conditions and typical experimental constraints on excitation pulse shaping. In particular, for the latter optimal control techniques as demonstrated in this work presently provide the only suitable approach to generate complex many-body states in an efficient and experimentally viable fashion.

We have placed particular focus on limitations and imperfections, such as lattice defects, that are practically unavoidable in experiments. Our optimized control pulses are robust against lattice defects, in the sense that they yield high preparation fidelities for nearly every randomly sampled spatial configuration and a high average fidelity with a small statistical spread. For example, comparing the average fidelities from two sets of $50$ random samples we found a difference of less than $10^{-3}$ for all three studied target states. 

\begin{table}[h!]
\begin{center}
\begin{tabular}{ |p{1.5cm}||c|c||c|c||c|c| } 
%\hline
% & \multicolumn{4}{ |c|| }{Output} & \multicolumn{4}{ |c| }{Input} \\
\hline
 & $ F$ & $F_s$ & $F_s-F$ & $P_d(\tau)$ & $\langle N_e\rangle$ & $\tau/\mu s$ \\ 
\hline
\hline
 Crystalline state & 0.85 & 0.923 & 0.073 & 0.1 & 3 & 4  \\ 
\hline
 GHZ state & 0.921 & 0.99991 & 0.079 & 0.07 & 2 & 3 \\  
\hline
$\ket{\psi_a}$ & 0.975 & 0.9995 & 0.025 & 0.02 & 1.4 & 3  \\  
\hline
\end{tabular}
\end{center}
\label{table:summarize}
\caption{Properties of the preparation of the crystalline state, GHZ state and the arbitrary superposition state. The second and third columns are respectively the average fidelity over $50$ defective realizations $F$ and the single-shot fidelity for perfect lattice $F_s$. The fourth column measures how much infidelity is caused by the lattice defects. The fifth column, $P_d(\tau)$, lists the decay probability at the final excitation time. The last two columns show the expectation value of the number of excitations, as well as the excitation times, respectively, for the three examples. }
\end{table}

Table \ref{table:summarize} summarises the overall performance of the dCRAB optimal control method for the three example states. As one can see the major limitation on achievable fidelities in all three cases stems from the finite fraction of lattice defects and spontaneous decay of the Rydberg state.  While we have considered here a filling fraction of $0.9$ \cite{Schauss15}, recent experiments have already reached considerably higher values in optical lattices \cite{Zeiher16,Zeiher17} and optical dipole trap arrays \cite{Barredo16,Endres16}. Equally important, spontaneous Rydberg state decay ultimately limits achievable preparation fidelities, which is why we have chosen relatively short pulse durations of a few $\mu$s (see table \ref{table:summarize}).

Within our optimisation approach, it should be possible to further reduce the total evolution time without a significant degradation of the preparation fidelity until reaching the quantum speed limit. Since the preparation time eventually determines the extend of undesired decoherence effects, the detailed exploration of the quantum speed limit in Rydberg lattices presents both a fundamentally interesting and practically important problem for future studies. In view of the recent advances in optically controlling the many-body dynamics of Rydberg atom lattices, the control techniques demonstrated in this work will enhance the capabilities of such systems for quantum simulations as well as the collective preparation of complex nonclassical many-body states for quantum information applications. We hope that the first theoretical steps in this direction, as presented in this article, will initiate further experimental and theoretical work to tap the full potential of optimal control techniques for Rydberg-atom many-body physics.

{ Acknowledgement.---} 
We thank Tommaso Macr\`{\i}, Victor Mukherjee, Johannes Zeiher, Christian Gross and Immanuel Bloch for valuable discussions. TP is supported by the DNRF through a Niels Bohr Professorship. SM gratefully acknowledges the support of the DFG via a Heisenberg fellowship. This work made use of the High Performance Computing Resource BwUniCluster and JUSUTS cluster. This work is supported by the European Commission funded FET project “RySQ” with Grant No. 640378, German Research Foundation (DFG) Priority Program GiRyd, DFG via the SFB/TRR21, and the Federal Ministry of Education and Research (BMBF) funded project Q.COM.

\bibliography{reference_update}

\begin{thebibliography}{99}%
\makeatletter
\providecommand \@ifxundefined [1]{%
 \@ifx{#1\undefined}
}%
\providecommand \@ifnum [1]{%
 \ifnum #1\expandafter \@firstoftwo
 \else \expandafter \@secondoftwo
 \fi
}%
\providecommand \@ifx [1]{%
 \ifx #1\expandafter \@firstoftwo
 \else \expandafter \@secondoftwo
 \fi
}%
\providecommand \natexlab [1]{#1}%
\providecommand \enquote  [1]{``#1''}%
\providecommand \bibnamefont  [1]{#1}%
\providecommand \bibfnamefont [1]{#1}%
\providecommand \citenamefont [1]{#1}%
\providecommand \href@noop [0]{\@secondoftwo}%
\providecommand \href [0]{\begingroup \@sanitize@url \@href}%
\providecommand \@href[1]{\@@startlink{#1}\@@href}%
\providecommand \@@href[1]{\endgroup#1\@@endlink}%
\providecommand \@sanitize@url [0]{\catcode `\\12\catcode `\$12\catcode
  `\&12\catcode `\#12\catcode `\^12\catcode `\_12\catcode `\%12\relax}%
\providecommand \@@startlink[1]{}%
\providecommand \@@endlink[0]{}%
\providecommand \url  [0]{\begingroup\@sanitize@url \@url }%
\providecommand \@url [1]{\endgroup\@href {#1}{\urlprefix }}%
\providecommand \urlprefix  [0]{URL }%
\providecommand \Eprint [0]{\href }%
\providecommand \doibase [0]{http://dx.doi.org/}%
\providecommand \selectlanguage [0]{\@gobble}%
\providecommand \bibinfo  [0]{\@secondoftwo}%
\providecommand \bibfield  [0]{\@secondoftwo}%
\providecommand \translation [1]{[#1]}%
\providecommand \BibitemOpen [0]{}%
\providecommand \bibitemStop [0]{}%
\providecommand \bibitemNoStop [0]{.\EOS\space}%
\providecommand \EOS [0]{\spacefactor3000\relax}%
\providecommand \BibitemShut  [1]{\csname bibitem#1\endcsname}%
\let\auto@bib@innerbib\@empty
%</preamble>
\bibitem [{\citenamefont {Bloch}\ \emph {et~al.}(2008)\citenamefont {Bloch},
  \citenamefont {Dalibard},\ and\ \citenamefont {Zwerger}}]{Bloch08}%
  \BibitemOpen
  \bibfield  {author} {\bibinfo {author} {\bibfnamefont {I.}~\bibnamefont
  {Bloch}}, \bibinfo {author} {\bibfnamefont {J.}~\bibnamefont {Dalibard}}, \
  and\ \bibinfo {author} {\bibfnamefont {W.}~\bibnamefont {Zwerger}},\ }\href
  {\doibase 10.1103/RevModPhys.80.885} {\bibfield  {journal} {\bibinfo
  {journal} {Rev. Mod. Phys.}\ }\textbf {\bibinfo {volume} {80}},\ \bibinfo
  {pages} {885} (\bibinfo {year} {2008})}\BibitemShut {NoStop}%
\bibitem [{\citenamefont {Nogrette}\ \emph {et~al.}(2014)\citenamefont
  {Nogrette}, \citenamefont {Labuhn}, \citenamefont {Ravets}, \citenamefont
  {Barredo}, \citenamefont {B\'eguin}, \citenamefont {Vernier}, \citenamefont
  {Lahaye},\ and\ \citenamefont {Browaeys}}]{Nogrette14}%
  \BibitemOpen
  \bibfield  {author} {\bibinfo {author} {\bibfnamefont {F.}~\bibnamefont
  {Nogrette}}, \bibinfo {author} {\bibfnamefont {H.}~\bibnamefont {Labuhn}},
  \bibinfo {author} {\bibfnamefont {S.}~\bibnamefont {Ravets}}, \bibinfo
  {author} {\bibfnamefont {D.}~\bibnamefont {Barredo}}, \bibinfo {author}
  {\bibfnamefont {L.}~\bibnamefont {B\'eguin}}, \bibinfo {author}
  {\bibfnamefont {A.}~\bibnamefont {Vernier}}, \bibinfo {author} {\bibfnamefont
  {T.}~\bibnamefont {Lahaye}}, \ and\ \bibinfo {author} {\bibfnamefont
  {A.}~\bibnamefont {Browaeys}},\ }\href {\doibase 10.1103/PhysRevX.4.021034}
  {\bibfield  {journal} {\bibinfo  {journal} {Phys. Rev. X}\ }\textbf {\bibinfo
  {volume} {4}},\ \bibinfo {pages} {021034} (\bibinfo {year}
  {2014})}\BibitemShut {NoStop}%
\bibitem [{\citenamefont {Barredo}\ \emph {et~al.}(2016)\citenamefont
  {Barredo}, \citenamefont {de~L{\'e}s{\'e}leuc}, \citenamefont {Lienhard},
  \citenamefont {Lahaye},\ and\ \citenamefont {Browaeys}}]{Barredo16}%
  \BibitemOpen
  \bibfield  {author} {\bibinfo {author} {\bibfnamefont {D.}~\bibnamefont
  {Barredo}}, \bibinfo {author} {\bibfnamefont {S.}~\bibnamefont
  {de~L{\'e}s{\'e}leuc}}, \bibinfo {author} {\bibfnamefont {V.}~\bibnamefont
  {Lienhard}}, \bibinfo {author} {\bibfnamefont {T.}~\bibnamefont {Lahaye}}, \
  and\ \bibinfo {author} {\bibfnamefont {A.}~\bibnamefont {Browaeys}},\ }\href
  {\doibase 10.1126/science.aah3778} {\bibfield  {journal} {\bibinfo  {journal}
  {Science}\ }\textbf {\bibinfo {volume} {354}},\ \bibinfo {pages} {1021}
  (\bibinfo {year} {2016})}\BibitemShut {NoStop}%
\bibitem [{\citenamefont {Endres}\ \emph {et~al.}(2016)\citenamefont {Endres},
  \citenamefont {Bernien}, \citenamefont {Keesling}, \citenamefont {Levine},
  \citenamefont {Anschuetz}, \citenamefont {Krajenbrink}, \citenamefont
  {Senko}, \citenamefont {Vuletic}, \citenamefont {Greiner},\ and\
  \citenamefont {Lukin}}]{Endres16}%
  \BibitemOpen
  \bibfield  {author} {\bibinfo {author} {\bibfnamefont {M.}~\bibnamefont
  {Endres}}, \bibinfo {author} {\bibfnamefont {H.}~\bibnamefont {Bernien}},
  \bibinfo {author} {\bibfnamefont {A.}~\bibnamefont {Keesling}}, \bibinfo
  {author} {\bibfnamefont {H.}~\bibnamefont {Levine}}, \bibinfo {author}
  {\bibfnamefont {E.~R.}\ \bibnamefont {Anschuetz}}, \bibinfo {author}
  {\bibfnamefont {A.}~\bibnamefont {Krajenbrink}}, \bibinfo {author}
  {\bibfnamefont {C.}~\bibnamefont {Senko}}, \bibinfo {author} {\bibfnamefont
  {V.}~\bibnamefont {Vuletic}}, \bibinfo {author} {\bibfnamefont
  {M.}~\bibnamefont {Greiner}}, \ and\ \bibinfo {author} {\bibfnamefont
  {M.~D.}\ \bibnamefont {Lukin}},\ }\href {\doibase 10.1126/science.aah3752}
  {\bibfield  {journal} {\bibinfo  {journal} {Science}\ }\textbf {\bibinfo
  {volume} {354}},\ \bibinfo {pages} {1024} (\bibinfo {year}
  {2016})}\BibitemShut {NoStop}%
\bibitem [{\citenamefont {Ott}(2016)}]{Qgas}%
  \BibitemOpen
  \bibfield  {author} {\bibinfo {author} {\bibfnamefont {H.}~\bibnamefont
  {Ott}},\ }\href {http://stacks.iop.org/0034-4885/79/i=5/a=054401} {\bibfield
  {journal} {\bibinfo  {journal} {Reports on Progress in Physics}\ }\textbf
  {\bibinfo {volume} {79}},\ \bibinfo {pages} {054401} (\bibinfo {year}
  {2016})}\BibitemShut {NoStop}%
\bibitem [{\citenamefont {Saffman}\ \emph {et~al.}(2010)\citenamefont
  {Saffman}, \citenamefont {Walker},\ and\ \citenamefont
  {M\o{}lmer}}]{SaffmanReview}%
  \BibitemOpen
  \bibfield  {author} {\bibinfo {author} {\bibfnamefont {M.}~\bibnamefont
  {Saffman}}, \bibinfo {author} {\bibfnamefont {T.~G.}\ \bibnamefont {Walker}},
  \ and\ \bibinfo {author} {\bibfnamefont {K.}~\bibnamefont {M\o{}lmer}},\
  }\href {\doibase 10.1103/RevModPhys.82.2313} {\bibfield  {journal} {\bibinfo
  {journal} {Rev. Mod. Phys.}\ }\textbf {\bibinfo {volume} {82}},\ \bibinfo
  {pages} {2313} (\bibinfo {year} {2010})}\BibitemShut {NoStop}%
\bibitem [{\citenamefont {L\"ow}\ \emph {et~al.}(2012)\citenamefont {L\"ow},
  \citenamefont {Weimer}, \citenamefont {Nipper}, \citenamefont {Balewski},
  \citenamefont {Butscher}, \citenamefont {B\"uchler},\ and\ \citenamefont
  {Pfau}}]{LoewReview}%
  \BibitemOpen
  \bibfield  {author} {\bibinfo {author} {\bibfnamefont {R.}~\bibnamefont
  {L\"ow}}, \bibinfo {author} {\bibfnamefont {H.}~\bibnamefont {Weimer}},
  \bibinfo {author} {\bibfnamefont {J.}~\bibnamefont {Nipper}}, \bibinfo
  {author} {\bibfnamefont {J.~B.}\ \bibnamefont {Balewski}}, \bibinfo {author}
  {\bibfnamefont {B.}~\bibnamefont {Butscher}}, \bibinfo {author}
  {\bibfnamefont {H.~P.}\ \bibnamefont {B\"uchler}}, \ and\ \bibinfo {author}
  {\bibfnamefont {T.}~\bibnamefont {Pfau}},\ }\href
  {http://stacks.iop.org/0953-4075/45/i=11/a=113001} {\bibfield  {journal}
  {\bibinfo  {journal} {Journal of Physics B: Atomic, Molecular and Optical
  Physics}\ }\textbf {\bibinfo {volume} {45}},\ \bibinfo {pages} {113001}
  (\bibinfo {year} {2012})}\BibitemShut {NoStop}%
\bibitem [{\citenamefont {Jaksch}\ \emph {et~al.}(2000)\citenamefont {Jaksch},
  \citenamefont {Cirac}, \citenamefont {Zoller}, \citenamefont {Rolston},
  \citenamefont {C\^ot\'e},\ and\ \citenamefont {Lukin}}]{Jaksch00}%
  \BibitemOpen
  \bibfield  {author} {\bibinfo {author} {\bibfnamefont {D.}~\bibnamefont
  {Jaksch}}, \bibinfo {author} {\bibfnamefont {J.~I.}\ \bibnamefont {Cirac}},
  \bibinfo {author} {\bibfnamefont {P.}~\bibnamefont {Zoller}}, \bibinfo
  {author} {\bibfnamefont {S.~L.}\ \bibnamefont {Rolston}}, \bibinfo {author}
  {\bibfnamefont {R.}~\bibnamefont {C\^ot\'e}}, \ and\ \bibinfo {author}
  {\bibfnamefont {M.~D.}\ \bibnamefont {Lukin}},\ }\href {\doibase
  10.1103/PhysRevLett.85.2208} {\bibfield  {journal} {\bibinfo  {journal}
  {Phys. Rev. Lett.}\ }\textbf {\bibinfo {volume} {85}},\ \bibinfo {pages}
  {2208} (\bibinfo {year} {2000})}\BibitemShut {NoStop}%
\bibitem [{\citenamefont {Lukin}\ \emph {et~al.}(2001)\citenamefont {Lukin},
  \citenamefont {Fleischhauer}, \citenamefont {Cote}, \citenamefont {Duan},
  \citenamefont {Jaksch}, \citenamefont {Cirac},\ and\ \citenamefont
  {Zoller}}]{Lukin01}%
  \BibitemOpen
  \bibfield  {author} {\bibinfo {author} {\bibfnamefont {M.~D.}\ \bibnamefont
  {Lukin}}, \bibinfo {author} {\bibfnamefont {M.}~\bibnamefont {Fleischhauer}},
  \bibinfo {author} {\bibfnamefont {R.}~\bibnamefont {Cote}}, \bibinfo {author}
  {\bibfnamefont {L.~M.}\ \bibnamefont {Duan}}, \bibinfo {author}
  {\bibfnamefont {D.}~\bibnamefont {Jaksch}}, \bibinfo {author} {\bibfnamefont
  {J.~I.}\ \bibnamefont {Cirac}}, \ and\ \bibinfo {author} {\bibfnamefont
  {P.}~\bibnamefont {Zoller}},\ }\href {\doibase 10.1103/PhysRevLett.87.037901}
  {\bibfield  {journal} {\bibinfo  {journal} {Phys. Rev. Lett.}\ }\textbf
  {\bibinfo {volume} {87}},\ \bibinfo {pages} {037901} (\bibinfo {year}
  {2001})}\BibitemShut {NoStop}%
\bibitem [{\citenamefont {Gaetan}\ \emph {et~al.}(2009)\citenamefont {Gaetan},
  \citenamefont {Miroshnychenko}, \citenamefont {Wilk}, \citenamefont {Chotia},
  \citenamefont {Viteau}, \citenamefont {Comparat}, \citenamefont {Pillet},
  \citenamefont {Browaeys},\ and\ \citenamefont {Grangier}}]{Gaetan09}%
  \BibitemOpen
  \bibfield  {author} {\bibinfo {author} {\bibfnamefont {A.}~\bibnamefont
  {Gaetan}}, \bibinfo {author} {\bibfnamefont {Y.}~\bibnamefont
  {Miroshnychenko}}, \bibinfo {author} {\bibfnamefont {T.}~\bibnamefont
  {Wilk}}, \bibinfo {author} {\bibfnamefont {A.}~\bibnamefont {Chotia}},
  \bibinfo {author} {\bibfnamefont {M.}~\bibnamefont {Viteau}}, \bibinfo
  {author} {\bibfnamefont {D.}~\bibnamefont {Comparat}}, \bibinfo {author}
  {\bibfnamefont {P.}~\bibnamefont {Pillet}}, \bibinfo {author} {\bibfnamefont
  {A.}~\bibnamefont {Browaeys}}, \ and\ \bibinfo {author} {\bibfnamefont
  {P.}~\bibnamefont {Grangier}},\ }\href {http://dx.doi.org/10.1038/nphys1183}
  {\bibfield  {journal} {\bibinfo  {journal} {Nat Phys}\ }\textbf {\bibinfo
  {volume} {5}},\ \bibinfo {pages} {115} (\bibinfo {year} {2009})}\BibitemShut
  {NoStop}%
\bibitem [{\citenamefont {Urban}\ \emph {et~al.}(2009)\citenamefont {Urban},
  \citenamefont {Johnson}, \citenamefont {Henage}, \citenamefont {Isenhower},
  \citenamefont {Yavuz}, \citenamefont {Walker},\ and\ \citenamefont
  {Saffman}}]{Urban09}%
  \BibitemOpen
  \bibfield  {author} {\bibinfo {author} {\bibfnamefont {E.}~\bibnamefont
  {Urban}}, \bibinfo {author} {\bibfnamefont {T.~A.}\ \bibnamefont {Johnson}},
  \bibinfo {author} {\bibfnamefont {T.}~\bibnamefont {Henage}}, \bibinfo
  {author} {\bibfnamefont {L.}~\bibnamefont {Isenhower}}, \bibinfo {author}
  {\bibfnamefont {D.~D.}\ \bibnamefont {Yavuz}}, \bibinfo {author}
  {\bibfnamefont {T.~G.}\ \bibnamefont {Walker}}, \ and\ \bibinfo {author}
  {\bibfnamefont {M.}~\bibnamefont {Saffman}},\ }\href
  {http://dx.doi.org/10.1038/nphys1178} {\bibfield  {journal} {\bibinfo
  {journal} {Nat Phys}\ }\textbf {\bibinfo {volume} {5}},\ \bibinfo {pages}
  {110} (\bibinfo {year} {2009})}\BibitemShut {NoStop}%
\bibitem [{\citenamefont {Jau}\ \emph {et~al.}(2016)\citenamefont {Jau},
  \citenamefont {Hankin}, \citenamefont {Keating}, \citenamefont {Deutsch},\
  and\ \citenamefont {Biedermann}}]{Jau16}%
  \BibitemOpen
  \bibfield  {author} {\bibinfo {author} {\bibfnamefont {Y.-Y.}\ \bibnamefont
  {Jau}}, \bibinfo {author} {\bibfnamefont {A.~M.}\ \bibnamefont {Hankin}},
  \bibinfo {author} {\bibfnamefont {T.}~\bibnamefont {Keating}}, \bibinfo
  {author} {\bibfnamefont {I.~H.}\ \bibnamefont {Deutsch}}, \ and\ \bibinfo
  {author} {\bibfnamefont {G.~W.}\ \bibnamefont {Biedermann}},\ }\href
  {\doibase 10.1038/nphys3487} {\bibfield  {journal} {\bibinfo  {journal}
  {Nature Physics}\ }\textbf {\bibinfo {volume} {12}},\ \bibinfo {pages} {71}
  (\bibinfo {year} {2016})}\BibitemShut {NoStop}%
\bibitem [{\citenamefont {Heidemann}\ \emph {et~al.}(2007)\citenamefont
  {Heidemann}, \citenamefont {Raitzsch}, \citenamefont {Bendkowsky},
  \citenamefont {Butscher}, \citenamefont {L\"ow}, \citenamefont {Santos},\
  and\ \citenamefont {Pfau}}]{Heidemann07}%
  \BibitemOpen
  \bibfield  {author} {\bibinfo {author} {\bibfnamefont {R.}~\bibnamefont
  {Heidemann}}, \bibinfo {author} {\bibfnamefont {U.}~\bibnamefont {Raitzsch}},
  \bibinfo {author} {\bibfnamefont {V.}~\bibnamefont {Bendkowsky}}, \bibinfo
  {author} {\bibfnamefont {B.}~\bibnamefont {Butscher}}, \bibinfo {author}
  {\bibfnamefont {R.}~\bibnamefont {L\"ow}}, \bibinfo {author} {\bibfnamefont
  {L.}~\bibnamefont {Santos}}, \ and\ \bibinfo {author} {\bibfnamefont
  {T.}~\bibnamefont {Pfau}},\ }\href {\doibase 10.1103/PhysRevLett.99.163601}
  {\bibfield  {journal} {\bibinfo  {journal} {Phys. Rev. Lett.}\ }\textbf
  {\bibinfo {volume} {99}},\ \bibinfo {pages} {163601} (\bibinfo {year}
  {2007})}\BibitemShut {NoStop}%
\bibitem [{\citenamefont {Dudin}\ and\ \citenamefont
  {Kuzmich}(2012)}]{Dudin12}%
  \BibitemOpen
  \bibfield  {author} {\bibinfo {author} {\bibfnamefont {Y.~O.}\ \bibnamefont
  {Dudin}}\ and\ \bibinfo {author} {\bibfnamefont {A.}~\bibnamefont
  {Kuzmich}},\ }\href {\doibase 10.1126/science.1217901} {\bibfield  {journal}
  {\bibinfo  {journal} {Science}\ }\textbf {\bibinfo {volume} {336}},\ \bibinfo
  {pages} {887} (\bibinfo {year} {2012})}\BibitemShut {NoStop}%
\bibitem [{\citenamefont {Ebert}\ \emph {et~al.}(2014)\citenamefont {Ebert},
  \citenamefont {Gill}, \citenamefont {Gibbons}, \citenamefont {Zhang},
  \citenamefont {Saffman},\ and\ \citenamefont {Walker}}]{Ebert14}%
  \BibitemOpen
  \bibfield  {author} {\bibinfo {author} {\bibfnamefont {M.}~\bibnamefont
  {Ebert}}, \bibinfo {author} {\bibfnamefont {A.}~\bibnamefont {Gill}},
  \bibinfo {author} {\bibfnamefont {M.}~\bibnamefont {Gibbons}}, \bibinfo
  {author} {\bibfnamefont {X.}~\bibnamefont {Zhang}}, \bibinfo {author}
  {\bibfnamefont {M.}~\bibnamefont {Saffman}}, \ and\ \bibinfo {author}
  {\bibfnamefont {T.~G.}\ \bibnamefont {Walker}},\ }\href {\doibase
  10.1103/PhysRevLett.112.043602} {\bibfield  {journal} {\bibinfo  {journal}
  {Phys. Rev. Lett.}\ }\textbf {\bibinfo {volume} {112}},\ \bibinfo {pages}
  {043602} (\bibinfo {year} {2014})}\BibitemShut {NoStop}%
\bibitem [{\citenamefont {Ebert}\ \emph {et~al.}(2015)\citenamefont {Ebert},
  \citenamefont {Kwon}, \citenamefont {Walker},\ and\ \citenamefont
  {Saffman}}]{Ebert15}%
  \BibitemOpen
  \bibfield  {author} {\bibinfo {author} {\bibfnamefont {M.}~\bibnamefont
  {Ebert}}, \bibinfo {author} {\bibfnamefont {M.}~\bibnamefont {Kwon}},
  \bibinfo {author} {\bibfnamefont {T.~G.}\ \bibnamefont {Walker}}, \ and\
  \bibinfo {author} {\bibfnamefont {M.}~\bibnamefont {Saffman}},\ }\href
  {\doibase 10.1103/PhysRevLett.115.093601} {\bibfield  {journal} {\bibinfo
  {journal} {Phys. Rev. Lett.}\ }\textbf {\bibinfo {volume} {115}},\ \bibinfo
  {pages} {093601} (\bibinfo {year} {2015})}\BibitemShut {NoStop}%
\bibitem [{\citenamefont {Weber}\ \emph
  {et~al.}(2015{\natexlab{a}})\citenamefont {Weber}, \citenamefont {Honing},
  \citenamefont {Niederprum}, \citenamefont {Manthey}, \citenamefont {Thomas},
  \citenamefont {Guarrera}, \citenamefont {Fleischhauer}, \citenamefont
  {Barontini},\ and\ \citenamefont {Ott}}]{Weber15}%
  \BibitemOpen
  \bibfield  {author} {\bibinfo {author} {\bibfnamefont {T.~M.}\ \bibnamefont
  {Weber}}, \bibinfo {author} {\bibfnamefont {M.}~\bibnamefont {Honing}},
  \bibinfo {author} {\bibfnamefont {T.}~\bibnamefont {Niederprum}}, \bibinfo
  {author} {\bibfnamefont {T.}~\bibnamefont {Manthey}}, \bibinfo {author}
  {\bibfnamefont {O.}~\bibnamefont {Thomas}}, \bibinfo {author} {\bibfnamefont
  {V.}~\bibnamefont {Guarrera}}, \bibinfo {author} {\bibfnamefont
  {M.}~\bibnamefont {Fleischhauer}}, \bibinfo {author} {\bibfnamefont
  {G.}~\bibnamefont {Barontini}}, \ and\ \bibinfo {author} {\bibfnamefont
  {H.}~\bibnamefont {Ott}},\ }\href {http://dx.doi.org/10.1038/nphys3214}
  {\bibfield  {journal} {\bibinfo  {journal} {Nat Phys}\ }\textbf {\bibinfo
  {volume} {11}},\ \bibinfo {pages} {157} (\bibinfo {year}
  {2015}{\natexlab{a}})}\BibitemShut {NoStop}%
\bibitem [{\citenamefont {Zeiher}\ \emph {et~al.}(2015)\citenamefont {Zeiher},
  \citenamefont {Schau\ss{}}, \citenamefont {Hild}, \citenamefont {Macr\`{\i}},
  \citenamefont {Bloch},\ and\ \citenamefont {Gross}}]{Zeiher15}%
  \BibitemOpen
  \bibfield  {author} {\bibinfo {author} {\bibfnamefont {J.}~\bibnamefont
  {Zeiher}}, \bibinfo {author} {\bibfnamefont {P.}~\bibnamefont {Schau\ss{}}},
  \bibinfo {author} {\bibfnamefont {S.}~\bibnamefont {Hild}}, \bibinfo {author}
  {\bibfnamefont {T.}~\bibnamefont {Macr\`{\i}}}, \bibinfo {author}
  {\bibfnamefont {I.}~\bibnamefont {Bloch}}, \ and\ \bibinfo {author}
  {\bibfnamefont {C.}~\bibnamefont {Gross}},\ }\href {\doibase
  10.1103/PhysRevX.5.031015} {\bibfield  {journal} {\bibinfo  {journal} {Phys.
  Rev. X}\ }\textbf {\bibinfo {volume} {5}},\ \bibinfo {pages} {031015}
  (\bibinfo {year} {2015})}\BibitemShut {NoStop}%
\bibitem [{\citenamefont {Labuhn}\ \emph
  {et~al.}(2016{\natexlab{a}})\citenamefont {Labuhn}, \citenamefont {Barredo},
  \citenamefont {Ravets}, \citenamefont {de~L\'es\'eleuc}, \citenamefont
  {Macr\`{\i}}, \citenamefont {Lahaye},\ and\ \citenamefont
  {Browaeys}}]{Labuhn16}%
  \BibitemOpen
  \bibfield  {author} {\bibinfo {author} {\bibfnamefont {H.}~\bibnamefont
  {Labuhn}}, \bibinfo {author} {\bibfnamefont {D.}~\bibnamefont {Barredo}},
  \bibinfo {author} {\bibfnamefont {S.}~\bibnamefont {Ravets}}, \bibinfo
  {author} {\bibfnamefont {S.}~\bibnamefont {de~L\'es\'eleuc}}, \bibinfo
  {author} {\bibfnamefont {T.}~\bibnamefont {Macr\`{\i}}}, \bibinfo {author}
  {\bibfnamefont {T.}~\bibnamefont {Lahaye}}, \ and\ \bibinfo {author}
  {\bibfnamefont {A.}~\bibnamefont {Browaeys}},\ }\href
  {http://dx.doi.org/10.1038/nature18274} {\bibfield  {journal} {\bibinfo
  {journal} {Nature}\ }\textbf {\bibinfo {volume} {534}},\ \bibinfo {pages}
  {667} (\bibinfo {year} {2016}{\natexlab{a}})}\BibitemShut {NoStop}%
\bibitem [{\citenamefont {Weimer}\ \emph {et~al.}(2008)\citenamefont {Weimer},
  \citenamefont {L\"ow}, \citenamefont {Pfau},\ and\ \citenamefont
  {B\"uchler}}]{Weimer08}%
  \BibitemOpen
  \bibfield  {author} {\bibinfo {author} {\bibfnamefont {H.}~\bibnamefont
  {Weimer}}, \bibinfo {author} {\bibfnamefont {R.}~\bibnamefont {L\"ow}},
  \bibinfo {author} {\bibfnamefont {T.}~\bibnamefont {Pfau}}, \ and\ \bibinfo
  {author} {\bibfnamefont {H.~P.}\ \bibnamefont {B\"uchler}},\ }\href {\doibase
  10.1103/PhysRevLett.101.250601} {\bibfield  {journal} {\bibinfo  {journal}
  {Phys. Rev. Lett.}\ }\textbf {\bibinfo {volume} {101}},\ \bibinfo {pages}
  {250601} (\bibinfo {year} {2008})}\BibitemShut {NoStop}%
\bibitem [{\citenamefont {Sela}\ \emph {et~al.}(2011)\citenamefont {Sela},
  \citenamefont {Punk},\ and\ \citenamefont {Garst}}]{Sela11}%
  \BibitemOpen
  \bibfield  {author} {\bibinfo {author} {\bibfnamefont {E.}~\bibnamefont
  {Sela}}, \bibinfo {author} {\bibfnamefont {M.}~\bibnamefont {Punk}}, \ and\
  \bibinfo {author} {\bibfnamefont {M.}~\bibnamefont {Garst}},\ }\href
  {\doibase 10.1103/PhysRevB.84.085434} {\bibfield  {journal} {\bibinfo
  {journal} {Phys. Rev. B}\ }\textbf {\bibinfo {volume} {84}},\ \bibinfo
  {pages} {085434} (\bibinfo {year} {2011})}\BibitemShut {NoStop}%
\bibitem [{\citenamefont {Pohl}\ \emph {et~al.}(2010)\citenamefont {Pohl},
  \citenamefont {Demler},\ and\ \citenamefont {Lukin}}]{Pohl10}%
  \BibitemOpen
  \bibfield  {author} {\bibinfo {author} {\bibfnamefont {T.}~\bibnamefont
  {Pohl}}, \bibinfo {author} {\bibfnamefont {E.}~\bibnamefont {Demler}}, \ and\
  \bibinfo {author} {\bibfnamefont {M.~D.}\ \bibnamefont {Lukin}},\ }\href
  {\doibase 10.1103/PhysRevLett.104.043002} {\bibfield  {journal} {\bibinfo
  {journal} {Phys. Rev. Lett.}\ }\textbf {\bibinfo {volume} {104}},\ \bibinfo
  {pages} {043002} (\bibinfo {year} {2010})}\BibitemShut {NoStop}%
\bibitem [{\citenamefont {Lesanovsky}(2011)}]{Lesanovsky11}%
  \BibitemOpen
  \bibfield  {author} {\bibinfo {author} {\bibfnamefont {I.}~\bibnamefont
  {Lesanovsky}},\ }\href {\doibase 10.1103/PhysRevLett.106.025301} {\bibfield
  {journal} {\bibinfo  {journal} {Phys. Rev. Lett.}\ }\textbf {\bibinfo
  {volume} {106}},\ \bibinfo {pages} {025301} (\bibinfo {year}
  {2011})}\BibitemShut {NoStop}%
\bibitem [{\citenamefont {Barredo}\ \emph {et~al.}(2014)\citenamefont
  {Barredo}, \citenamefont {Ravets}, \citenamefont {Labuhn}, \citenamefont
  {B\'eguin}, \citenamefont {Vernier}, \citenamefont {Nogrette}, \citenamefont
  {Lahaye},\ and\ \citenamefont {Browaeys}}]{Barredo14}%
  \BibitemOpen
  \bibfield  {author} {\bibinfo {author} {\bibfnamefont {D.}~\bibnamefont
  {Barredo}}, \bibinfo {author} {\bibfnamefont {S.}~\bibnamefont {Ravets}},
  \bibinfo {author} {\bibfnamefont {H.}~\bibnamefont {Labuhn}}, \bibinfo
  {author} {\bibfnamefont {L.}~\bibnamefont {B\'eguin}}, \bibinfo {author}
  {\bibfnamefont {A.}~\bibnamefont {Vernier}}, \bibinfo {author} {\bibfnamefont
  {F.}~\bibnamefont {Nogrette}}, \bibinfo {author} {\bibfnamefont
  {T.}~\bibnamefont {Lahaye}}, \ and\ \bibinfo {author} {\bibfnamefont
  {A.}~\bibnamefont {Browaeys}},\ }\href {\doibase
  10.1103/PhysRevLett.112.183002} {\bibfield  {journal} {\bibinfo  {journal}
  {Phys. Rev. Lett.}\ }\textbf {\bibinfo {volume} {112}},\ \bibinfo {pages}
  {183002} (\bibinfo {year} {2014})}\BibitemShut {NoStop}%
\bibitem [{\citenamefont {Zeiher}\ \emph {et~al.}(2016)\citenamefont {Zeiher},
  \citenamefont {van Bijnen}, \citenamefont {Schau\ss{}}, \citenamefont {Hild},
  \citenamefont {Choi}, \citenamefont {Pohl}, \citenamefont {Bloch},\ and\
  \citenamefont {Gross}}]{Zeiher16}%
  \BibitemOpen
  \bibfield  {author} {\bibinfo {author} {\bibfnamefont {J.}~\bibnamefont
  {Zeiher}}, \bibinfo {author} {\bibfnamefont {R.}~\bibnamefont {van Bijnen}},
  \bibinfo {author} {\bibfnamefont {P.}~\bibnamefont {Schau\ss{}}}, \bibinfo
  {author} {\bibfnamefont {S.}~\bibnamefont {Hild}}, \bibinfo {author}
  {\bibfnamefont {J.-y.}\ \bibnamefont {Choi}}, \bibinfo {author}
  {\bibfnamefont {T.}~\bibnamefont {Pohl}}, \bibinfo {author} {\bibfnamefont
  {I.}~\bibnamefont {Bloch}}, \ and\ \bibinfo {author} {\bibfnamefont
  {C.}~\bibnamefont {Gross}},\ }\href {http://dx.doi.org/10.1038/nphys3835}
  {\bibfield  {journal} {\bibinfo  {journal} {Nat. Phys.}\ }\textbf {\bibinfo
  {volume} {12}},\ \bibinfo {pages} {1095} (\bibinfo {year}
  {2016})}\BibitemShut {NoStop}%
\bibitem [{\citenamefont {van Ditzhuijzen}\ \emph {et~al.}(2008)\citenamefont
  {van Ditzhuijzen}, \citenamefont {Koenderink}, \citenamefont {Hern\'andez},
  \citenamefont {Robicheaux}, \citenamefont {Noordam},\ and\ \citenamefont
  {van~den Heuvell}}]{Ditzhuijzen08}%
  \BibitemOpen
  \bibfield  {author} {\bibinfo {author} {\bibfnamefont {C.~S.~E.}\
  \bibnamefont {van Ditzhuijzen}}, \bibinfo {author} {\bibfnamefont {A.~F.}\
  \bibnamefont {Koenderink}}, \bibinfo {author} {\bibfnamefont {J.~V.}\
  \bibnamefont {Hern\'andez}}, \bibinfo {author} {\bibfnamefont
  {F.}~\bibnamefont {Robicheaux}}, \bibinfo {author} {\bibfnamefont {L.~D.}\
  \bibnamefont {Noordam}}, \ and\ \bibinfo {author} {\bibfnamefont {H.~B.
  v.~L.}\ \bibnamefont {van~den Heuvell}},\ }\href {\doibase
  10.1103/PhysRevLett.100.243201} {\bibfield  {journal} {\bibinfo  {journal}
  {Phys. Rev. Lett.}\ }\textbf {\bibinfo {volume} {100}},\ \bibinfo {pages}
  {243201} (\bibinfo {year} {2008})}\BibitemShut {NoStop}%
\bibitem [{\citenamefont {Ravets}\ \emph {et~al.}(2014)\citenamefont {Ravets},
  \citenamefont {Labuhn}, \citenamefont {Barredo}, \citenamefont {Beguin},
  \citenamefont {Lahaye},\ and\ \citenamefont {Browaeys}}]{Ravets14}%
  \BibitemOpen
  \bibfield  {author} {\bibinfo {author} {\bibfnamefont {S.}~\bibnamefont
  {Ravets}}, \bibinfo {author} {\bibfnamefont {H.}~\bibnamefont {Labuhn}},
  \bibinfo {author} {\bibfnamefont {D.}~\bibnamefont {Barredo}}, \bibinfo
  {author} {\bibfnamefont {L.}~\bibnamefont {Beguin}}, \bibinfo {author}
  {\bibfnamefont {T.}~\bibnamefont {Lahaye}}, \ and\ \bibinfo {author}
  {\bibfnamefont {A.}~\bibnamefont {Browaeys}},\ }\href
  {http://dx.doi.org/10.1038/nphys3119} {\bibfield  {journal} {\bibinfo
  {journal} {Nat. Phys.}\ }\textbf {\bibinfo {volume} {10}},\ \bibinfo {pages}
  {914} (\bibinfo {year} {2014})}\BibitemShut {NoStop}%
\bibitem [{\citenamefont {Fahey}\ \emph {et~al.}(2015)\citenamefont {Fahey},
  \citenamefont {Carroll},\ and\ \citenamefont {Noel}}]{Fahey15}%
  \BibitemOpen
  \bibfield  {author} {\bibinfo {author} {\bibfnamefont {D.~P.}\ \bibnamefont
  {Fahey}}, \bibinfo {author} {\bibfnamefont {T.~J.}\ \bibnamefont {Carroll}},
  \ and\ \bibinfo {author} {\bibfnamefont {M.~W.}\ \bibnamefont {Noel}},\
  }\href {\doibase 10.1103/PhysRevA.91.062702} {\bibfield  {journal} {\bibinfo
  {journal} {Phys. Rev. A}\ }\textbf {\bibinfo {volume} {91}},\ \bibinfo
  {pages} {062702} (\bibinfo {year} {2015})}\BibitemShut {NoStop}%
\bibitem [{\citenamefont {Barredo}\ \emph {et~al.}(2015)\citenamefont
  {Barredo}, \citenamefont {Labuhn}, \citenamefont {Ravets}, \citenamefont
  {Lahaye}, \citenamefont {Browaeys},\ and\ \citenamefont {Adams}}]{Barredo15}%
  \BibitemOpen
  \bibfield  {author} {\bibinfo {author} {\bibfnamefont {D.}~\bibnamefont
  {Barredo}}, \bibinfo {author} {\bibfnamefont {H.}~\bibnamefont {Labuhn}},
  \bibinfo {author} {\bibfnamefont {S.}~\bibnamefont {Ravets}}, \bibinfo
  {author} {\bibfnamefont {T.}~\bibnamefont {Lahaye}}, \bibinfo {author}
  {\bibfnamefont {A.}~\bibnamefont {Browaeys}}, \ and\ \bibinfo {author}
  {\bibfnamefont {C.~S.}\ \bibnamefont {Adams}},\ }\href {\doibase
  10.1103/PhysRevLett.114.113002} {\bibfield  {journal} {\bibinfo  {journal}
  {Phys. Rev. Lett.}\ }\textbf {\bibinfo {volume} {114}},\ \bibinfo {pages}
  {113002} (\bibinfo {year} {2015})}\BibitemShut {NoStop}%
\bibitem [{\citenamefont {G{\"u}nter}\ \emph {et~al.}(2013)\citenamefont
  {G{\"u}nter}, \citenamefont {Schempp}, \citenamefont {Robert-de
  Saint-Vincent}, \citenamefont {Gavryusev}, \citenamefont {Helmrich},
  \citenamefont {Hofmann}, \citenamefont {Whitlock},\ and\ \citenamefont
  {Weidem{\"u}ller}}]{Guenter13}%
  \BibitemOpen
  \bibfield  {author} {\bibinfo {author} {\bibfnamefont {G.}~\bibnamefont
  {G{\"u}nter}}, \bibinfo {author} {\bibfnamefont {H.}~\bibnamefont {Schempp}},
  \bibinfo {author} {\bibfnamefont {M.}~\bibnamefont {Robert-de
  Saint-Vincent}}, \bibinfo {author} {\bibfnamefont {V.}~\bibnamefont
  {Gavryusev}}, \bibinfo {author} {\bibfnamefont {S.}~\bibnamefont {Helmrich}},
  \bibinfo {author} {\bibfnamefont {C.~S.}\ \bibnamefont {Hofmann}}, \bibinfo
  {author} {\bibfnamefont {S.}~\bibnamefont {Whitlock}}, \ and\ \bibinfo
  {author} {\bibfnamefont {M.}~\bibnamefont {Weidem{\"u}ller}},\ }\href
  {\doibase 10.1126/science.1244843} {\bibfield  {journal} {\bibinfo  {journal}
  {Science}\ }\textbf {\bibinfo {volume} {342}},\ \bibinfo {pages} {954}
  (\bibinfo {year} {2013})}\BibitemShut {NoStop}%
\bibitem [{\citenamefont {Glaetzle}\ \emph {et~al.}(2015)\citenamefont
  {Glaetzle}, \citenamefont {Dalmonte}, \citenamefont {Nath}, \citenamefont
  {Gross}, \citenamefont {Bloch},\ and\ \citenamefont {Zoller}}]{Glaetzle15}%
  \BibitemOpen
  \bibfield  {author} {\bibinfo {author} {\bibfnamefont {A.~W.}\ \bibnamefont
  {Glaetzle}}, \bibinfo {author} {\bibfnamefont {M.}~\bibnamefont {Dalmonte}},
  \bibinfo {author} {\bibfnamefont {R.}~\bibnamefont {Nath}}, \bibinfo {author}
  {\bibfnamefont {C.}~\bibnamefont {Gross}}, \bibinfo {author} {\bibfnamefont
  {I.}~\bibnamefont {Bloch}}, \ and\ \bibinfo {author} {\bibfnamefont
  {P.}~\bibnamefont {Zoller}},\ }\href {\doibase
  10.1103/PhysRevLett.114.173002} {\bibfield  {journal} {\bibinfo  {journal}
  {Phys. Rev. Lett.}\ }\textbf {\bibinfo {volume} {114}},\ \bibinfo {pages}
  {173002} (\bibinfo {year} {2015})}\BibitemShut {NoStop}%
\bibitem [{\citenamefont {van Bijnen}\ and\ \citenamefont
  {Pohl}(2015)}]{Bijnen15}%
  \BibitemOpen
  \bibfield  {author} {\bibinfo {author} {\bibfnamefont {R.~M.~W.}\
  \bibnamefont {van Bijnen}}\ and\ \bibinfo {author} {\bibfnamefont
  {T.}~\bibnamefont {Pohl}},\ }\href {\doibase 10.1103/PhysRevLett.114.243002}
  {\bibfield  {journal} {\bibinfo  {journal} {Phys. Rev. Lett.}\ }\textbf
  {\bibinfo {volume} {114}},\ \bibinfo {pages} {243002} (\bibinfo {year}
  {2015})}\BibitemShut {NoStop}%
\bibitem [{\citenamefont {Lee}\ \emph {et~al.}(2011)\citenamefont {Lee},
  \citenamefont {H\"affner},\ and\ \citenamefont {Cross}}]{Lee11}%
  \BibitemOpen
  \bibfield  {author} {\bibinfo {author} {\bibfnamefont {T.~E.}\ \bibnamefont
  {Lee}}, \bibinfo {author} {\bibfnamefont {H.}~\bibnamefont {H\"affner}}, \
  and\ \bibinfo {author} {\bibfnamefont {M.~C.}\ \bibnamefont {Cross}},\ }\href
  {\doibase 10.1103/PhysRevA.84.031402} {\bibfield  {journal} {\bibinfo
  {journal} {Phys. Rev. A}\ }\textbf {\bibinfo {volume} {84}},\ \bibinfo
  {pages} {031402} (\bibinfo {year} {2011})}\BibitemShut {NoStop}%
\bibitem [{\citenamefont {H\"oning}\ \emph {et~al.}(2013)\citenamefont
  {H\"oning}, \citenamefont {Muth}, \citenamefont {Petrosyan},\ and\
  \citenamefont {Fleischhauer}}]{Hoening13}%
  \BibitemOpen
  \bibfield  {author} {\bibinfo {author} {\bibfnamefont {M.}~\bibnamefont
  {H\"oning}}, \bibinfo {author} {\bibfnamefont {D.}~\bibnamefont {Muth}},
  \bibinfo {author} {\bibfnamefont {D.}~\bibnamefont {Petrosyan}}, \ and\
  \bibinfo {author} {\bibfnamefont {M.}~\bibnamefont {Fleischhauer}},\ }\href
  {\doibase 10.1103/PhysRevA.87.023401} {\bibfield  {journal} {\bibinfo
  {journal} {Phys. Rev. A}\ }\textbf {\bibinfo {volume} {87}},\ \bibinfo
  {pages} {023401} (\bibinfo {year} {2013})}\BibitemShut {NoStop}%
\bibitem [{\citenamefont {Marcuzzi}\ \emph {et~al.}(2014)\citenamefont
  {Marcuzzi}, \citenamefont {Levi}, \citenamefont {Diehl}, \citenamefont
  {Garrahan},\ and\ \citenamefont {Lesanovsky}}]{Marcuzzi14}%
  \BibitemOpen
  \bibfield  {author} {\bibinfo {author} {\bibfnamefont {M.}~\bibnamefont
  {Marcuzzi}}, \bibinfo {author} {\bibfnamefont {E.}~\bibnamefont {Levi}},
  \bibinfo {author} {\bibfnamefont {S.}~\bibnamefont {Diehl}}, \bibinfo
  {author} {\bibfnamefont {J.~P.}\ \bibnamefont {Garrahan}}, \ and\ \bibinfo
  {author} {\bibfnamefont {I.}~\bibnamefont {Lesanovsky}},\ }\href {\doibase
  10.1103/PhysRevLett.113.210401} {\bibfield  {journal} {\bibinfo  {journal}
  {Phys. Rev. Lett.}\ }\textbf {\bibinfo {volume} {113}},\ \bibinfo {pages}
  {210401} (\bibinfo {year} {2014})}\BibitemShut {NoStop}%
\bibitem [{\citenamefont {Sanders}\ \emph {et~al.}(2014)\citenamefont
  {Sanders}, \citenamefont {van Bijnen}, \citenamefont {Vredenbregt},\ and\
  \citenamefont {Kokkelmans}}]{Sanders14}%
  \BibitemOpen
  \bibfield  {author} {\bibinfo {author} {\bibfnamefont {J.}~\bibnamefont
  {Sanders}}, \bibinfo {author} {\bibfnamefont {R.}~\bibnamefont {van Bijnen}},
  \bibinfo {author} {\bibfnamefont {E.}~\bibnamefont {Vredenbregt}}, \ and\
  \bibinfo {author} {\bibfnamefont {S.}~\bibnamefont {Kokkelmans}},\ }\href
  {\doibase 10.1103/PhysRevLett.112.163001} {\bibfield  {journal} {\bibinfo
  {journal} {Phys. Rev. Lett.}\ }\textbf {\bibinfo {volume} {112}},\ \bibinfo
  {pages} {163001} (\bibinfo {year} {2014})}\BibitemShut {NoStop}%
\bibitem [{\citenamefont {Hoening}\ \emph {et~al.}(2014)\citenamefont
  {Hoening}, \citenamefont {Abdussalam}, \citenamefont {Fleischhauer},\ and\
  \citenamefont {Pohl}}]{Hoening14}%
  \BibitemOpen
  \bibfield  {author} {\bibinfo {author} {\bibfnamefont {M.}~\bibnamefont
  {Hoening}}, \bibinfo {author} {\bibfnamefont {W.}~\bibnamefont {Abdussalam}},
  \bibinfo {author} {\bibfnamefont {M.}~\bibnamefont {Fleischhauer}}, \ and\
  \bibinfo {author} {\bibfnamefont {T.}~\bibnamefont {Pohl}},\ }\href {\doibase
  10.1103/PhysRevA.90.021603} {\bibfield  {journal} {\bibinfo  {journal} {Phys.
  Rev. A}\ }\textbf {\bibinfo {volume} {90}},\ \bibinfo {pages} {021603}
  (\bibinfo {year} {2014})}\BibitemShut {NoStop}%
\bibitem [{\citenamefont {{Helmrich}}\ \emph {et~al.}(2016)\citenamefont
  {{Helmrich}}, \citenamefont {{Arias}},\ and\ \citenamefont
  {{Whitlock}}}]{Helmrich16}%
  \BibitemOpen
  \bibfield  {author} {\bibinfo {author} {\bibfnamefont {S.}~\bibnamefont
  {{Helmrich}}}, \bibinfo {author} {\bibfnamefont {A.}~\bibnamefont {{Arias}}},
  \ and\ \bibinfo {author} {\bibfnamefont {S.}~\bibnamefont {{Whitlock}}},\
  }\href@noop {} {\bibfield  {journal} {\bibinfo  {journal} {ArXiv e-prints}\ }
  (\bibinfo {year} {2016})},\ \Eprint {http://arxiv.org/abs/1605.08609}
  {arXiv:1605.08609 [physics.atom-ph]} \BibitemShut {NoStop}%
\bibitem [{\citenamefont {Overbeck}\ \emph {et~al.}(2017)\citenamefont
  {Overbeck}, \citenamefont {Maghrebi}, \citenamefont {Gorshkov},\ and\
  \citenamefont {Weimer}}]{Overbeck16}%
  \BibitemOpen
  \bibfield  {author} {\bibinfo {author} {\bibfnamefont {V.~R.}\ \bibnamefont
  {Overbeck}}, \bibinfo {author} {\bibfnamefont {M.~F.}\ \bibnamefont
  {Maghrebi}}, \bibinfo {author} {\bibfnamefont {A.~V.}\ \bibnamefont
  {Gorshkov}}, \ and\ \bibinfo {author} {\bibfnamefont {H.}~\bibnamefont
  {Weimer}},\ }\href {\doibase 10.1103/PhysRevA.95.042133} {\bibfield
  {journal} {\bibinfo  {journal} {Phys. Rev. A}\ }\textbf {\bibinfo {volume}
  {95}},\ \bibinfo {pages} {042133} (\bibinfo {year} {2017})}\BibitemShut
  {NoStop}%
\bibitem [{\citenamefont {{Roghani}}\ and\ \citenamefont
  {{Weimer}}(2016)}]{Roghani16}%
  \BibitemOpen
  \bibfield  {author} {\bibinfo {author} {\bibfnamefont {M.}~\bibnamefont
  {{Roghani}}}\ and\ \bibinfo {author} {\bibfnamefont {H.}~\bibnamefont
  {{Weimer}}},\ }\href@noop {} {\bibfield  {journal} {\bibinfo  {journal}
  {ArXiv e-prints}\ } (\bibinfo {year} {2016})},\ \Eprint
  {http://arxiv.org/abs/1611.09612} {arXiv:1611.09612 [quant-ph]} \BibitemShut
  {NoStop}%
\bibitem [{\citenamefont {Weimer}\ \emph {et~al.}(2010)\citenamefont {Weimer},
  \citenamefont {Muller}, \citenamefont {Lesanovsky}, \citenamefont {Zoller},\
  and\ \citenamefont {Buchler}}]{Weimer10qsim}%
  \BibitemOpen
  \bibfield  {author} {\bibinfo {author} {\bibfnamefont {H.}~\bibnamefont
  {Weimer}}, \bibinfo {author} {\bibfnamefont {M.}~\bibnamefont {Muller}},
  \bibinfo {author} {\bibfnamefont {I.}~\bibnamefont {Lesanovsky}}, \bibinfo
  {author} {\bibfnamefont {P.}~\bibnamefont {Zoller}}, \ and\ \bibinfo {author}
  {\bibfnamefont {H.~P.}\ \bibnamefont {Buchler}},\ }\href
  {http://dx.doi.org/10.1038/nphys1614} {\bibfield  {journal} {\bibinfo
  {journal} {Nat Phys}\ }\textbf {\bibinfo {volume} {6}},\ \bibinfo {pages}
  {382} (\bibinfo {year} {2010})}\BibitemShut {NoStop}%
\bibitem [{\citenamefont {{Glaetzle}}\ \emph {et~al.}(2016)\citenamefont
  {{Glaetzle}}, \citenamefont {{van Bijnen}}, \citenamefont {{Zoller}},\ and\
  \citenamefont {{Lechner}}}]{Annealer}%
  \BibitemOpen
  \bibfield  {author} {\bibinfo {author} {\bibfnamefont {A.~W.}\ \bibnamefont
  {{Glaetzle}}}, \bibinfo {author} {\bibfnamefont {R.~M.~W.}\ \bibnamefont
  {{van Bijnen}}}, \bibinfo {author} {\bibfnamefont {P.}~\bibnamefont
  {{Zoller}}}, \ and\ \bibinfo {author} {\bibfnamefont {W.}~\bibnamefont
  {{Lechner}}},\ }\href@noop {} {\bibfield  {journal} {\bibinfo  {journal}
  {ArXiv e-prints}\ } (\bibinfo {year} {2016})},\ \Eprint
  {http://arxiv.org/abs/1611.02594} {arXiv:1611.02594 [quant-ph]} \BibitemShut
  {NoStop}%
\bibitem [{\citenamefont {Weimer}\ and\ \citenamefont
  {B\"uchler}(2010)}]{Weimer10}%
  \BibitemOpen
  \bibfield  {author} {\bibinfo {author} {\bibfnamefont {H.}~\bibnamefont
  {Weimer}}\ and\ \bibinfo {author} {\bibfnamefont {H.~P.}\ \bibnamefont
  {B\"uchler}},\ }\href {\doibase 10.1103/PhysRevLett.105.230403} {\bibfield
  {journal} {\bibinfo  {journal} {Phys. Rev. Lett.}\ }\textbf {\bibinfo
  {volume} {105}},\ \bibinfo {pages} {230403} (\bibinfo {year}
  {2010})}\BibitemShut {NoStop}%
\bibitem [{\citenamefont {Glaetzle}\ \emph {et~al.}(2014)\citenamefont
  {Glaetzle}, \citenamefont {Dalmonte}, \citenamefont {Nath}, \citenamefont
  {Rousochatzakis}, \citenamefont {Moessner},\ and\ \citenamefont
  {Zoller}}]{Glaetzle14}%
  \BibitemOpen
  \bibfield  {author} {\bibinfo {author} {\bibfnamefont {A.~W.}\ \bibnamefont
  {Glaetzle}}, \bibinfo {author} {\bibfnamefont {M.}~\bibnamefont {Dalmonte}},
  \bibinfo {author} {\bibfnamefont {R.}~\bibnamefont {Nath}}, \bibinfo {author}
  {\bibfnamefont {I.}~\bibnamefont {Rousochatzakis}}, \bibinfo {author}
  {\bibfnamefont {R.}~\bibnamefont {Moessner}}, \ and\ \bibinfo {author}
  {\bibfnamefont {P.}~\bibnamefont {Zoller}},\ }\href {\doibase
  10.1103/PhysRevX.4.041037} {\bibfield  {journal} {\bibinfo  {journal} {Phys.
  Rev. X}\ }\textbf {\bibinfo {volume} {4}},\ \bibinfo {pages} {041037}
  (\bibinfo {year} {2014})}\BibitemShut {NoStop}%
\bibitem [{\citenamefont {Beterov}\ \emph
  {et~al.}(2009{\natexlab{a}})\citenamefont {Beterov}, \citenamefont
  {Ryabtsev}, \citenamefont {Tretyakov},\ and\ \citenamefont
  {Entin}}]{Beterov09}%
  \BibitemOpen
  \bibfield  {author} {\bibinfo {author} {\bibfnamefont {I.~I.}\ \bibnamefont
  {Beterov}}, \bibinfo {author} {\bibfnamefont {I.~I.}\ \bibnamefont
  {Ryabtsev}}, \bibinfo {author} {\bibfnamefont {D.~B.}\ \bibnamefont
  {Tretyakov}}, \ and\ \bibinfo {author} {\bibfnamefont {V.~M.}\ \bibnamefont
  {Entin}},\ }\href {\doibase 10.1103/PhysRevA.79.052504} {\bibfield  {journal}
  {\bibinfo  {journal} {Phys. Rev. A}\ }\textbf {\bibinfo {volume} {79}},\
  \bibinfo {pages} {052504} (\bibinfo {year} {2009}{\natexlab{a}})}\BibitemShut
  {NoStop}%
\bibitem [{\citenamefont {Schauss}\ \emph {et~al.}(2015)\citenamefont
  {Schauss}, \citenamefont {Zeiher}, \citenamefont {Fukuhara}, \citenamefont
  {Hild}, \citenamefont {Cheneau}, \citenamefont {Macr{\`i}}, \citenamefont
  {Pohl}, \citenamefont {Bloch},\ and\ \citenamefont {Gross}}]{Schauss15}%
  \BibitemOpen
  \bibfield  {author} {\bibinfo {author} {\bibfnamefont {P.}~\bibnamefont
  {Schauss}}, \bibinfo {author} {\bibfnamefont {J.}~\bibnamefont {Zeiher}},
  \bibinfo {author} {\bibfnamefont {T.}~\bibnamefont {Fukuhara}}, \bibinfo
  {author} {\bibfnamefont {S.}~\bibnamefont {Hild}}, \bibinfo {author}
  {\bibfnamefont {M.}~\bibnamefont {Cheneau}}, \bibinfo {author} {\bibfnamefont
  {T.}~\bibnamefont {Macr{\`i}}}, \bibinfo {author} {\bibfnamefont
  {T.}~\bibnamefont {Pohl}}, \bibinfo {author} {\bibfnamefont {I.}~\bibnamefont
  {Bloch}}, \ and\ \bibinfo {author} {\bibfnamefont {C.}~\bibnamefont
  {Gross}},\ }\href {\doibase 10.1126/science.1258351} {\bibfield  {journal}
  {\bibinfo  {journal} {Science}\ }\textbf {\bibinfo {volume} {347}},\ \bibinfo
  {pages} {1455} (\bibinfo {year} {2015})}\BibitemShut {NoStop}%
\bibitem [{\citenamefont {Bak}\ and\ \citenamefont
  {Bruinsma}(1982)}]{BakBruinsma}%
  \BibitemOpen
  \bibfield  {author} {\bibinfo {author} {\bibfnamefont {P.}~\bibnamefont
  {Bak}}\ and\ \bibinfo {author} {\bibfnamefont {R.}~\bibnamefont {Bruinsma}},\
  }\href {\doibase 10.1103/PhysRevLett.49.249} {\bibfield  {journal} {\bibinfo
  {journal} {Phys. Rev. Lett.}\ }\textbf {\bibinfo {volume} {49}},\ \bibinfo
  {pages} {249} (\bibinfo {year} {1982})}\BibitemShut {NoStop}%
\bibitem [{\citenamefont {Schachenmayer}\ \emph
  {et~al.}(2010{\natexlab{a}})\citenamefont {Schachenmayer}, \citenamefont
  {Lesanovsky}, \citenamefont {Micheli},\ and\ \citenamefont
  {Daley}}]{Schachenmayer10}%
  \BibitemOpen
  \bibfield  {author} {\bibinfo {author} {\bibfnamefont {J.}~\bibnamefont
  {Schachenmayer}}, \bibinfo {author} {\bibfnamefont {I.}~\bibnamefont
  {Lesanovsky}}, \bibinfo {author} {\bibfnamefont {A.}~\bibnamefont {Micheli}},
  \ and\ \bibinfo {author} {\bibfnamefont {A.~J.}\ \bibnamefont {Daley}},\
  }\href {http://stacks.iop.org/1367-2630/12/i=10/a=103044} {\bibfield
  {journal} {\bibinfo  {journal} {New Journal of Physics}\ }\textbf {\bibinfo
  {volume} {12}},\ \bibinfo {pages} {103044} (\bibinfo {year}
  {2010}{\natexlab{a}})}\BibitemShut {NoStop}%
\bibitem [{\citenamefont {van Bijnen}\ \emph {et~al.}(2011)\citenamefont {van
  Bijnen}, \citenamefont {Smit}, \citenamefont {van Leeuwen}, \citenamefont
  {Vredenbregt},\ and\ \citenamefont {Kokkelmans}}]{Bijnen11}%
  \BibitemOpen
  \bibfield  {author} {\bibinfo {author} {\bibfnamefont {R.~M.~W.}\
  \bibnamefont {van Bijnen}}, \bibinfo {author} {\bibfnamefont
  {S.}~\bibnamefont {Smit}}, \bibinfo {author} {\bibfnamefont {K.~A.~H.}\
  \bibnamefont {van Leeuwen}}, \bibinfo {author} {\bibfnamefont {E.~J.~D.}\
  \bibnamefont {Vredenbregt}}, \ and\ \bibinfo {author} {\bibfnamefont {S.~J.
  J. M.~F.}\ \bibnamefont {Kokkelmans}},\ }\href
  {http://stacks.iop.org/0953-4075/44/i=18/a=184008} {\bibfield  {journal}
  {\bibinfo  {journal} {Journal of Physics B: Atomic, Molecular and Optical
  Physics}\ }\textbf {\bibinfo {volume} {44}},\ \bibinfo {pages} {184008}
  (\bibinfo {year} {2011})}\BibitemShut {NoStop}%
\bibitem [{\citenamefont {Petrosyan}\ \emph {et~al.}(2016)\citenamefont
  {Petrosyan}, \citenamefont {Mølmer},\ and\ \citenamefont
  {Fleischhauer}}]{Petrosyan16}%
  \BibitemOpen
  \bibfield  {author} {\bibinfo {author} {\bibfnamefont {D.}~\bibnamefont
  {Petrosyan}}, \bibinfo {author} {\bibfnamefont {K.}~\bibnamefont {Mølmer}},
  \ and\ \bibinfo {author} {\bibfnamefont {M.}~\bibnamefont {Fleischhauer}},\
  }\href {http://stacks.iop.org/0953-4075/49/i=8/a=084003} {\bibfield
  {journal} {\bibinfo  {journal} {Journal of Physics B: Atomic, Molecular and
  Optical Physics}\ }\textbf {\bibinfo {volume} {49}},\ \bibinfo {pages}
  {084003} (\bibinfo {year} {2016})}\BibitemShut {NoStop}%
\bibitem [{\citenamefont {Khaneja}\ \emph {et~al.}(2005)\citenamefont
  {Khaneja}, \citenamefont {Reiss}, \citenamefont {Kehlet}, \citenamefont
  {Schulte-Herbr{\"u}ggen},\ and\ \citenamefont {Glaser}}]{Khaneja2005296}%
  \BibitemOpen
  \bibfield  {author} {\bibinfo {author} {\bibfnamefont {N.}~\bibnamefont
  {Khaneja}}, \bibinfo {author} {\bibfnamefont {T.}~\bibnamefont {Reiss}},
  \bibinfo {author} {\bibfnamefont {C.}~\bibnamefont {Kehlet}}, \bibinfo
  {author} {\bibfnamefont {T.}~\bibnamefont {Schulte-Herbr{\"u}ggen}}, \ and\
  \bibinfo {author} {\bibfnamefont {S.}~\bibnamefont {Glaser}},\ }\href
  {\doibase http://dx.doi.org/10.1016/j.jmr.2004.11.004} {\bibfield  {journal}
  {\bibinfo  {journal} {Journal of Magnetic Resonance}\ }\textbf {\bibinfo
  {volume} {172}},\ \bibinfo {pages} {296 } (\bibinfo {year}
  {2005})}\BibitemShut {NoStop}%
\bibitem [{\citenamefont {Krotov}(1995)}]{krotov}%
  \BibitemOpen
  \bibfield  {author} {\bibinfo {author} {\bibfnamefont {V.}~\bibnamefont
  {Krotov}},\ }\href {https://books.google.de/books?id=SbWdfKZtvj0C} {\emph
  {\bibinfo {title} {Global Methods in Optimal Control Theory}}},\ Chapman \&
  Hall/CRC Pure and Applied Mathematics\ (\bibinfo  {publisher} {Taylor \&
  Francis},\ \bibinfo {year} {1995})\BibitemShut {NoStop}%
\bibitem [{\citenamefont {Sola}\ \emph {et~al.}(1998)\citenamefont {Sola},
  \citenamefont {Santamaria},\ and\ \citenamefont {Tannor}}]{controlmethod}%
  \BibitemOpen
  \bibfield  {author} {\bibinfo {author} {\bibfnamefont {I.~R.}\ \bibnamefont
  {Sola}}, \bibinfo {author} {\bibfnamefont {J.}~\bibnamefont {Santamaria}}, \
  and\ \bibinfo {author} {\bibfnamefont {D.~J.}\ \bibnamefont {Tannor}},\
  }\href {\doibase 10.1021/jp980281l} {\bibfield  {journal} {\bibinfo
  {journal} {The Journal of Physical Chemistry A}\ }\textbf {\bibinfo {volume}
  {102}},\ \bibinfo {pages} {4301} (\bibinfo {year} {1998})}\BibitemShut
  {NoStop}%
\bibitem [{\citenamefont {Werschnik}\ and\ \citenamefont
  {Gross}(2007)}]{OC_tutorial}%
  \BibitemOpen
  \bibfield  {author} {\bibinfo {author} {\bibfnamefont {J.}~\bibnamefont
  {Werschnik}}\ and\ \bibinfo {author} {\bibfnamefont {E.~K.~U.}\ \bibnamefont
  {Gross}},\ }\href {http://stacks.iop.org/0953-4075/40/i=18/a=R01} {\bibfield
  {journal} {\bibinfo  {journal} {Journal of Physics B: Atomic, Molecular and
  Optical Physics}\ }\textbf {\bibinfo {volume} {40}},\ \bibinfo {pages} {R175}
  (\bibinfo {year} {2007})}\BibitemShut {NoStop}%
\bibitem [{\citenamefont {Glaser}\ \emph {et~al.}(2015)\citenamefont {Glaser},
  \citenamefont {Boscain}, \citenamefont {Calarco}, \citenamefont {Koch},
  \citenamefont {K{\"o}ckenberger}, \citenamefont {Kosloff}, \citenamefont
  {Kuprov}, \citenamefont {Luy}, \citenamefont {Schirmer}, \citenamefont
  {Schulte-Herbr{\"u}ggen}, \citenamefont {Sugny},\ and\ \citenamefont
  {Wilhelm}}]{reviewOC}%
  \BibitemOpen
  \bibfield  {author} {\bibinfo {author} {\bibfnamefont {S.~J.}\ \bibnamefont
  {Glaser}}, \bibinfo {author} {\bibfnamefont {U.}~\bibnamefont {Boscain}},
  \bibinfo {author} {\bibfnamefont {T.}~\bibnamefont {Calarco}}, \bibinfo
  {author} {\bibfnamefont {C.~P.}\ \bibnamefont {Koch}}, \bibinfo {author}
  {\bibfnamefont {W.}~\bibnamefont {K{\"o}ckenberger}}, \bibinfo {author}
  {\bibfnamefont {R.}~\bibnamefont {Kosloff}}, \bibinfo {author} {\bibfnamefont
  {I.}~\bibnamefont {Kuprov}}, \bibinfo {author} {\bibfnamefont
  {B.}~\bibnamefont {Luy}}, \bibinfo {author} {\bibfnamefont {S.}~\bibnamefont
  {Schirmer}}, \bibinfo {author} {\bibfnamefont {T.}~\bibnamefont
  {Schulte-Herbr{\"u}ggen}}, \bibinfo {author} {\bibfnamefont {D.}~\bibnamefont
  {Sugny}}, \ and\ \bibinfo {author} {\bibfnamefont {K.~F.}\ \bibnamefont
  {Wilhelm}},\ }\href {\doibase 10.1140/epjd/e2015-60464-1} {\bibfield
  {journal} {\bibinfo  {journal} {The European Physical Journal D}\ }\textbf
  {\bibinfo {volume} {69}},\ \bibinfo {pages} {1} (\bibinfo {year}
  {2015})}\BibitemShut {NoStop}%
\bibitem [{\citenamefont {Palao}\ and\ \citenamefont
  {Kosloff}(2002)}]{OC_quantuminformation}%
  \BibitemOpen
  \bibfield  {author} {\bibinfo {author} {\bibfnamefont {J.~P.}\ \bibnamefont
  {Palao}}\ and\ \bibinfo {author} {\bibfnamefont {R.}~\bibnamefont
  {Kosloff}},\ }\href {\doibase 10.1103/PhysRevLett.89.188301} {\bibfield
  {journal} {\bibinfo  {journal} {Phys. Rev. Lett.}\ }\textbf {\bibinfo
  {volume} {89}},\ \bibinfo {pages} {188301} (\bibinfo {year}
  {2002})}\BibitemShut {NoStop}%
\bibitem [{\citenamefont {Grace}\ \emph {et~al.}(2007)\citenamefont {Grace},
  \citenamefont {Brif}, \citenamefont {Rabitz}, \citenamefont {Walmsley},
  \citenamefont {Kosut},\ and\ \citenamefont {Lidar}}]{Gate}%
  \BibitemOpen
  \bibfield  {author} {\bibinfo {author} {\bibfnamefont {M.}~\bibnamefont
  {Grace}}, \bibinfo {author} {\bibfnamefont {C.}~\bibnamefont {Brif}},
  \bibinfo {author} {\bibfnamefont {H.}~\bibnamefont {Rabitz}}, \bibinfo
  {author} {\bibfnamefont {I.~A.}\ \bibnamefont {Walmsley}}, \bibinfo {author}
  {\bibfnamefont {R.~L.}\ \bibnamefont {Kosut}}, \ and\ \bibinfo {author}
  {\bibfnamefont {D.~A.}\ \bibnamefont {Lidar}},\ }\href
  {http://stacks.iop.org/0953-4075/40/i=9/a=S06} {\bibfield  {journal}
  {\bibinfo  {journal} {Journal of Physics B: Atomic, Molecular and Optical
  Physics}\ }\textbf {\bibinfo {volume} {40}},\ \bibinfo {pages} {S103}
  (\bibinfo {year} {2007})}\BibitemShut {NoStop}%
\bibitem [{\citenamefont {Montangero}\ \emph {et~al.}(2007)\citenamefont
  {Montangero}, \citenamefont {Calarco},\ and\ \citenamefont
  {Fazio}}]{Montangero2007}%
  \BibitemOpen
  \bibfield  {author} {\bibinfo {author} {\bibfnamefont {S.}~\bibnamefont
  {Montangero}}, \bibinfo {author} {\bibfnamefont {T.}~\bibnamefont {Calarco}},
  \ and\ \bibinfo {author} {\bibfnamefont {R.}~\bibnamefont {Fazio}},\ }\href
  {\doibase 10.1103/PhysRevLett.99.170501} {\bibfield  {journal} {\bibinfo
  {journal} {Phys. Rev. Lett.}\ }\textbf {\bibinfo {volume} {99}},\ \bibinfo
  {pages} {170501} (\bibinfo {year} {2007})}\BibitemShut {NoStop}%
\bibitem [{\citenamefont {Calarco}\ \emph {et~al.}(2004)\citenamefont
  {Calarco}, \citenamefont {Dorner}, \citenamefont {Julienne}, \citenamefont
  {Williams},\ and\ \citenamefont {Zoller}}]{PhysRevA.70.012306}%
  \BibitemOpen
  \bibfield  {author} {\bibinfo {author} {\bibfnamefont {T.}~\bibnamefont
  {Calarco}}, \bibinfo {author} {\bibfnamefont {U.}~\bibnamefont {Dorner}},
  \bibinfo {author} {\bibfnamefont {P.~S.}\ \bibnamefont {Julienne}}, \bibinfo
  {author} {\bibfnamefont {C.~J.}\ \bibnamefont {Williams}}, \ and\ \bibinfo
  {author} {\bibfnamefont {P.}~\bibnamefont {Zoller}},\ }\href {\doibase
  10.1103/PhysRevA.70.012306} {\bibfield  {journal} {\bibinfo  {journal} {Phys.
  Rev. A}\ }\textbf {\bibinfo {volume} {70}},\ \bibinfo {pages} {012306}
  (\bibinfo {year} {2004})}\BibitemShut {NoStop}%
\bibitem [{\citenamefont {Wang}\ \emph {et~al.}(2010)\citenamefont {Wang},
  \citenamefont {Bayat}, \citenamefont {Schirmer},\ and\ \citenamefont
  {Bose}}]{BoseSpinChain}%
  \BibitemOpen
  \bibfield  {author} {\bibinfo {author} {\bibfnamefont {X.}~\bibnamefont
  {Wang}}, \bibinfo {author} {\bibfnamefont {A.}~\bibnamefont {Bayat}},
  \bibinfo {author} {\bibfnamefont {S.~G.}\ \bibnamefont {Schirmer}}, \ and\
  \bibinfo {author} {\bibfnamefont {S.}~\bibnamefont {Bose}},\ }\href {\doibase
  10.1103/PhysRevA.81.032312} {\bibfield  {journal} {\bibinfo  {journal} {Phys.
  Rev. A}\ }\textbf {\bibinfo {volume} {81}},\ \bibinfo {pages} {032312}
  (\bibinfo {year} {2010})}\BibitemShut {NoStop}%
\bibitem [{\citenamefont {Cui}\ and\ \citenamefont
  {Mintert}(2015)}]{Cui_Mintert}%
  \BibitemOpen
  \bibfield  {author} {\bibinfo {author} {\bibfnamefont {J.}~\bibnamefont
  {Cui}}\ and\ \bibinfo {author} {\bibfnamefont {F.}~\bibnamefont {Mintert}},\
  }\href {http://stacks.iop.org/1367-2630/17/i=9/a=093014} {\bibfield
  {journal} {\bibinfo  {journal} {New Journal of Physics}\ }\textbf {\bibinfo
  {volume} {17}},\ \bibinfo {pages} {093014} (\bibinfo {year}
  {2015})}\BibitemShut {NoStop}%
\bibitem [{\citenamefont {Caneva}\ \emph {et~al.}(2012)\citenamefont {Caneva},
  \citenamefont {Calarco},\ and\ \citenamefont {Montangero}}]{ESU}%
  \BibitemOpen
  \bibfield  {author} {\bibinfo {author} {\bibfnamefont {T.}~\bibnamefont
  {Caneva}}, \bibinfo {author} {\bibfnamefont {T.}~\bibnamefont {Calarco}}, \
  and\ \bibinfo {author} {\bibfnamefont {S.}~\bibnamefont {Montangero}},\
  }\href {http://stacks.iop.org/1367-2630/14/i=9/a=093041} {\bibfield
  {journal} {\bibinfo  {journal} {New Journal of Physics}\ }\textbf {\bibinfo
  {volume} {14}},\ \bibinfo {pages} {093041} (\bibinfo {year}
  {2012})}\BibitemShut {NoStop}%
\bibitem [{\citenamefont {M\"uller}\ \emph {et~al.}(2011)\citenamefont
  {M\"uller}, \citenamefont {Reich}, \citenamefont {Murphy}, \citenamefont
  {Yuan}, \citenamefont {Vala}, \citenamefont {Whaley}, \citenamefont
  {Calarco},\ and\ \citenamefont {Koch}}]{Koch11}%
  \BibitemOpen
  \bibfield  {author} {\bibinfo {author} {\bibfnamefont {M.~M.}\ \bibnamefont
  {M\"uller}}, \bibinfo {author} {\bibfnamefont {D.~M.}\ \bibnamefont {Reich}},
  \bibinfo {author} {\bibfnamefont {M.}~\bibnamefont {Murphy}}, \bibinfo
  {author} {\bibfnamefont {H.}~\bibnamefont {Yuan}}, \bibinfo {author}
  {\bibfnamefont {J.}~\bibnamefont {Vala}}, \bibinfo {author} {\bibfnamefont
  {K.~B.}\ \bibnamefont {Whaley}}, \bibinfo {author} {\bibfnamefont
  {T.}~\bibnamefont {Calarco}}, \ and\ \bibinfo {author} {\bibfnamefont
  {C.~P.}\ \bibnamefont {Koch}},\ }\href {\doibase 10.1103/PhysRevA.84.042315}
  {\bibfield  {journal} {\bibinfo  {journal} {Phys. Rev. A}\ }\textbf {\bibinfo
  {volume} {84}},\ \bibinfo {pages} {042315} (\bibinfo {year}
  {2011})}\BibitemShut {NoStop}%
\bibitem [{\citenamefont {Goerz}\ \emph {et~al.}(2014)\citenamefont {Goerz},
  \citenamefont {Halperin}, \citenamefont {Aytac}, \citenamefont {Koch},\ and\
  \citenamefont {Whaley}}]{Koch14}%
  \BibitemOpen
  \bibfield  {author} {\bibinfo {author} {\bibfnamefont {M.~H.}\ \bibnamefont
  {Goerz}}, \bibinfo {author} {\bibfnamefont {E.~J.}\ \bibnamefont {Halperin}},
  \bibinfo {author} {\bibfnamefont {J.~M.}\ \bibnamefont {Aytac}}, \bibinfo
  {author} {\bibfnamefont {C.~P.}\ \bibnamefont {Koch}}, \ and\ \bibinfo
  {author} {\bibfnamefont {K.~B.}\ \bibnamefont {Whaley}},\ }\href {\doibase
  10.1103/PhysRevA.90.032329} {\bibfield  {journal} {\bibinfo  {journal} {Phys.
  Rev. A}\ }\textbf {\bibinfo {volume} {90}},\ \bibinfo {pages} {032329}
  (\bibinfo {year} {2014})}\BibitemShut {NoStop}%
\bibitem [{\citenamefont {Caneva}\ \emph {et~al.}(2009)\citenamefont {Caneva},
  \citenamefont {Murphy}, \citenamefont {Calarco}, \citenamefont {Fazio},
  \citenamefont {Montangero}, \citenamefont {Giovannetti},\ and\ \citenamefont
  {Santoro}}]{PhysRevLett.103.240501}%
  \BibitemOpen
  \bibfield  {author} {\bibinfo {author} {\bibfnamefont {T.}~\bibnamefont
  {Caneva}}, \bibinfo {author} {\bibfnamefont {M.}~\bibnamefont {Murphy}},
  \bibinfo {author} {\bibfnamefont {T.}~\bibnamefont {Calarco}}, \bibinfo
  {author} {\bibfnamefont {R.}~\bibnamefont {Fazio}}, \bibinfo {author}
  {\bibfnamefont {S.}~\bibnamefont {Montangero}}, \bibinfo {author}
  {\bibfnamefont {V.}~\bibnamefont {Giovannetti}}, \ and\ \bibinfo {author}
  {\bibfnamefont {G.~E.}\ \bibnamefont {Santoro}},\ }\href {\doibase
  10.1103/PhysRevLett.103.240501} {\bibfield  {journal} {\bibinfo  {journal}
  {Phys. Rev. Lett.}\ }\textbf {\bibinfo {volume} {103}},\ \bibinfo {pages}
  {240501} (\bibinfo {year} {2009})}\BibitemShut {NoStop}%
\bibitem [{\citenamefont {Doria}\ \emph {et~al.}(2011)\citenamefont {Doria},
  \citenamefont {Calarco},\ and\ \citenamefont {Montangero}}]{CRAB}%
  \BibitemOpen
  \bibfield  {author} {\bibinfo {author} {\bibfnamefont {P.}~\bibnamefont
  {Doria}}, \bibinfo {author} {\bibfnamefont {T.}~\bibnamefont {Calarco}}, \
  and\ \bibinfo {author} {\bibfnamefont {S.}~\bibnamefont {Montangero}},\
  }\href {\doibase 10.1103/PhysRevLett.106.190501} {\bibfield  {journal}
  {\bibinfo  {journal} {Phys. Rev. Lett.}\ }\textbf {\bibinfo {volume} {106}},\
  \bibinfo {pages} {190501} (\bibinfo {year} {2011})}\BibitemShut {NoStop}%
\bibitem [{\citenamefont {Caneva}\ \emph {et~al.}(2011)\citenamefont {Caneva},
  \citenamefont {Calarco}, \citenamefont {Fazio}, \citenamefont {Santoro},\
  and\ \citenamefont {Montangero}}]{PhysRevA.84.012312}%
  \BibitemOpen
  \bibfield  {author} {\bibinfo {author} {\bibfnamefont {T.}~\bibnamefont
  {Caneva}}, \bibinfo {author} {\bibfnamefont {T.}~\bibnamefont {Calarco}},
  \bibinfo {author} {\bibfnamefont {R.}~\bibnamefont {Fazio}}, \bibinfo
  {author} {\bibfnamefont {G.~E.}\ \bibnamefont {Santoro}}, \ and\ \bibinfo
  {author} {\bibfnamefont {S.}~\bibnamefont {Montangero}},\ }\href {\doibase
  10.1103/PhysRevA.84.012312} {\bibfield  {journal} {\bibinfo  {journal} {Phys.
  Rev. A}\ }\textbf {\bibinfo {volume} {84}},\ \bibinfo {pages} {012312}
  (\bibinfo {year} {2011})}\BibitemShut {NoStop}%
\bibitem [{\citenamefont {Rosi}\ \emph {et~al.}(2013)\citenamefont {Rosi},
  \citenamefont {Bernard}, \citenamefont {Fabbri}, \citenamefont {Fallani},
  \citenamefont {Fort}, \citenamefont {Inguscio}, \citenamefont {Calarco},\
  and\ \citenamefont {Montangero}}]{ExpClosedLoop}%
  \BibitemOpen
  \bibfield  {author} {\bibinfo {author} {\bibfnamefont {S.}~\bibnamefont
  {Rosi}}, \bibinfo {author} {\bibfnamefont {A.}~\bibnamefont {Bernard}},
  \bibinfo {author} {\bibfnamefont {N.}~\bibnamefont {Fabbri}}, \bibinfo
  {author} {\bibfnamefont {L.}~\bibnamefont {Fallani}}, \bibinfo {author}
  {\bibfnamefont {C.}~\bibnamefont {Fort}}, \bibinfo {author} {\bibfnamefont
  {M.}~\bibnamefont {Inguscio}}, \bibinfo {author} {\bibfnamefont
  {T.}~\bibnamefont {Calarco}}, \ and\ \bibinfo {author} {\bibfnamefont
  {S.}~\bibnamefont {Montangero}},\ }\href {\doibase
  10.1103/PhysRevA.88.021601} {\bibfield  {journal} {\bibinfo  {journal} {Phys.
  Rev. A}\ }\textbf {\bibinfo {volume} {88}},\ \bibinfo {pages} {021601}
  (\bibinfo {year} {2013})}\BibitemShut {NoStop}%
\bibitem [{\citenamefont {Lovecchio}\ \emph {et~al.}(2016)\citenamefont
  {Lovecchio}, \citenamefont {Sch\"afer}, \citenamefont {Cherukattil},
  \citenamefont {Al\`{\i}~Khan}, \citenamefont {Herrera}, \citenamefont
  {Cataliotti}, \citenamefont {Calarco}, \citenamefont {Montangero},\ and\
  \citenamefont {Caruso}}]{Exp_atom-chip}%
  \BibitemOpen
  \bibfield  {author} {\bibinfo {author} {\bibfnamefont {C.}~\bibnamefont
  {Lovecchio}}, \bibinfo {author} {\bibfnamefont {F.}~\bibnamefont
  {Sch\"afer}}, \bibinfo {author} {\bibfnamefont {S.}~\bibnamefont
  {Cherukattil}}, \bibinfo {author} {\bibfnamefont {M.}~\bibnamefont
  {Al\`{\i}~Khan}}, \bibinfo {author} {\bibfnamefont {I.}~\bibnamefont
  {Herrera}}, \bibinfo {author} {\bibfnamefont {F.~S.}\ \bibnamefont
  {Cataliotti}}, \bibinfo {author} {\bibfnamefont {T.}~\bibnamefont {Calarco}},
  \bibinfo {author} {\bibfnamefont {S.}~\bibnamefont {Montangero}}, \ and\
  \bibinfo {author} {\bibfnamefont {F.}~\bibnamefont {Caruso}},\ }\href
  {\doibase 10.1103/PhysRevA.93.010304} {\bibfield  {journal} {\bibinfo
  {journal} {Phys. Rev. A}\ }\textbf {\bibinfo {volume} {93}},\ \bibinfo
  {pages} {010304} (\bibinfo {year} {2016})}\BibitemShut {NoStop}%
\bibitem [{\citenamefont {N\"obauer}\ \emph {et~al.}(2015)\citenamefont
  {N\"obauer}, \citenamefont {Angerer}, \citenamefont {Bartels}, \citenamefont
  {Trupke}, \citenamefont {Rotter}, \citenamefont {Schmiedmayer}, \citenamefont
  {Mintert},\ and\ \citenamefont {Majer}}]{Exp_Mintert}%
  \BibitemOpen
  \bibfield  {author} {\bibinfo {author} {\bibfnamefont {T.}~\bibnamefont
  {N\"obauer}}, \bibinfo {author} {\bibfnamefont {A.}~\bibnamefont {Angerer}},
  \bibinfo {author} {\bibfnamefont {B.}~\bibnamefont {Bartels}}, \bibinfo
  {author} {\bibfnamefont {M.}~\bibnamefont {Trupke}}, \bibinfo {author}
  {\bibfnamefont {S.}~\bibnamefont {Rotter}}, \bibinfo {author} {\bibfnamefont
  {J.}~\bibnamefont {Schmiedmayer}}, \bibinfo {author} {\bibfnamefont
  {F.}~\bibnamefont {Mintert}}, \ and\ \bibinfo {author} {\bibfnamefont
  {J.}~\bibnamefont {Majer}},\ }\href {\doibase 10.1103/PhysRevLett.115.190801}
  {\bibfield  {journal} {\bibinfo  {journal} {Phys. Rev. Lett.}\ }\textbf
  {\bibinfo {volume} {115}},\ \bibinfo {pages} {190801} (\bibinfo {year}
  {2015})}\BibitemShut {NoStop}%
\bibitem [{\citenamefont {van Frank}\ \emph {et~al.}(2016)\citenamefont {van
  Frank}, \citenamefont {Bonneau}, \citenamefont {Schmiedmayer}, \citenamefont
  {Hild}, \citenamefont {Gross}, \citenamefont {Cheneau}, \citenamefont
  {Bloch}, \citenamefont {Pichler}, \citenamefont {Negretti}, \citenamefont
  {Calarco},\ and\ \citenamefont {Montangero}}]{Exp_atom}%
  \BibitemOpen
  \bibfield  {author} {\bibinfo {author} {\bibfnamefont {S.}~\bibnamefont {van
  Frank}}, \bibinfo {author} {\bibfnamefont {M.}~\bibnamefont {Bonneau}},
  \bibinfo {author} {\bibfnamefont {J.}~\bibnamefont {Schmiedmayer}}, \bibinfo
  {author} {\bibfnamefont {S.}~\bibnamefont {Hild}}, \bibinfo {author}
  {\bibfnamefont {C.}~\bibnamefont {Gross}}, \bibinfo {author} {\bibfnamefont
  {M.}~\bibnamefont {Cheneau}}, \bibinfo {author} {\bibfnamefont
  {I.}~\bibnamefont {Bloch}}, \bibinfo {author} {\bibfnamefont
  {T.}~\bibnamefont {Pichler}}, \bibinfo {author} {\bibfnamefont
  {A.}~\bibnamefont {Negretti}}, \bibinfo {author} {\bibfnamefont
  {T.}~\bibnamefont {Calarco}}, \ and\ \bibinfo {author} {\bibfnamefont
  {S.}~\bibnamefont {Montangero}},\ }\href
  {http://dx.doi.org/10.1038/srep34187} {\bibfield  {journal} {\bibinfo
  {journal} {Scientific Reports}\ }\textbf {\bibinfo {volume} {6}},\ \bibinfo
  {pages} {34187} (\bibinfo {year} {2016})}\BibitemShut {NoStop}%
\bibitem [{\citenamefont {M{\"u}ller}\ \emph {et~al.}(2011)\citenamefont
  {M{\"u}ller}, \citenamefont {Haakh}, \citenamefont {Calarco}, \citenamefont
  {Koch},\ and\ \citenamefont {Henkel}}]{RydbergOC2011}%
  \BibitemOpen
  \bibfield  {author} {\bibinfo {author} {\bibfnamefont {M.~M.}\ \bibnamefont
  {M{\"u}ller}}, \bibinfo {author} {\bibfnamefont {H.~R.}\ \bibnamefont
  {Haakh}}, \bibinfo {author} {\bibfnamefont {T.}~\bibnamefont {Calarco}},
  \bibinfo {author} {\bibfnamefont {C.~P.}\ \bibnamefont {Koch}}, \ and\
  \bibinfo {author} {\bibfnamefont {C.}~\bibnamefont {Henkel}},\ }\href
  {\doibase 10.1007/s11128-011-0296-0} {\bibfield  {journal} {\bibinfo
  {journal} {Quantum Information Processing}\ }\textbf {\bibinfo {volume}
  {10}},\ \bibinfo {pages} {771} (\bibinfo {year} {2011})}\BibitemShut
  {NoStop}%
\bibitem [{\citenamefont {Keating}\ \emph {et~al.}(2016)\citenamefont
  {Keating}, \citenamefont {Baldwin}, \citenamefont {Jau}, \citenamefont {Lee},
  \citenamefont {Biedermann},\ and\ \citenamefont
  {Deutsch}}]{RydbergOptimalControl16}%
  \BibitemOpen
  \bibfield  {author} {\bibinfo {author} {\bibfnamefont {T.}~\bibnamefont
  {Keating}}, \bibinfo {author} {\bibfnamefont {C.~H.}\ \bibnamefont
  {Baldwin}}, \bibinfo {author} {\bibfnamefont {Y.-Y.}\ \bibnamefont {Jau}},
  \bibinfo {author} {\bibfnamefont {J.}~\bibnamefont {Lee}}, \bibinfo {author}
  {\bibfnamefont {G.~W.}\ \bibnamefont {Biedermann}}, \ and\ \bibinfo {author}
  {\bibfnamefont {I.~H.}\ \bibnamefont {Deutsch}},\ }\href {\doibase
  10.1103/PhysRevLett.117.213601} {\bibfield  {journal} {\bibinfo  {journal}
  {Phys. Rev. Lett.}\ }\textbf {\bibinfo {volume} {117}},\ \bibinfo {pages}
  {213601} (\bibinfo {year} {2016})}\BibitemShut {NoStop}%
\bibitem [{\citenamefont {M{\"u}ller}\ \emph {et~al.}(2016)\citenamefont
  {M{\"u}ller}, \citenamefont {Pichler}, \citenamefont {Montangero},\ and\
  \citenamefont {Calarco}}]{RydbergOC2016}%
  \BibitemOpen
  \bibfield  {author} {\bibinfo {author} {\bibfnamefont {M.~M.}\ \bibnamefont
  {M{\"u}ller}}, \bibinfo {author} {\bibfnamefont {T.}~\bibnamefont {Pichler}},
  \bibinfo {author} {\bibfnamefont {S.}~\bibnamefont {Montangero}}, \ and\
  \bibinfo {author} {\bibfnamefont {T.}~\bibnamefont {Calarco}},\ }\href
  {\doibase 10.1007/s00340-016-6383-2} {\bibfield  {journal} {\bibinfo
  {journal} {Applied Physics B}\ }\textbf {\bibinfo {volume} {122}},\ \bibinfo
  {pages} {104} (\bibinfo {year} {2016})}\BibitemShut {NoStop}%
\bibitem [{\citenamefont {Lloyd}\ and\ \citenamefont
  {Montangero}(2014)}]{Lloyd2014a}%
  \BibitemOpen
  \bibfield  {author} {\bibinfo {author} {\bibfnamefont {S.}~\bibnamefont
  {Lloyd}}\ and\ \bibinfo {author} {\bibfnamefont {S.}~\bibnamefont
  {Montangero}},\ }\href {\doibase 10.1103/PhysRevLett.113.010502} {\bibfield
  {journal} {\bibinfo  {journal} {Phys. Rev. Lett.}\ }\textbf {\bibinfo
  {volume} {113}},\ \bibinfo {pages} {010502} (\bibinfo {year}
  {2014})}\BibitemShut {NoStop}%
\bibitem [{\citenamefont {Rach}\ \emph {et~al.}(2015)\citenamefont {Rach},
  \citenamefont {M\"uller}, \citenamefont {Calarco},\ and\ \citenamefont
  {Montangero}}]{dCRAB}%
  \BibitemOpen
  \bibfield  {author} {\bibinfo {author} {\bibfnamefont {N.}~\bibnamefont
  {Rach}}, \bibinfo {author} {\bibfnamefont {M.~M.}\ \bibnamefont {M\"uller}},
  \bibinfo {author} {\bibfnamefont {T.}~\bibnamefont {Calarco}}, \ and\
  \bibinfo {author} {\bibfnamefont {S.}~\bibnamefont {Montangero}},\ }\href
  {\doibase 10.1103/PhysRevA.92.062343} {\bibfield  {journal} {\bibinfo
  {journal} {Phys. Rev. A}\ }\textbf {\bibinfo {volume} {92}},\ \bibinfo
  {pages} {062343} (\bibinfo {year} {2015})}\BibitemShut {NoStop}%
\bibitem [{\citenamefont {Gottesman}\ and\ \citenamefont
  {Chuang}(1999)}]{GHZ_teleportation}%
  \BibitemOpen
  \bibfield  {author} {\bibinfo {author} {\bibfnamefont {D.}~\bibnamefont
  {Gottesman}}\ and\ \bibinfo {author} {\bibfnamefont {I.~L.}\ \bibnamefont
  {Chuang}},\ }\href {http://dx.doi.org/10.1038/46503} {\bibfield  {journal}
  {\bibinfo  {journal} {Nature}\ }\textbf {\bibinfo {volume} {402}},\ \bibinfo
  {pages} {390} (\bibinfo {year} {1999})}\BibitemShut {NoStop}%
\bibitem [{\citenamefont {Zhao}\ \emph {et~al.}(2004)\citenamefont {Zhao},
  \citenamefont {Chen}, \citenamefont {Zhang}, \citenamefont {Yang},
  \citenamefont {Briegel},\ and\ \citenamefont {Pan}}]{GHZ_teleportationExp}%
  \BibitemOpen
  \bibfield  {author} {\bibinfo {author} {\bibfnamefont {Z.}~\bibnamefont
  {Zhao}}, \bibinfo {author} {\bibfnamefont {Y.-A.}\ \bibnamefont {Chen}},
  \bibinfo {author} {\bibfnamefont {A.-N.}\ \bibnamefont {Zhang}}, \bibinfo
  {author} {\bibfnamefont {T.}~\bibnamefont {Yang}}, \bibinfo {author}
  {\bibfnamefont {H.~J.}\ \bibnamefont {Briegel}}, \ and\ \bibinfo {author}
  {\bibfnamefont {J.-W.}\ \bibnamefont {Pan}},\ }\href
  {http://dx.doi.org/10.1038/nature02643} {\bibfield  {journal} {\bibinfo
  {journal} {Nature}\ }\textbf {\bibinfo {volume} {430}},\ \bibinfo {pages}
  {54} (\bibinfo {year} {2004})}\BibitemShut {NoStop}%
\bibitem [{\citenamefont {Yeo}\ and\ \citenamefont
  {Chua}(2006)}]{GHZ_teleportationanddensecoding}%
  \BibitemOpen
  \bibfield  {author} {\bibinfo {author} {\bibfnamefont {Y.}~\bibnamefont
  {Yeo}}\ and\ \bibinfo {author} {\bibfnamefont {W.~K.}\ \bibnamefont {Chua}},\
  }\href {\doibase 10.1103/PhysRevLett.96.060502} {\bibfield  {journal}
  {\bibinfo  {journal} {Phys. Rev. Lett.}\ }\textbf {\bibinfo {volume} {96}},\
  \bibinfo {pages} {060502} (\bibinfo {year} {2006})}\BibitemShut {NoStop}%
\bibitem [{\citenamefont {Allati}\ \emph {et~al.}(2011)\citenamefont {Allati},
  \citenamefont {Baz},\ and\ \citenamefont {Hassouni}}]{GHZ_QKD}%
  \BibitemOpen
  \bibfield  {author} {\bibinfo {author} {\bibfnamefont {A.~E.}\ \bibnamefont
  {Allati}}, \bibinfo {author} {\bibfnamefont {M.~E.}\ \bibnamefont {Baz}}, \
  and\ \bibinfo {author} {\bibfnamefont {Y.}~\bibnamefont {Hassouni}},\ }\href
  {\doibase 10.1007/s11128-010-0213-y} {\bibfield  {journal} {\bibinfo
  {journal} {Quantum Information Processing}\ }\textbf {\bibinfo {volume}
  {10}},\ \bibinfo {pages} {589} (\bibinfo {year} {2011})}\BibitemShut
  {NoStop}%
\bibitem [{\citenamefont {Bose}\ \emph {et~al.}(1998)\citenamefont {Bose},
  \citenamefont {Vedral},\ and\ \citenamefont {Knight}}]{GHZ_swapping}%
  \BibitemOpen
  \bibfield  {author} {\bibinfo {author} {\bibfnamefont {S.}~\bibnamefont
  {Bose}}, \bibinfo {author} {\bibfnamefont {V.}~\bibnamefont {Vedral}}, \ and\
  \bibinfo {author} {\bibfnamefont {P.~L.}\ \bibnamefont {Knight}},\ }\href
  {\doibase 10.1103/PhysRevA.57.822} {\bibfield  {journal} {\bibinfo  {journal}
  {Phys. Rev. A}\ }\textbf {\bibinfo {volume} {57}},\ \bibinfo {pages} {822}
  (\bibinfo {year} {1998})}\BibitemShut {NoStop}%
\bibitem [{\citenamefont {Ostmann}\ \emph {et~al.}(2017)\citenamefont
  {Ostmann}, \citenamefont {Min\'a\v{r}}, \citenamefont {Marcuzzi},
  \citenamefont {Levi},\ and\ \citenamefont {Lesanovsky}}]{}%
  \BibitemOpen
  \bibfield  {author} {\bibinfo {author} {\bibfnamefont {M.}~\bibnamefont
  {Ostmann}}, \bibinfo {author} {\bibfnamefont {J.}~\bibnamefont
  {Min\'a\v{r}}}, \bibinfo {author} {\bibfnamefont {M.}~\bibnamefont
  {Marcuzzi}}, \bibinfo {author} {\bibfnamefont {E.}~\bibnamefont {Levi}}, \
  and\ \bibinfo {author} {\bibfnamefont {I.}~\bibnamefont {Lesanovsky}},\
  }\href {https://arxiv.org/abs/1707.02203} {\bibfield  {journal} {\bibinfo
  {journal} {ArXiv e-prints}\ } (\bibinfo {year} {2017})},\ \Eprint
  {http://arxiv.org/abs/1707.02203} {arXiv:1707.02203 [quant-ph]} \BibitemShut
  {NoStop}%
\bibitem [{\citenamefont {Schauss}\ \emph {et~al.}(2012)\citenamefont
  {Schauss}, \citenamefont {Cheneau}, \citenamefont {Endres}, \citenamefont
  {Fukuhara}, \citenamefont {Hild}, \citenamefont {Omran}, \citenamefont
  {Pohl}, \citenamefont {Gross}, \citenamefont {Kuhr},\ and\ \citenamefont
  {Bloch}}]{NatExp}%
  \BibitemOpen
  \bibfield  {author} {\bibinfo {author} {\bibfnamefont {P.}~\bibnamefont
  {Schauss}}, \bibinfo {author} {\bibfnamefont {M.}~\bibnamefont {Cheneau}},
  \bibinfo {author} {\bibfnamefont {M.}~\bibnamefont {Endres}}, \bibinfo
  {author} {\bibfnamefont {T.}~\bibnamefont {Fukuhara}}, \bibinfo {author}
  {\bibfnamefont {S.}~\bibnamefont {Hild}}, \bibinfo {author} {\bibfnamefont
  {A.}~\bibnamefont {Omran}}, \bibinfo {author} {\bibfnamefont
  {T.}~\bibnamefont {Pohl}}, \bibinfo {author} {\bibfnamefont {C.}~\bibnamefont
  {Gross}}, \bibinfo {author} {\bibfnamefont {S.}~\bibnamefont {Kuhr}}, \ and\
  \bibinfo {author} {\bibfnamefont {I.}~\bibnamefont {Bloch}},\ }\href
  {\doibase 10.1038/nature11596} {\bibfield  {journal} {\bibinfo  {journal}
  {Nature}\ }\textbf {\bibinfo {volume} {491}},\ \bibinfo {pages} {87}
  (\bibinfo {year} {2012})}\BibitemShut {NoStop}%
\bibitem [{\citenamefont {van Bijnen}\ \emph {et~al.}(2015)\citenamefont {van
  Bijnen}, \citenamefont {Ravensbergen}, \citenamefont {Bakker}, \citenamefont
  {Dijk}, \citenamefont {Kokkelmans},\ and\ \citenamefont
  {Vredenbregt}}]{Bijnen15b}%
  \BibitemOpen
  \bibfield  {author} {\bibinfo {author} {\bibfnamefont {R.~M.~W.}\
  \bibnamefont {van Bijnen}}, \bibinfo {author} {\bibfnamefont
  {C.}~\bibnamefont {Ravensbergen}}, \bibinfo {author} {\bibfnamefont {D.~J.}\
  \bibnamefont {Bakker}}, \bibinfo {author} {\bibfnamefont {G.~J.}\
  \bibnamefont {Dijk}}, \bibinfo {author} {\bibfnamefont {S.~J. J. M.~F.}\
  \bibnamefont {Kokkelmans}}, \ and\ \bibinfo {author} {\bibfnamefont
  {E.~J.~D.}\ \bibnamefont {Vredenbregt}},\ }\href
  {http://stacks.iop.org/1367-2630/17/i=2/a=023045} {\bibfield  {journal}
  {\bibinfo  {journal} {New Journal of Physics}\ }\textbf {\bibinfo {volume}
  {17}},\ \bibinfo {pages} {023045} (\bibinfo {year} {2015})}\BibitemShut
  {NoStop}%
\bibitem [{\citenamefont {Younge}\ \emph {et~al.}(2009)\citenamefont {Younge},
  \citenamefont {Reinhard}, \citenamefont {Pohl}, \citenamefont {Berman},\ and\
  \citenamefont {Raithel}}]{Younge09}%
  \BibitemOpen
  \bibfield  {author} {\bibinfo {author} {\bibfnamefont {K.~C.}\ \bibnamefont
  {Younge}}, \bibinfo {author} {\bibfnamefont {A.}~\bibnamefont {Reinhard}},
  \bibinfo {author} {\bibfnamefont {T.}~\bibnamefont {Pohl}}, \bibinfo {author}
  {\bibfnamefont {P.~R.}\ \bibnamefont {Berman}}, \ and\ \bibinfo {author}
  {\bibfnamefont {G.}~\bibnamefont {Raithel}},\ }\href {\doibase
  10.1103/PhysRevA.79.043420} {\bibfield  {journal} {\bibinfo  {journal} {Phys.
  Rev. A}\ }\textbf {\bibinfo {volume} {79}},\ \bibinfo {pages} {043420}
  (\bibinfo {year} {2009})}\BibitemShut {NoStop}%
\bibitem [{\citenamefont {Mukherjee}\ \emph {et~al.}(2011)\citenamefont
  {Mukherjee}, \citenamefont {Millen}, \citenamefont {Nath}, \citenamefont
  {Jones},\ and\ \citenamefont {Pohl}}]{RydbergGHZ_Pohl}%
  \BibitemOpen
  \bibfield  {author} {\bibinfo {author} {\bibfnamefont {R.}~\bibnamefont
  {Mukherjee}}, \bibinfo {author} {\bibfnamefont {J.}~\bibnamefont {Millen}},
  \bibinfo {author} {\bibfnamefont {R.}~\bibnamefont {Nath}}, \bibinfo {author}
  {\bibfnamefont {M.~P.~A.}\ \bibnamefont {Jones}}, \ and\ \bibinfo {author}
  {\bibfnamefont {T.}~\bibnamefont {Pohl}},\ }\href
  {http://stacks.iop.org/0953-4075/44/i=18/a=184010} {\bibfield  {journal}
  {\bibinfo  {journal} {Journal of Physics B: Atomic, Molecular and Optical
  Physics}\ }\textbf {\bibinfo {volume} {44}},\ \bibinfo {pages} {184010}
  (\bibinfo {year} {2011})}\BibitemShut {NoStop}%
\bibitem [{\citenamefont {Reinhard}\ \emph {et~al.}(2008)\citenamefont
  {Reinhard}, \citenamefont {Cubel~Liebisch}, \citenamefont {Younge},
  \citenamefont {Berman},\ and\ \citenamefont {Raithel}}]{Reinhard08}%
  \BibitemOpen
  \bibfield  {author} {\bibinfo {author} {\bibfnamefont {A.}~\bibnamefont
  {Reinhard}}, \bibinfo {author} {\bibfnamefont {T.}~\bibnamefont
  {Cubel~Liebisch}}, \bibinfo {author} {\bibfnamefont {K.~C.}\ \bibnamefont
  {Younge}}, \bibinfo {author} {\bibfnamefont {P.~R.}\ \bibnamefont {Berman}},
  \ and\ \bibinfo {author} {\bibfnamefont {G.}~\bibnamefont {Raithel}},\ }\href
  {\doibase 10.1103/PhysRevLett.100.123007} {\bibfield  {journal} {\bibinfo
  {journal} {Phys. Rev. Lett.}\ }\textbf {\bibinfo {volume} {100}},\ \bibinfo
  {pages} {123007} (\bibinfo {year} {2008})}\BibitemShut {NoStop}%
\bibitem [{\citenamefont {Singer}\ \emph {et~al.}(2005)\citenamefont {Singer},
  \citenamefont {Stanojevic}, \citenamefont {Weidemüller},\ and\ \citenamefont
  {Côté}}]{C6}%
  \BibitemOpen
  \bibfield  {author} {\bibinfo {author} {\bibfnamefont {K.}~\bibnamefont
  {Singer}}, \bibinfo {author} {\bibfnamefont {J.}~\bibnamefont {Stanojevic}},
  \bibinfo {author} {\bibfnamefont {M.}~\bibnamefont {Weidemüller}}, \ and\
  \bibinfo {author} {\bibfnamefont {R.}~\bibnamefont {Côté}},\ }\href
  {http://stacks.iop.org/0953-4075/38/i=2/a=021} {\bibfield  {journal}
  {\bibinfo  {journal} {Journal of Physics B: Atomic, Molecular and Optical
  Physics}\ }\textbf {\bibinfo {volume} {38}},\ \bibinfo {pages} {S295}
  (\bibinfo {year} {2005})}\BibitemShut {NoStop}%
\bibitem [{\citenamefont {Comparat}\ and\ \citenamefont
  {Pillet}(2010)}]{InteractionComparat}%
  \BibitemOpen
  \bibfield  {author} {\bibinfo {author} {\bibfnamefont {D.}~\bibnamefont
  {Comparat}}\ and\ \bibinfo {author} {\bibfnamefont {P.}~\bibnamefont
  {Pillet}},\ }\href {\doibase 10.1364/JOSAB.27.00A208} {\bibfield  {journal}
  {\bibinfo  {journal} {J. Opt. Soc. Am. B}\ }\textbf {\bibinfo {volume}
  {27}},\ \bibinfo {pages} {A208} (\bibinfo {year} {2010})}\BibitemShut
  {NoStop}%
\bibitem [{\citenamefont {Browaeys}\ \emph {et~al.}(2016)\citenamefont
  {Browaeys}, \citenamefont {Barredo},\ and\ \citenamefont
  {Lahaye}}]{Browaeys2016_reviewRydExp}%
  \BibitemOpen
  \bibfield  {author} {\bibinfo {author} {\bibfnamefont {A.}~\bibnamefont
  {Browaeys}}, \bibinfo {author} {\bibfnamefont {D.}~\bibnamefont {Barredo}}, \
  and\ \bibinfo {author} {\bibfnamefont {T.}~\bibnamefont {Lahaye}},\ }\href
  {http://stacks.iop.org/0953-4075/49/i=15/a=152001} {\bibfield  {journal}
  {\bibinfo  {journal} {Journal of Physics B: Atomic, Molecular and Optical
  Physics}\ }\textbf {\bibinfo {volume} {49}},\ \bibinfo {pages} {152001}
  (\bibinfo {year} {2016})}\BibitemShut {NoStop}%
\bibitem [{\citenamefont {Schachenmayer}\ \emph
  {et~al.}(2010{\natexlab{b}})\citenamefont {Schachenmayer}, \citenamefont
  {Lesanovsky}, \citenamefont {Micheli},\ and\ \citenamefont
  {Daley}}]{Schachenmeyer10}%
  \BibitemOpen
  \bibfield  {author} {\bibinfo {author} {\bibfnamefont {J.}~\bibnamefont
  {Schachenmayer}}, \bibinfo {author} {\bibfnamefont {I.}~\bibnamefont
  {Lesanovsky}}, \bibinfo {author} {\bibfnamefont {A.}~\bibnamefont {Micheli}},
  \ and\ \bibinfo {author} {\bibfnamefont {A.~J.}\ \bibnamefont {Daley}},\
  }\href {http://stacks.iop.org/1367-2630/12/i=10/a=103044} {\bibfield
  {journal} {\bibinfo  {journal} {New Journal of Physics}\ }\textbf {\bibinfo
  {volume} {12}},\ \bibinfo {pages} {103044} (\bibinfo {year}
  {2010}{\natexlab{b}})}\BibitemShut {NoStop}%
\bibitem [{\citenamefont {Beterov}\ \emph
  {et~al.}(2009{\natexlab{b}})\citenamefont {Beterov}, \citenamefont
  {Ryabtsev}, \citenamefont {Tretyakov},\ and\ \citenamefont
  {Entin}}]{decayrate}%
  \BibitemOpen
  \bibfield  {author} {\bibinfo {author} {\bibfnamefont {I.~I.}\ \bibnamefont
  {Beterov}}, \bibinfo {author} {\bibfnamefont {I.~I.}\ \bibnamefont
  {Ryabtsev}}, \bibinfo {author} {\bibfnamefont {D.~B.}\ \bibnamefont
  {Tretyakov}}, \ and\ \bibinfo {author} {\bibfnamefont {V.~M.}\ \bibnamefont
  {Entin}},\ }\href {\doibase 10.1103/PhysRevA.79.052504} {\bibfield  {journal}
  {\bibinfo  {journal} {Phys. Rev. A}\ }\textbf {\bibinfo {volume} {79}},\
  \bibinfo {pages} {052504} (\bibinfo {year} {2009}{\natexlab{b}})}\BibitemShut
  {NoStop}%
\bibitem [{\citenamefont {Weitenberg}\ \emph {et~al.}(2011)\citenamefont
  {Weitenberg}, \citenamefont {Endres}, \citenamefont {Sherson}, \citenamefont
  {Cheneau}, \citenamefont {Schausz}, \citenamefont {Fukuhara}, \citenamefont
  {Bloch},\ and\ \citenamefont {Kuhr}}]{Weitenberg11}%
  \BibitemOpen
  \bibfield  {author} {\bibinfo {author} {\bibfnamefont {C.}~\bibnamefont
  {Weitenberg}}, \bibinfo {author} {\bibfnamefont {M.}~\bibnamefont {Endres}},
  \bibinfo {author} {\bibfnamefont {J.~F.}\ \bibnamefont {Sherson}}, \bibinfo
  {author} {\bibfnamefont {M.}~\bibnamefont {Cheneau}}, \bibinfo {author}
  {\bibfnamefont {P.}~\bibnamefont {Schausz}}, \bibinfo {author} {\bibfnamefont
  {T.}~\bibnamefont {Fukuhara}}, \bibinfo {author} {\bibfnamefont
  {I.}~\bibnamefont {Bloch}}, \ and\ \bibinfo {author} {\bibfnamefont
  {S.}~\bibnamefont {Kuhr}},\ }\href {http://dx.doi.org/10.1038/nature09827}
  {\bibfield  {journal} {\bibinfo  {journal} {Nature}\ }\textbf {\bibinfo
  {volume} {471}},\ \bibinfo {pages} {319} (\bibinfo {year}
  {2011})}\BibitemShut {NoStop}%
\bibitem [{\citenamefont {G\"arttner}\ \emph {et~al.}(2014)\citenamefont
  {G\"arttner}, \citenamefont {Whitlock}, \citenamefont {Sch\"onleber},\ and\
  \citenamefont {Evers}}]{SA2014}%
  \BibitemOpen
  \bibfield  {author} {\bibinfo {author} {\bibfnamefont {M.}~\bibnamefont
  {G\"arttner}}, \bibinfo {author} {\bibfnamefont {S.}~\bibnamefont
  {Whitlock}}, \bibinfo {author} {\bibfnamefont {D.~W.}\ \bibnamefont
  {Sch\"onleber}}, \ and\ \bibinfo {author} {\bibfnamefont {J.}~\bibnamefont
  {Evers}},\ }\href {\doibase 10.1103/PhysRevLett.113.233002} {\bibfield
  {journal} {\bibinfo  {journal} {Phys. Rev. Lett.}\ }\textbf {\bibinfo
  {volume} {113}},\ \bibinfo {pages} {233002} (\bibinfo {year}
  {2014})}\BibitemShut {NoStop}%
\bibitem [{\citenamefont {Weber}\ \emph
  {et~al.}(2015{\natexlab{b}})\citenamefont {Weber}, \citenamefont {Honing},
  \citenamefont {Niederprum}, \citenamefont {Manthey}, \citenamefont {Thomas},
  \citenamefont {Guarrera}, \citenamefont {Fleischhauer}, \citenamefont
  {Barontini},\ and\ \citenamefont {Ott}}]{SA2015Fleischhauer}%
  \BibitemOpen
  \bibfield  {author} {\bibinfo {author} {\bibfnamefont {T.~M.}\ \bibnamefont
  {Weber}}, \bibinfo {author} {\bibfnamefont {M.}~\bibnamefont {Honing}},
  \bibinfo {author} {\bibfnamefont {T.}~\bibnamefont {Niederprum}}, \bibinfo
  {author} {\bibfnamefont {T.}~\bibnamefont {Manthey}}, \bibinfo {author}
  {\bibfnamefont {O.}~\bibnamefont {Thomas}}, \bibinfo {author} {\bibfnamefont
  {V.}~\bibnamefont {Guarrera}}, \bibinfo {author} {\bibfnamefont
  {M.}~\bibnamefont {Fleischhauer}}, \bibinfo {author} {\bibfnamefont
  {G.}~\bibnamefont {Barontini}}, \ and\ \bibinfo {author} {\bibfnamefont
  {H.}~\bibnamefont {Ott}},\ }\href {\doibase 10.1038/nphys3214} {\bibfield
  {journal} {\bibinfo  {journal} {Nat Phys}\ }\textbf {\bibinfo {volume}
  {11}},\ \bibinfo {pages} {157} (\bibinfo {year}
  {2015}{\natexlab{b}})}\BibitemShut {NoStop}%
\bibitem [{\citenamefont {Labuhn}\ \emph
  {et~al.}(2016{\natexlab{b}})\citenamefont {Labuhn}, \citenamefont {Barredo},
  \citenamefont {Ravets}, \citenamefont {de~L\'es\'eleuc}, \citenamefont
  {Macr\`{\i}}, \citenamefont {Lahaye},\ and\ \citenamefont
  {Browaeys}}]{SA2016Browaeys}%
  \BibitemOpen
  \bibfield  {author} {\bibinfo {author} {\bibfnamefont {H.}~\bibnamefont
  {Labuhn}}, \bibinfo {author} {\bibfnamefont {D.}~\bibnamefont {Barredo}},
  \bibinfo {author} {\bibfnamefont {S.}~\bibnamefont {Ravets}}, \bibinfo
  {author} {\bibfnamefont {S.}~\bibnamefont {de~L\'es\'eleuc}}, \bibinfo
  {author} {\bibfnamefont {T.}~\bibnamefont {Macr\`{\i}}}, \bibinfo {author}
  {\bibfnamefont {T.}~\bibnamefont {Lahaye}}, \ and\ \bibinfo {author}
  {\bibfnamefont {A.}~\bibnamefont {Browaeys}},\ }\href {\doibase
  10.1038/nature18274} {\bibfield  {journal} {\bibinfo  {journal} {Nature}\
  }\textbf {\bibinfo {volume} {534}},\ \bibinfo {pages} {667} (\bibinfo {year}
  {2016}{\natexlab{b}})}\BibitemShut {NoStop}%
\bibitem [{\citenamefont {Lovecchio}\ \emph {et~al.}(2015)\citenamefont
  {Lovecchio}, \citenamefont {Cherukattil}, \citenamefont {Cilenti},
  \citenamefont {Herrera}, \citenamefont {Cataliotti}, \citenamefont
  {Montangero}, \citenamefont {Calarco},\ and\ \citenamefont
  {Caruso}}]{FlorenceDetection}%
  \BibitemOpen
  \bibfield  {author} {\bibinfo {author} {\bibfnamefont {C.}~\bibnamefont
  {Lovecchio}}, \bibinfo {author} {\bibfnamefont {S.}~\bibnamefont
  {Cherukattil}}, \bibinfo {author} {\bibfnamefont {B.}~\bibnamefont
  {Cilenti}}, \bibinfo {author} {\bibfnamefont {I.}~\bibnamefont {Herrera}},
  \bibinfo {author} {\bibfnamefont {F.~S.}\ \bibnamefont {Cataliotti}},
  \bibinfo {author} {\bibfnamefont {S.}~\bibnamefont {Montangero}}, \bibinfo
  {author} {\bibfnamefont {T.}~\bibnamefont {Calarco}}, \ and\ \bibinfo
  {author} {\bibfnamefont {F.}~\bibnamefont {Caruso}},\ }\href
  {http://stacks.iop.org/1367-2630/17/i=9/a=093024} {\bibfield  {journal}
  {\bibinfo  {journal} {New Journal of Physics}\ }\textbf {\bibinfo {volume}
  {17}},\ \bibinfo {pages} {093024} (\bibinfo {year} {2015})}\BibitemShut
  {NoStop}%
\bibitem [{\citenamefont {Islam}\ \emph {et~al.}(2015)\citenamefont {Islam},
  \citenamefont {Ma}, \citenamefont {Preiss}, \citenamefont {Tai},
  \citenamefont {Lukin}, \citenamefont {Rispoli},\ and\ \citenamefont
  {Greiner}}]{Exp_purity}%
  \BibitemOpen
  \bibfield  {author} {\bibinfo {author} {\bibfnamefont {R.}~\bibnamefont
  {Islam}}, \bibinfo {author} {\bibfnamefont {R.}~\bibnamefont {Ma}}, \bibinfo
  {author} {\bibfnamefont {P.~M.}\ \bibnamefont {Preiss}}, \bibinfo {author}
  {\bibfnamefont {M.~E.}\ \bibnamefont {Tai}}, \bibinfo {author} {\bibfnamefont
  {A.}~\bibnamefont {Lukin}}, \bibinfo {author} {\bibfnamefont
  {M.}~\bibnamefont {Rispoli}}, \ and\ \bibinfo {author} {\bibfnamefont
  {M.}~\bibnamefont {Greiner}},\ }\href {http://dx.doi.org/10.1038/nature15750}
  {\bibfield  {journal} {\bibinfo  {journal} {Nature}\ }\textbf {\bibinfo
  {volume} {528}},\ \bibinfo {pages} {77} (\bibinfo {year} {2015})}\BibitemShut
  {NoStop}%
\bibitem [{\citenamefont {Zeiher}\ \emph {et~al.}(2017)\citenamefont {Zeiher},
  \citenamefont {Choi}, \citenamefont {Rubio-Abadal}, \citenamefont {Pohl},
  \citenamefont {Bijnen}, \citenamefont {Bloch},\ and\ \citenamefont
  {Gross}}]{Zeiher17}%
  \BibitemOpen
  \bibfield  {author} {\bibinfo {author} {\bibfnamefont {J.}~\bibnamefont
  {Zeiher}}, \bibinfo {author} {\bibfnamefont {J.-y.}\ \bibnamefont {Choi}},
  \bibinfo {author} {\bibfnamefont {A.}~\bibnamefont {Rubio-Abadal}}, \bibinfo
  {author} {\bibfnamefont {T.}~\bibnamefont {Pohl}}, \bibinfo {author}
  {\bibfnamefont {R.~v.}\ \bibnamefont {Bijnen}}, \bibinfo {author}
  {\bibfnamefont {I.}~\bibnamefont {Bloch}}, \ and\ \bibinfo {author}
  {\bibfnamefont {C.}~\bibnamefont {Gross}},\ }\href
  {https://arxiv.org/abs/1705.08372} {\bibfield  {journal} {\bibinfo  {journal}
  {ArXiv e-prints}\ } (\bibinfo {year} {2017})},\ \Eprint
  {http://arxiv.org/abs/1705.08372} {arXiv:1705.08372} \BibitemShut {NoStop}%
\end{thebibliography}%

\clearpage

\end{document}